\documentclass[11pt,a4paper]{article}
\pdfoutput=1

\usepackage[latin1]{inputenc}
\usepackage[english]{babel}
\usepackage{myway}
\usepackage{mathdots}
\usepackage{bigstrut}
\usepackage{dsfont}
\usepackage{bbold}
\usepackage{ulem}
\usepackage{hhline}
\usepackage[percent]{overpic}

\hypersetup{pdftitle={D7-Brane Moduli Space in Axion Monodromy and Fluxbrane Inflation}, 
pdfauthor={Maximilian Arends, Arthur Hebecker, Konrad Heimpel, Sebastian Kraus, Dieter L{\"u}st, Christoph Mayrhofer, Christoph Schick, Timo Weigand}, 
pdfsubject={String theory model building},
bookmarksopen, bookmarksnumbered, bookmarksopenlevel=2}

\newcommand{\Pp}{\mathds{P}}
\newcommand{\bft}{\mathbf{t}}

\newcommand{\bT}{\boldsymbol{T}}

\newcommand{\ccA}{\mathcal{A}}
\newcommand{\ccD}{\mathcal{D}}

\newcommand{\ccF}{\mathcal{F}}
\newcommand{\ccB}{\mathcal{B}}
\newcommand{\ccV}{\mathcal{V}}
\newcommand{\ccK}{\mathcal{K}}
\newcommand{\ccC}{\mathcal{C}}
\newcommand{\ccO}{\mathcal{O}}

\newcommand{\ccN}{\mathcal{N}}

\renewcommand{\Im}{\operatorname{Im}}
\renewcommand{\Re}{\operatorname{Re}}

\newcommand{\vp}{\varphi}
\def\ov{\overline}
\newcommand{\mkk}{m_{\mbox{\tiny KK}}}
\newcommand{\tK}{\widetilde{K3}}
\newcommand{\vol}{{\cal V}}
\newcommand{\tw}{\widetilde{\omega}}

\newcommand{\tPp}{\widetilde{\mathds{P}}}

\newcommand{\tE}{\widetilde{E}}

\newcommand{\tP}{\widetilde{\Pi}}

\newcommand{\gym}{g_{\text{YM}}}
\newcommand{\vd}{\ccV_{\text{D7}}}

\newcommand{\wt}[1]{\widetilde{#1}}
\def\d{\rm d}

\newcommand{\secref}[1]{section~\ref{#1}}
\newcommand{\Secref}[1]{Section~\ref{#1}}

\newcommand{\appref}[1]{appendix~\ref{#1}}

\title{D7-Brane Moduli Space in Axion Monodromy and Fluxbrane Inflation}

\preprint{MPP-2014-172\\
LMU-ASC 23/14\\
}

\author[1]{Maximilian Arends,}
\author[1]{Arthur Hebecker\note[ ]
{\href{mailto:A.Hebecker@ThPhys.Uni-Heidelberg.de}
{A.Hebecker@ThPhys.Uni-Heidelberg.de}},}
\author[1]{Konrad Heimpel,}
\author[1]{Sebastian C. Kraus\note[ ]
{\href{mailto:S.Kraus@ThPhys.Uni-Heidelberg.de}
{S.Kraus@ThPhys.Uni-Heidelberg.de}},}
\author[2,3]{Dieter L{\"u}st\note[ ]
{\href{mailto:Dieter.Luest@lmu.de}
{Dieter.Luest@lmu.de}},}
\author[1,2]{Christoph Mayrhofer\note[ ]
{\href{mailto:C.Mayrhofer@ThPhys.Uni-Heidelberg.de}
{C.Mayrhofer@ThPhys.Uni-Heidelberg.de}},}
\author[1]{Christoph Schick,}
\author[1]{and Timo Weigand\note[ ]
{\href{mailto:T.Weigand@ThPhys.Uni-Heidelberg.de}
{T.Weigand@ThPhys.Uni-Heidelberg.de}}}

\affiliation[1]{Institut f\"ur Theoretische Physik, Universit\"at Heidelberg, 
Philosophenweg 19, D-69120 Heidelberg\vspace{0.1cm}}
\affiliation[2]{Arnold-Sommerfeld-Center, Ludwig-Maximilians-Universit\"at, 
Theresienstrasse 33, \\  D-80333 M{\"u}nchen\vspace{0.1cm}}
\affiliation[3]{Max-Planck-Institut f\"ur Physik, F\"ohringer Ring 6, 
D-80805 M\"unchen\vspace{0.1cm}}

\abstract{
We analyze the quantum-corrected moduli space of D7-brane position moduli with special emphasis on inflationary model building. D7-brane deformation moduli are key players in two recently proposed inflationary scenarios: The first, D7-brane chaotic inflation, is a variant of axion monodromy inflation which allows for an effective 4d supergravity description. The second, fluxbrane inflation, is a stringy version of $D$-term hybrid inflation. Both proposals 
rely on the fact that D7-brane coordinates enjoy a shift-symmetric K\"ahler potential at large complex structure of the Calabi-Yau threefold, making them naturally lighter than other fields. This shift symmetry is inherited from the mirror-dual Type IIA Wilson line on a D6-brane at large volume. The inflaton mass can be provided by a tree-level term in the flux superpotential. It induces a monodromy and, if tuned to a sufficiently small value, can give rise to a large-field model of inflation. Alternatively, by a sensible
flux choice one can completely avoid a tree-level mass term, in which case the inflaton potential is induced via loop corrections. The positive vacuum energy can then be provided by a $D$-term, leading to a small-field model of hybrid natural inflation. In the present paper, we continue to develop a detailed understanding of the D7-brane moduli space focusing among others on shift-symmetry-preserving flux choices, flux-induced superpotential in Type IIB/F-theory language, and loop 
corrections. While the inflationary applications represent our main physics motivation, we expect that some of our findings will be useful for other phenomenological issues involving 7-branes in Type IIB/F-theory constructions.
}

\begin{document}

\noindent July 7, 2014

\vspace*{-0.8 cm}
\maketitle

\section{Introduction}
\label{Introduction}
This paper discusses approximately flat directions in the moduli space of D7-branes. Our investigation can be motivated in two fairly independent ways -- one more physical, the other more geometric.

On the physics side, there are two broad classes of inflation models: First, in the so-called large-field models the inflaton traverses a super-planckian distance in field space during inflation, $\Delta \vp > M_p$. By contrast, in small-field models the field range remains sub-planckian,  $\Delta \vp < M_p$. These two classes have distinguished features, e.g., in small-field inflation the amplitude of gravitational waves relative to the amplitude of scalar perturbations is typically tiny \cite{Lyth:1996im,Boubekeur:2005zm}. On the other hand, some of the crucial model-building ingredients are similar: In both classes a specific structure of the K\"ahler potential (e.g.\ a shift-symmetric form) is required to avoid the supergravity $\eta$-problem. Furthermore, a detailed understanding of the inflaton-dependence in the superpotential is needed in order to analyze whether the phenomenologically required features are available in given string compactifications. 
In fact, it can be {\it the same} modulus which realizes either large-field or small-field inflation, depending on the details of the model: In `D7-brane chaotic inflation' \cite{Hebecker:2014eua} a monodromy in the moduli space of a D7-brane position is used to obtain a large-field model of inflation. On the other hand, `fluxbrane inflation' \cite{Hebecker:2011hk,Hebecker:2012aw} uses the distance of two D7-branes to realize a stringy version of hybrid natural inflation, a small-field inflation model. The shift-symmetric K\"ahler potential for the D7-brane modulus is inherited from the mirror-dual Type IIA picture of a Wilson line on a D6-brane. The merit of working with D7-branes in IIB lies in the existence of a rather detailed understanding of moduli stabilization in these compactifications \cite{Giddings:2001yu,Kachru:2003aw,Balasubramanian:2005zx,Westphal:2006tn,Rummel:2011cd}. In the following we outline the two D7-brane inflation proposals in slightly more detail.

In `D7-brane chaotic inflation' \cite{Hebecker:2014eua} the leading shift-symmetry-breaking term in the scalar potential arises due to the choice of a certain background flux which leads to an explicit appearance of the inflaton in the tree-level superpotential. This flux `unfolds' the a priori periodic field space of the D7-brane position modulus and allows for effectively super-planckian field excursions during inflation. This is along the lines of axion monodromy inflation \cite{Silverstein:2008sg,McAllister:2008hb,Berg:2009tg,Palti:2014kza},\footnote{For a field theory realization see \cite{Kaloper:2008fb,Kaloper:2011jz,Dubovsky:2011tu,Lawrence:2012ua,Kaloper:2014zba}. For large-field axion-type models without monodromy see~\cite{Kim:2004rp,Dimopoulos:2005ac,Kallosh:2007cc,Grimm:2007hs,Ashoorioon:2009wa,Ashoorioon:2011ki,Chatzistavrakidis:2012bb} and especially \cite{Cicoli:2014sva,Blumenhagen:2014gta,Grimm:2014vva,Choi:2014rja,Tye:2014tja,Kappl:2014lra,Bachlechner:2014hsa,Ben-Dayan:2014zsa,Long:2014dta} 
for recent developments. For a recent large-field proposal in the non-geometric context see \cite{Hassler:2014mla}.} however, with 
spontaneous 
rather than explicit supersymmetry breaking. The vacuum energy during inflation is provided by non-vanishing $F$-terms for the D7-brane modulus. As a result, in
D7-brane chaotic inflation the control issues due to the presence of anti-branes are absent. Similar ideas for $F$-term inflation using axion monodromy have been put 
forward simultaneously in \cite{Marchesano:2014mla,Blumenhagen:2014gta} (see also \cite{Ibanez:2014kia}). A particular strength of the D7-brane chaotic inflation model is that the coefficients of the inflaton in the superpotential can in principle be tuned small by a suitable flux choice. This is required in order to avoid destabilization in the K\"ahler moduli directions.

`Fluxbrane inflation' \cite{Hebecker:2011hk,Hebecker:2012aw}, on the other hand, is the attempt to build a non-fine-tuned model of inflation in string theory.\footnote{Of 
course, given that we have mostly accepted a fine-tuned cosmological constant and one may be forced to accept a fine-tuned weak scale in the not too distant future, it is hard to make a case against fine-tuned inflation in string theory \cite{Kachru:2003sx,Baumann:2008kq}. Furthermore, there are certainly suggestions to ensure a sufficiently flat potential by means other than a shift symmetry (see e.g.\ \cite{Conlon:2005jm,Cicoli:2008gp,Gaillard:1995az,Antusch:2008pn,Antusch:2011ei}). Nevertheless, implementing hybrid inflation with a shift symmetry in string theory appears to us to be a worthwhile endeavor.
} From a 4d field-theory perspective, the combination of hybrid inflation in supergravity \cite{Linde:1993cn,Dvali:1994ms,Copeland:1994vg,Halyo:1996pp,Binetruy:1996xj} with a shift symmetry \cite{Freese:1990rb} protecting the flatness of the potential appears particularly appealing \cite{Cohn:2000hc,Stewart:2000pa,Kaplan:2003aj,ArkaniHamed:2003mz,Ross:2009hg,Ross:2010fg,Hebecker:2013jm}. In principle, this has a straightforward stringy realization in the form of (D6-brane) Wilson line inflation \cite{Avgoustidis:2006zp,Avgoustidis:2008zu}. Again, given that moduli stabilization is much better understood in Type IIB, we are naturally lead to consider the mirror-dual setting, using D7-brane position moduli as the inflaton. In the fluxbrane inflation scenario, the relative deformation of two homologous D7-branes plays the role of the inflaton. The energy density during inflation is provided by a $D$-term which appears due to the choice of a supersymmetry-breaking flux on the D7-branes. Inflation ends,
 as in hybrid models, when a certain brane flux decays in a tachyonic instability. Hence, this can be viewed as a stringy version of `hybrid natural inflation' \cite{Cohn:2000hc,Stewart:2000pa,Kaplan:2003aj,ArkaniHamed:2003mz,Ross:2009hg,Ross:2010fg,Hebecker:2013jm}. No D3-branes  
are necessary in both models. It is apparent that the study of all kinds of corrections which affect the flat leading-order D7-brane potential is mandatory to establish or dismiss these scenarios.

From a more geometric perspective of the general study of string compactifications, our investigation can be viewed as follows: As is well-known, the brane fluxes in Type IIB Calabi-Yau orientifolds can be chosen in such a way that certain D7-brane positions are left unfixed \cite{Jockers:2005zy}. The resulting light brane scalars are then by no means massless but are rather stabilized by the interplay of a number of sub-leading effects. These include the mirror dual of Type IIA open-string instantons \cite{Kachru:2000ih}, gauge theory loops, and the indirect effects of closed-string fluxes on brane positions \cite{Denef:2008wq,Gomis:2005wc}. Developing a consistent overall picture of the resulting brane stabilization is certainly interesting and may have applications beyond the specific inflationary scenarios advertised earlier. For example, light brane scalars may play some other role in cosmology (as a curvaton field or during reheating) or they may even be part of the visible sector (e.g.\ 
in the form 
of 
a shift-symmetric Higgs field \cite{Hebecker:2012qp,Ibanez:2012zg,Hebecker:2013lha}).

\subsection{D7-Brane Inflation in Light of the \mbox{BICEP2} Results}
Recently, the BICEP2 collaboration has reported the measurement of B-mode polarization \cite{Ade:2014xna}. They claim that the measurement is well fit by the B-mode spectrum sourced by primordial gravitational waves which are produced during an epoch of slow-roll inflation. The corresponding amplitude of primordial tensor perturbations relative to the amplitude of scalar perturbations is given by $r = 0.2_{-0.05}^{+0.07}$.

B-modes are sourced by various effects (see e.g.\ \cite{Cook:2011hg,Senatore:2011sp,Barnaby:2011qe,Barnaby:2012xt,Carney:2012pk}). For example, it was shown in \cite{Moss:2014cra} that the conclusion of the BICEP2 team that $r=0$ is ruled out with high significance is altered if one includes cosmic strings in the model (see, however, \cite{Lizarraga:2014eaa}).\footnote{Another interesting issue has been raised in \cite{Liu:2014mpa}, where it was stated that `radio loops' may dominate over the primordial B-mode signal in some regions of the sky.} We believe that, while the attribution of the B-mode signal to primordial tensor modes is tempting, it will take additional time and effort to prove this claim and reliably exclude other possible sources.

The predicted value of the tensor-to-scalar ratio $r$ in the two D7-brane inflation scenarios outlined above is certainly one important feature which phenomenologically distinguishes the models from each other. In particular, if the measurement in \cite{Ade:2014xna} and its attribution to primordial gravitational waves is correct, this would imply that models of small-field inflation, such as fluxbrane inflation \cite{Hebecker:2011hk,Hebecker:2012aw}, are ruled out \cite{Lyth:1996im,Boubekeur:2005zm} (see, however, \cite{Hui:2001ce,Ashoorioon:2013eia,Collins:2014yua,Aravind:2014axa,Choudhury:2013iaa,Choudhury:2014kma,Antusch:2014cpa} and references therein). On the other hand, in such a situation D7-brane chaotic inflation \cite{Hebecker:2014eua} looks very promising: The leading order inflaton potential in this model takes a quadratic form, well-known since the early proposal of chaotic inflation \cite{Linde:1983gd}. Correspondingly, the tensor-to-scalar ratio is large, $r\simeq 0.16$, in reasonable 
agreement with the BICEP2 results. 
Confirmation or rejection of the gravitational wave signal is thus crucial to be able to tell whether D7-brane chaotic inflation in the form proposed in \cite{Hebecker:2014eua} is phenomenologically viable.

Let us finally put forward a further consideration: Suppose there are string models which realize small-field inflation (or, more generally, $r \ll 0.1$) in a non-fine-tuned manner, possible candidates including the fluxbrane inflation model and some K\"ahler moduli inflation models \cite{Cicoli:2008gp}. Further let us assume that stringy realizations of large-field inflation which also give a large $r$ are fine-tuned (as far as we understand, this is the case for the so far proposed models, including our D7-brane chaotic inflation scenario). A confirmation of the measurement of $r$ would then tell us that, from the various available models in the string landscape, nature chose a tuned one. This observation may lead to interesting implications for the landscape of flux vacua.

We take all these arguments as motivation to investigate in detail the moduli space of D7-branes. As emphasized before, the mechanisms investigated here, such as the shift symmetry or the `extended no-scale' structure, are crucial ingredients in both models of D7-brane inflation and, furthermore, universal features of Type IIB string compactifications and are therefore of importance from different perspectives.

\subsection{Brief Summary of our Results}
The viability of the D7-brane inflation models rests on three pillars: The existence of a shift symmetry in the K\"ahler potential for the D7-brane moduli, the ability to choose fluxes in order for the D7-brane position moduli either not to show up at all or have a small coefficient in the superpotential, and the fact that loop corrections respect the `extended no-scale' structure, even if an additional light degree of freedom, namely the D7-brane modulus, is included in the effective theory after stabilization of axio-dilaton and complex structure moduli. We will briefly summarize our view on those three pillars.

In the low energy effective theory arising from compactifying Type IIB string theory to four dimensions, let $c$ denote a complex scalar which describes a D7-brane deformation modulus. Such deformation modes are known to enter the K\"ahler potential in the form $K \supset -\ln\left(-i(S-\ov S) - k_{\text{D7}}(z,\ov z; c, \ov c)\right)$ \cite{Jockers:2004yj,Jockers:2005zy}. Here, $S$ is the axio-dilaton and $z$ collectively denotes complex structure moduli of the Calabi-Yau threefold. The existence of a shift symmetry in the moduli space of brane deformations is manifest if the K\"ahler potential takes the form
\begin{equation}
 K \supset -\ln\left(-i(S-\ov S) - k_{\text{D7}}(z,\ov z; c-\ov c)\right).
\end{equation}
If this is the case, the K\"ahler potential will be invariant under $c \to c + \delta, \ \delta \in \mathds{R}$, and the inflaton will be associated with the real part of $c$, i.e.\ $\varphi \sim \Re (c)$. The existence of this shift symmetry is crucial in order to avoid the well-known supergravity $\eta$-problem. The presence of the shift symmetry is expected in the vicinity of the `large complex structure' point and can be understood from different viewpoints: Via mirror symmetry the brane deformation modulus, corresponding to the inflaton, is mapped to a Wilson line on a D6-brane in a Type IIA string compactification. At leading order the IIA K\"ahler potential does not depend on the Wilson line, a structure which is broken by non-perturbative effects due to worldsheet instantons. These effects are expected to be small at large volume on the Type IIA side, which corresponds to the large complex structure limit on the Type IIB side. On the other hand, the K\"ahler potential of an F-theory compactification 
exhibits a shift symmetry for the fourfold complex structure moduli in the vicinity of the large complex structure point of the fourfold. In the weak coupling limit, this shift symmetry persists as a symmetry in the D7-brane sector. These issues are illustrated in the explicit example of a compactification of F-theory on $\tK \times K3 $ in \secref{Flat Directions} and are discussed more generally in \secref{Mirror-Symmetry}.

The second crucial requirement of D7-brane inflation is the choice of a non-generic flux which either leads to a very weak stabilization of some brane moduli, or leaves some of them unstabilized completely. In this context we make a more general contribution to the discussion of brane stabilization: Starting from the well-known expression for the F-theory superpotential we go to the weak coupling limit and try to reproduce the bulk and brane superpotential for the corresponding IIB compactification. We find in \secref{ModStabGen} that, {\it in addition} to the term
\begin{equation}\label{eq:StandardBranePotential}
 \wt{W}_{\text{D7}} = \int_{\Gamma_5} \ccF_2 \wedge \Omega_3 , \quad \ccF_2 := F_2 - B_2,
\end{equation}
there is an explicit appearance of the brane coordinate in the superpotential, which is non-zero even if the flux $\ccF$ is of type $(1,1)$ and therefore the term in \eqref{eq:StandardBranePotential} vanishes identically. In the above expression, $\Gamma_5$ is the five-chain swept out by two D7-branes as they are deformed into each other, $\Omega_3$ is the pullback of the holomorphic three-form to the brane cycle, $B_2$ is the Kalb-Ramond field and $F_2$ is the brane flux.

As a result, even if no brane flux is present the D7-brane coordinates may appear in the superpotential, leading to a stabilization of the branes via leading order $F$-terms. This is well-known in explicit examples, e.g.\ the compactification of F-theory on $\tK \times K3 $ \cite{Lust:2005bd} and its corresponding Type IIB limit. It is due to the fact that the fourfold periods, which, in the orientifold limit reduce to bulk complex structure moduli and the axio-dilaton, have a brane moduli dependence as soon as the D7-branes are pulled off the O-plane. Accordingly, part of \secref{Flat Directions} is devoted to investigating this explicit example and specifying a flux which does not stabilize the D7-branes. In the same spirit, we analyze implications for the `open-string landscape' \cite{Gomis:2005wc} which arise due to the observed `brane backreaction'. We conclude that it is certainly possible to choose a brane flux preserving the masslessness of D7-brane deformations in the tree-level effective action. 
Though, 
we did not manage to find an explicit flux in the $\tK \times K3 $ example which, at the same time, stabilizes both the axio-dilaton and all complex structure moduli.

The shift symmetry of the K\"ahler potential for the brane coordinate will certainly be broken by loop corrections due to interactions of the inflaton with other open-string states (e.g.\ waterfall fields in the case of fluxbrane inflation) in the superpotential. Loop corrections to the K\"ahler potential have been computed in \cite{vonGersdorff:2005bf,Berg:2005ja}, and their use in models of K\"ahler moduli stabilization has been analyzed in \cite{Berg:2005yu,Berg:2007wt,Cicoli:2007xp}. The usual assumption in these works 
is that all moduli except for the K\"ahler moduli are stabilized supersymmetrically by their respective $F$-terms. If this is the case, it is well-known that loop corrections feature the `extended no-scale' structure, which makes them subleading with respect to the $\alpha'^3$-corrections \cite{Becker:2002nn}. This is an important prerequisite for K\"ahler moduli stabilization in the Large Volume Scenario to work. It is thus crucial to ensure, that this 
structure is not spoiled if one includes an additional 
light degree of freedom, the inflaton, in the effective 
theory below the scale where complex structure moduli and the axio-dilaton are stabilized. Explicitly demonstrating this will be the subject of \secref{String Loop Corrections}.

The focus of the phenomenological sections \ref{d7d7} and \ref{sec:pheno} in this paper will be on the fluxbrane inflation model \cite{Hebecker:2011hk,Hebecker:2012aw}, the main reason being that, to date, a consistent overall picture of this model, including the parametric size of loop corrections in a scenario where moduli stabilization is taken into account, is still missing.
By contrast, the parametric situation in D7-brane chaotic inflation has already been analyzed in some detail in \cite{Hebecker:2014eua}, including moduli stabilization and corrections (referring to some of the results obtained in the present paper). Thus, besides providing a more detailed investigation of the ingredients used in \cite{Hebecker:2014eua}, we aim towards a 
parametrically controlled realization of fluxbrane inflation.
The latter is achieved in \secref{sec:pheno}, where the phenomenological implications of the fluxbrane inflation model are thoroughly discussed. Our focus is on the size of the loop-induced inflaton dependence of the scalar potential relative to the (constant) potential energy density of the waterfall fields during inflation. We find that the suppression of loop corrections due to the exchange of Kaluza-Klein modes between the two D7-branes is sufficient to be able to reproduce the required relative size. On the other hand, loop corrections due to winding modes of the strings around potential one-cycles of D7-brane intersections are on the verge of being too large. It is then a matter of model-dependent $\ccO(1)$-factors (which we neglect in this paper) whether or not a given scenario is viable. To be on the safe side, we also consider compactifications in which the self-intersection of the divisor wrapped by the D7-branes 
responsible for inflation is empty or contains no
non-trivial one-cycle. The fluxbrane inflation scenario is able to reproduce the correct value for the spectral index, the number of $e$-foldings, and the amplitude of curvature 
perturbations. It satisfies the cosmic string bound and the running of the spectral index is moderately small, $n_s' \lesssim 10^{-2}$. The tensor-to-scalar ratio is tiny, $r \lesssim 2.6 \cdot 10^{-5}$.

In the fluxbrane inflation model, we assume that relative sizes of four-cycles are stabilized by the condition of vanishing $D$-terms. The uplift of the AdS vacuum obtained in the Large Volume Scenario to a Minkowski vacuum is achieved by a further, non-vanishing $D$-term as in \cite{Hebecker:2012aw}. This $D$-term has to be tuned to a small value (since it is generically dominant in the scalar potential of the Large Volume Scenario), which is realized by tuning the position in K\"ahler moduli space. Thus, the $D$-term tuning is part of our version of the $D$-term uplifting proposal. This tuning is only slightly worsened by insisting on an inflationary model which relies on a $D$-term for a different U(1) theory, but involving the same combination of K\"ahler moduli.\footnote{Realizing inflation with a $D$-term involving a {\it different} combination of K\"ahler moduli does not help since it requires a further tuning.} Importantly, in fluxbrane inflation no additional fine-tuning is needed in order to have a 
small $\eta$-parameter.

\Secref{statistics} contains a brief comment on the statistics of vacua at large complex structure. It was argued \cite{Denef:2004dm} that such vacua are statistically heavily disfavored. While we do not question this general result, we disagree with the pessimistic view regarding the existence of any vacua in the vicinity of the large complex structure point, in particular for models with few complex structure moduli. Furthermore, including the (generally) huge number of possible brane fluxes, we expect the number of vacua at large complex structure to be sizable, potentially leaving enough room for tuning the cosmological constant along the lines of \cite{Bousso:2000xa}.

Before getting into a detailed discussion of the issues outlined above, we start in \secref{d7d7} by recalling the basic setting for fluxbrane inflation
and the demands it has for the dynamics of the two relevant D7-branes. In particular, in \secref{pheno} we discuss the D7-D7 moduli space and its generic scalar potential in more detail than in our previous publications on the subject~\cite{Hebecker:2011hk,Hebecker:2012aw}. The corresponding discussion for the case of D7-brane chaotic inflation can be found in \cite{Hebecker:2014eua}.

\section{Supersymmetric Hybrid Natural Inflation and its Stringy Embedding}
\label{d7d7}
Fluxbrane inflation \cite{Hebecker:2011hk,Hebecker:2012aw} is a stringy version of supersymmetric hybrid natural inflation \cite{Ross:2009hg,Ross:2010fg}. In this model, a certain relative deformation modulus of two D7-branes is associated with the inflaton. One crucial feature of hybrid natural inflation is a shift symmetry which forbids dangerous mass terms for the inflaton. Such a shift symmetry can arise if the holomorphic variable $c$, which describes the D7-brane deformation associated with the inflaton, enters the K\"ahler potential only in the form $(c - \ov c)$. In addition, the fluxes have to be chosen such that there is no explicit dependence on $c$ in the superpotential. Consequently, the scalar potential will be invariant under
\begin{equation}
 c \to c + \delta, \quad \delta \in \mathds{R}.
\end{equation}
The shift symmetry will be broken by couplings of the inflaton to other sectors of the theory such as, for example, zero modes of open strings which stretch between those D7-branes (`waterfall fields'). Additionally, non-perturbative effects will break the continuous symmetry. Supersymmetry is required in order to keep the size of the perturbative quantum corrections under control. Since the field-space of the D7-brane modulus is periodic, the resulting potential will be periodic (i.e.\ the shift symmetry is broken to a discrete subgroup) and can be parametrized, at leading order, as
\begin{equation}\label{cosPot}
 V(\varphi) = V_0 \left(1- \alpha \cos \left(\frac{\varphi}{f}\right)\right),
\end{equation}
where $\varphi \sim \Re(c)$ is the canonically normalized inflaton.

\subsection{Phenomenological Constraints}
\label{PhenoConstraints}
The potential \eqref{cosPot} has to satisfy a number of phenomenological constraints. From \eqref{cosPot} one easily computes the slow-roll parameters at the beginning of the last $N$ $e$-folds in the limit $|\alpha| \ll 1$ as
\begin{align}\label{eq:slow-roll-parameters}
\begin{split}
 \epsilon &\equiv \frac{1}{2}\left(\frac{V'}{V}\right)^2 \simeq \frac{1}{2}\left(\frac{\alpha}{f}\right)^2 \sin^2 \left(\frac{\varphi_N}{f}\right),\\
 \eta &\equiv \frac{V''}{V} \simeq \frac{\alpha}{f^2} \cos \left(\frac{\varphi_N}{f}\right),\\
 \tilde\xi^2 &\equiv - \frac{V'V'''}{V^2} \simeq \frac{2\epsilon}{f^2}.
\end{split}
\end{align}
While the inflaton rolls from $\varphi_N$ to $\varphi_0$ the universe undergoes an accelerated expansion with the number of $e$-folds given by
\begin{equation}\label{eq:e-folds}
 N \equiv \int_{\varphi_0}^{\varphi_N} \frac{{\d}\varphi}{\sqrt{2\epsilon}} \simeq \frac{f^2}{\alpha} \ln\left(\frac{\tan\left(\frac{\varphi_N}{2f}\right)}{\tan\left(\frac{\varphi_0}{2f}\right)}\right).
\end{equation}

\begin{figure}
\centering
 \begin{overpic}[width=0.5\textwidth,tics=10]{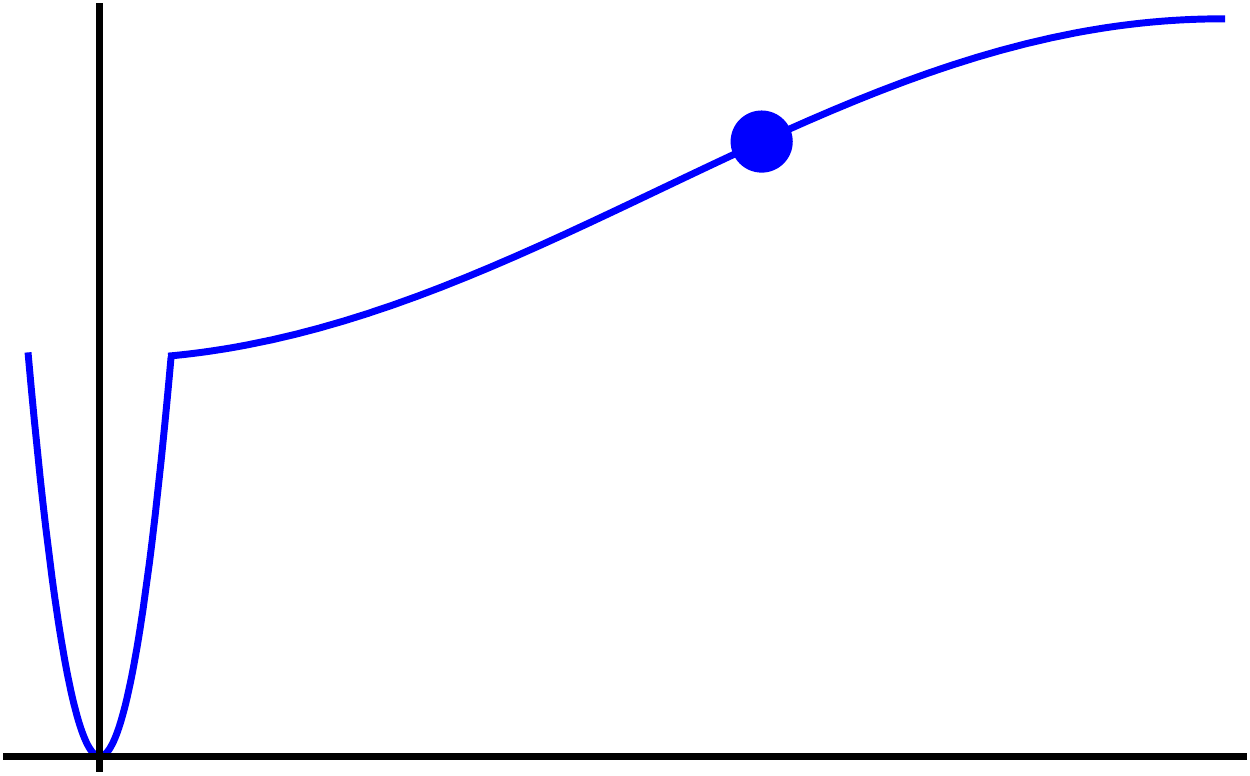}
 \put (102,1) {$\varphi$} \put (3,65) {$V\left(\varphi\right)$}
\end{overpic}
\caption{Plot of the potential \eqref{cosPot}. For illustrative purposes the relative size of the variation with respect to the constant is exaggerated.}\label{CubicPot}
\end{figure}

Being a variant of hybrid inflation, fluxbrane inflation ends when the mass-square of the waterfall field becomes tachyonic. Figure \ref{CubicPot} shows a schematic plot of the potential, including the waterfall transition. The tachyon exists due to the presence of supersymmetry-breaking brane flux which leads to a $D$-term in the effective theory. During inflation, when the waterfall fields are stabilized at zero vev, the $D$-term is given by
\begin{equation}\label{eq:D-termFieldTheory}
 V_D = \frac{g^2_{\text{YM}} \xi^2}{2} \equiv V_0,
\end{equation}
where $g_{\text{YM}}$ is the coupling of the gauge theory living on the branes and $\xi$ is the Fayet-Iliopoulos parameter.
In the present subsection we will assume that the system enters the waterfall regime at $\varphi_0$, i.e., at this point in field space the tachyon appears in the spectrum. This can be achieved by adjusting the coupling of the inflaton to the waterfall fields appropriately. Note, however, that in our stringy realization of the hybrid natural inflation model there is a relation between this superpotential coupling and the gauge coupling constant. This relation is a remnant of an underlying $\mathcal N =2$ supersymmetry \cite{Kallosh:2001tm,Kallosh:2003ux,Hebecker:2011hk}. As a consequence, $\varphi_0$ will eventually be set by the FI-parameter $\xi$, with no further model building freedom. We will further discuss this in the next subsection.

The model, as described above, can thus be characterized by the parameters $\alpha$, $f$, $V_0$, $\varphi_0$, and $g^2_{\text{YM}}$. The quantity $\varphi_N$ is then adjusted in order to satisfy phenomenological requirements. The model parameters are constrained by experiment \cite{Ade:2013uln} as
\begin{align}\label{phenoConstr}
\begin{split}
 N &\simeq 60,\\
 n_s  &\simeq 1-6\epsilon + 2 \eta \simeq 0.9603 \pm 0.0073,\\
 \tilde \zeta &= \frac{V^{3/2}}{V'} \simeq \sqrt{\frac{V_0}{2\epsilon}} \simeq 5.10\times 10^{-4},\\
 n_s' &\equiv \frac{{\d} n_s}{{\d} \ln k} =  16 \epsilon \eta - 24 \epsilon^2 + 2 \tilde\xi^2  = -0.0134 \pm 0.0090.
\end{split}
\end{align}

Generically, during tachyon condensation, cosmic strings will form. If existent, those topological defects will leave an imprint on the CMB spectrum which can, in principle, be measured. The fact that no such signal has been observed yet constrains the (dimensionless) cosmic string tension $G\mu = \xi/4$ as
\begin{equation}\label{CSbound}
 \frac{\xi}{4} \lesssim 1.3 \times 10^{-7}, \hspace{0.3cm} \text{(see \cite{Ade:2013xla}).}
\end{equation}
Note that this bound depends on various things, such as the way in which the cosmic string network is modeled as well as the dataset used for constraining the string tension. In \eqref{CSbound} we quote the most stringent bound from \cite{Ade:2013xla}.

\subsection{Embedding Hybrid Natural Inflation in String Theory}\label{pheno}
Given the model parameters in the field theory description, how do they relate to the parameters of the underlying string embedding? In order to answer this question, let us give the following intuitive general picture of how we think fluxbrane inflation works:
\begin{figure}
\centering
 \begin{overpic}[width=0.6\textwidth,tics=10]{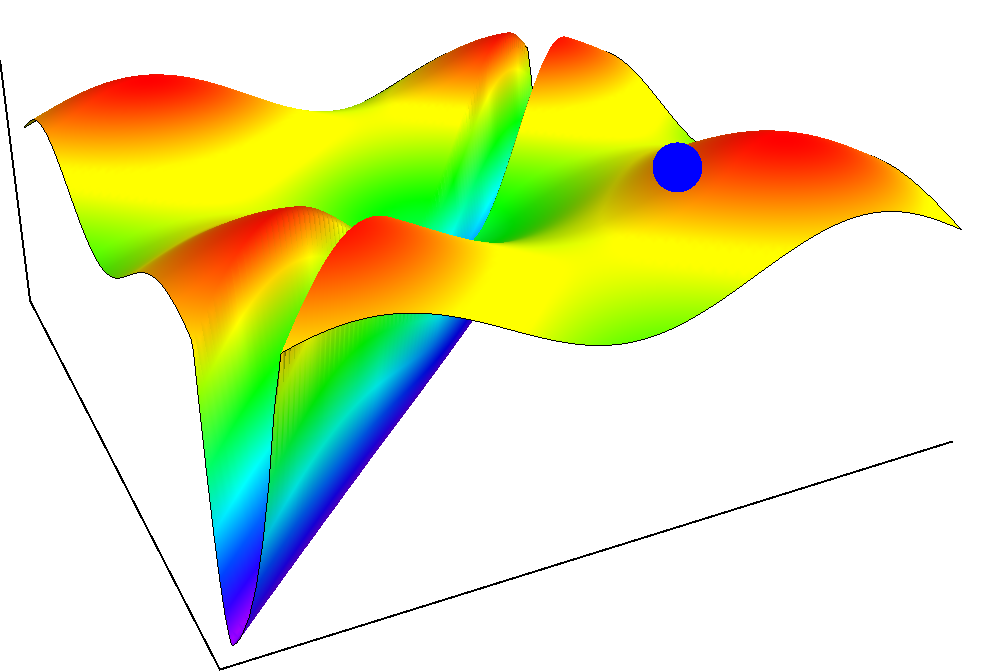}
 \put (5,18) {$\varphi_2$} \put (62,8) {$\varphi_1$} \put (-19,50) {$V\left(\varphi_1 , \varphi_2\right)$}
\end{overpic}
\caption{Plot of the combined potential $\delta V = \sum_{i}\tilde\alpha_i \sin( \varphi_i/f)$}\label{genPot}
\end{figure}
In the simplest setup two D7-branes, whose positions are encoded by the vevs of two fields $c_i$, $i=1,2$, will move in the transverse space along a $S^1$ with circumference $R$.\footnote{We choose the convention $\ell_s = 2\pi\sqrt{\alpha'}$ and measure all lengths in the ten-dimensional Einstein frame, i.e.\ after rescaling the metric $g_{MN}^s = e^{\phi/2}g_{MN}^E$, where $g_s = \langle e^{\phi}\rangle$.} This $S^1$ corresponds to the directions $\Re(c_i)$ in field space, along which the leading order potential is flat. At subleading order, this potential receives periodic corrections which we assume to be, at lowest order, $\delta V_i = \tilde\alpha_i \sin(\varphi_i/f)$. Here, the $\varphi_i \sim \Re(c_i)$ are canonically normalized fields and the field displacement $\Delta \varphi_i = 2\pi f$ corresponds to shifting one of the D7-branes once around the $S^1$. The total potential, which is displayed in figure~\ref{genPot}, is then a function of both $\varphi_i$, $i=1,2$. It is thus clear that a `generic' 
trajectory of the canonically normalized inflaton $\varphi$, corresponding to the distance of the two D7-branes, can be parametrized, at leading order, by \eqref{cosPot} (see figure~\ref{CubicPot}).

The D7-branes will wrap a four-cycle whose volume we denote by $\ccV_{\text{D7}}$. Furthermore, $\ccV$ will be the volume of the whole Calabi-Yau. The circumference $R$ of the transverse periodic direction, along which the D7-branes are separated, can be translated into the size of the field space for the canonically normalized inflaton \cite{Hebecker:2011hk}:
\begin{equation}\label{phimax}
2\pi f \simeq R\sqrt{\frac{g_s}{4}\frac{\ccV_{\text{D7}}}{\ccV}} \equiv \frac{1}{2}\sqrt{\frac{g_s}{z}}.
\end{equation}
This equation defines the `complex structure modulus' $z = \ccV /(\ccV_{\text{D7}} R^{2})$.

In fluxbrane inflation \cite{Hebecker:2011hk,Hebecker:2012aw}, a supersymmetry-breaking flux configuration on the D7-branes leads to the appearance of a $D$-term in the effective action. This $D$-term gives rise to a tachyonic mass term for the waterfall field. It reads
\begin{equation}\label{D-termPot}
 V_D = \frac{g_{\text{YM}}^2 \xi^2}{2}, \quad\text{where}\quad g_{\text{YM}}^2 = \frac{2\pi}{\ccV_{\text{D7}}}, \quad   \xi = \frac{1}{4\pi}\frac{\int J\wedge \ccF }{\ccV} \equiv \frac{x}{4\pi}\frac{\sqrt{\ccV_{\text{D7}}}}{\ccV}.
\end{equation}
The last equation defines $x = \int J\wedge \ccF /\sqrt{\ccV_{\text{D7}}}$.
Finally, the point where the tachyon condensation sets in is given by \cite{Hebecker:2011hk}
\begin{equation}\label{eq:phi0}
 \varphi_0 = \frac{\sqrt{\xi}}{2^{1/4}} = \frac{1}{2^{1/4}}\sqrt{\frac{x}{4\pi}}\sqrt{\frac{\sqrt{\ccV_{\text{D7}}}}{\ccV}}.
\end{equation}

We have thus identified $f$, $V_0$, $\varphi_0$, and $g^2_{\text{YM}}$ in terms of quantities which generically parametrize the fluxbrane inflation scenario. The crucial and much more involved issue is to derive an expression for $\alpha$ in terms of stringy model parameters. To obtain such an expression we need to understand terms which violate the shift symmetry. This is the topic of the subsequent sections. We will return to the phenomenology of the fluxbrane inflation model in \secref{sec:pheno}.

\section{Simple Toy Models}
\label{Flat Directions}
Before analyzing effects which break the shift symmetry we should make sure that this is actually a sensible thing to do, i.e.\ that there are examples in which there is a shift symmetry in the D7-brane moduli space which remains intact even after turning on fluxes.  Discussing such examples will be the subject of this section.

\subsection[Explicit Example with Flat Directions and $W \neq 0$]{Explicit Example with Flat Directions and {\boldmath $W \neq 0$}}\label{modulispacek3}
We start with F-theory on $\tK \times K3$. It is well-known that, in this example, the 7-brane coordinate enjoys a shift symmetry in the K\"ahler potential \cite{Gorlich:2004qm,Lust:2005bd,Dasgupta:1999ss}. We now demonstrate explicitly that, for a suitable flux choice, we can achieve $W \neq 0 $, $W$ being the tree-level superpotential, in such a way that the scalar potential actually respects the shift symmetry. In this section we closely follow the notation of \cite{Braun:2008ua,Braun:2008pz}. For a comprehensive review on $K3$ and for references to the mathematical literature we refer to \appref{k3surface} and~\cite{Aspinwall:1996mn}.

\subsubsection[Complex Structure Moduli Space of $\tK \times K3$]{Complex Structure Moduli Space of {\boldmath $\tK \times K3$}}
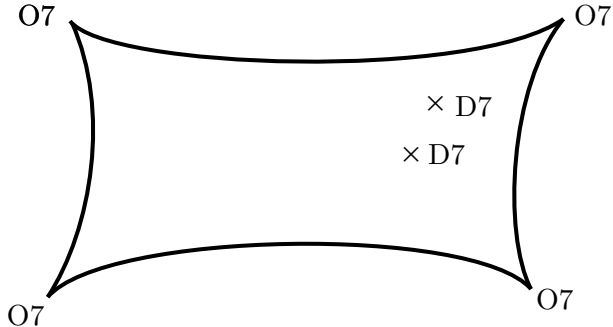
\begin{figure}
\centering
\begin{overpic}[width=0.47\textwidth,tics=10]{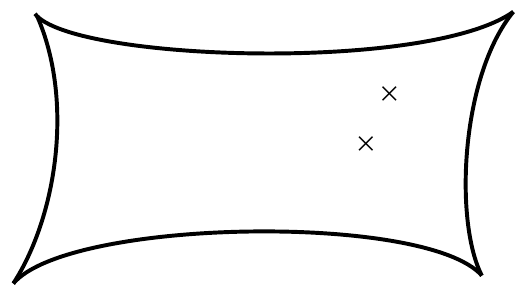}
 \put (-3,53) {O7}\put (-5,-3) {O7}\put (93,0) {O7} \put (100,53) {O7}  \put (-3,53) {O7}\put (73,27) {D7}\put (78,36) {D7}
\end{overpic}
\caption{Visualization of the base of $K3$, i.e.\ $T^2/\mathds{Z}_2$, in the orientifold limit. Only two of the sixteen D7-branes are shown.}\label{K3Illustration}
\end{figure}

We take one of the $K3$s to be an elliptic fibration of $T^2$ over the complex projective plane $\mathds{CP}^1$. The complex structure of the $T^2$-fiber corresponds to the axio-dilaton and points in the base where the fiber degenerates (i.e.\ a linear combination of the two one-cycles of $T^2$ shrinks to zero size) correspond to locations of 7-branes. The orientifold limit of F-theory on this space is then described by Type IIB with constant axio-dilaton compactified on $T^2/\mathds{Z}_2$. This leads to the pillow depicted in figure~\ref{K3Illustration}, where the corners of the pillow correspond to fixed points of the involution. Each of those fixed points denotes the location of one O7-plane and four D7-branes. The other $K3$, which for definiteness we denote by $\tK$, is not affected when taking the orientifold limit. It is completely wrapped by the D7-branes and O7-planes. We will assign a tilde to all quantities associated with $\tK$.

The K\"ahler potential for the complex structure moduli space is given by ${\cal K}_{\text{CS}} = -\log \left(i \int \Omega_4 \wedge \ov \Omega_4 \right)$, where $\Omega_4$ is the holomorphic four-form on $\tK \times K3$. The latter splits into a product of the holomorphic two-forms on $K3$ according to $\Omega_4 = \widetilde \Omega_2 \wedge \Omega_2$. We do not write down the K\"ahler potential for the K\"ahler moduli since their stabilization will not be discussed in this section.

The K\"ahler potential can be conveniently expressed in terms of the periods (i.e.\ integrals of the holomorphic two-form $\Omega_2$ over integral two-cycles) $\Pi$ and $\tP$ as
\begin{equation}\label{kaehlerk3-2}
 {\cal K}_{\text{CS}} = -\ln\left( \left( \tP \cdot \ov\tP  \right)\left( \Pi \cdot \ov\Pi  \right)\right).
\end{equation}
The first factor in \eqref{kaehlerk3-2} corresponds to the contribution from $\tK$ and will not be of interest here. As $K3$ is elliptically fibered, there are 18 possible complex structure deformations which are commonly denoted by $S$, $U$ and $C^a$, $a = 1,\ldots,16$. $S$ is the complex structure of the fiber torus and will be identified with the axio-dilaton of the Type IIB orientifold. $U$ is the complex structure of the base and thus determines the shape of the pillow in figure~\ref{K3Illustration}. The $C^a$ describe positions of D7-branes. In the orientifold limit all the $C^a$ vanish, which corresponds to a situation in which there are, on top of each of the four O7-planes, four D7-branes. Non-vanishing values for the $C^a$ parametrize the position of the D7-branes relative to the O7-planes. The quantities $S$, $U$ and $C^a$ arise by integrating $\Omega_2$ over two-cycles of $K3$ in a certain basis\footnote{For more details see \appref{k3surface}.} and assemble into the period vector as
\cite{Angelantonj:2003zx,Lust:2005bd,Braun:2008ua,Braun:2008pz}\footnote{The original results were derived from a $\mathcal{N}=2$ prepotential. We follow the notation of \cite{Braun:2008ua,Braun:2008pz}. The missing factor of 2 in \cite{Angelantonj:2003zx,Lust:2005bd} is presumably due to different normalizations.}
\begin{equation} \label{bigperiodk3}
 \Pi = \frac{1}{2}\begin{pmatrix} 1\\ C^2-SU \\ S \\ U \\ 2C^a \end{pmatrix} .
\end{equation}
In this basis the intersection matrix for the two-cycles reads
\begin{equation} \label{intersectionsk3}
 M = \begin{pmatrix} 0 & 2 & & & \\ 2 & 0 & & &  \\   &  & 0 & 2 & \\  &  & 2 & 0 & \\  &  &  & & -\mathds{1}_{16 \times 16} \end{pmatrix}  .
\end{equation}
One then finds
\begin{equation}\label{explicitPeriods}
 \Pi \cdot \ov\Pi := \sum_{i,j} \Pi_i M_{ij} \ov \Pi_j =  -\frac{1}{2}\left(\left(S -\ov{S}\right)\left(U-\ov{U}\right) - \sum_a \left(C^a -\ov{C}^a\right)^2   \right) .
\end{equation}
It is now obvious that the K\"ahler potential is invariant under shifts of the real parts of the~$C^a$.

\subsubsection[Moduli Stabilization on $\tK \times K3$]{Moduli Stabilization on {\boldmath $\tK \times K3$}}
\label{fluxk3}
In order to stabilize the moduli of an F-theory compactification on $Y=\tK \times K3$ one can switch on $G_4$ flux (which we will denote by $G$).
We will be interested in the stabilization mechanism for the complex structure moduli of $\tK \times K3$ in this section. The minimization conditions are $D_I W=0$, where $W = \int G \wedge \Omega_4$ is the tree-level superpotential and $D_I$ are covariant derivatives with respect to the complex structure moduli labeled by $I$. This set of equations is in general rather difficult to solve. However, for F-theory on $\tK \times K3$ it was shown in \cite{Braun:2008pz} that it is possible to rewrite the scalar potential in a form where the minimization conditions are easier to solve:
\begin{equation} \label{Vgeo}
V=-\frac{2\pi}{\vol^{3}}\sum_{i=1}^3 \left( \left\| \tPp[G^a\omega_i] \right\|^2 
  + \left\| \Pp[G\,\tw_i] \vphantom{\tPp}\right\|^2 \right) .
\end{equation}
Recall that $\Omega_2$ and $J_{K3}$\footnote{$J_{K3}$ is the K\"ahler form on $K3$.} can be parametrized by three real two-forms $\omega_i$, $i=1,\ldots,3$ such that $\Omega_2 = \omega_1 + i \omega_2$, $J_{K3} \sim \omega_3$ and
\begin{equation}
\omega_i \cdot \omega_j = \delta_{ij},\quad (i,j=1,2,3),
\end{equation}
where the inner product is defined as $v \cdot w \equiv \int_{K3}v \wedge w$ for two-forms $v,w$ on $K3$. The $\omega_i$ span a three-dimensional subspace $\Sigma$ inside the space of two-forms on $K3$. For further details see \appref{k3surface}.
The flux $G$ can be naturally viewed as a linear map between the vector spaces of two-forms, \mbox{$G: H^2(K3) \rightarrow H^2(\tK)$} (and $G^a$ is the adjoint with respect to the inner product specified by the intersection matrix \eqref{intersectionsk3}). This map explicitly reads
\begin{equation}
 H^2(K3) \ni v \longmapsto G v = \int_{K3} G\wedge v \in H^2(\tK).
\end{equation}
Furthermore, $\Pp (\tPp)$ is the projector on the subspace orthogonal to the $\omega_i (\tw_i)$, $i=1,2,3$, and $\left\| \cdot \right\|$ denotes the (negative definite) norm on that subspace. As $K3$ is elliptically fibered there are two distinct integral $(1,1)$-forms which correspond to base and fiber. These have intersections only with each other and not with any of the other two-forms. $\Omega_2$ is orthogonal to those $(1,1)$-forms and thus the period vector has $20$ $(=h^{2}(K3) - 2)$ non-vanishing entries.

Let us now specify the flux we choose in order to stabilize the complex structure moduli.
If we want to take the F-theory limit and preserve four-dimensional Lorentz invariance, no flux component in the direction of the base or the fiber is allowed (c.f.\ \cite{Dasgupta:1999ss}). Therefore, the two corresponding columns of the flux matrix $G$ in (\ref{Vgeo}) have vanishing entries. For convenience we assume $\tK$ to be elliptically fibered as well and choose the same basis of two-forms as for $K3$. Furthermore, we do not turn on fluxes along the $\widetilde{\text{base}}$ and the $\widetilde{\text{fiber}}$, although those fluxes are allowed in principle. The flux matrix then is a $20\times 20$ matrix and the index $i$ in \eqref{Vgeo} runs from 1 to 2. The entries of this matrix obviously depend on the choice of the basis of two-cyles on $K3$. In the following we choose to work in the basis in which the period vector takes the form \eqref{bigperiodk3} (the basis of two-cycles on $\tK$ will not be of any importance). The analysis of \cite{Braun:2008pz} then shows:
\begin{itemize}
 \item Equation (\ref{Vgeo}) implies that the (positive definite) scalar potential is minimized (at $V=0$) if
$G(\Sigma) \subset \widetilde \Sigma$ and $G^a(\widetilde \Sigma) \subset \Sigma$. For the above flux choice this actually means that the planes spanned by $\omega_1,\omega_2$ and $\tw_1,\tw_2$ are mapped to each other. 
 \item If the conditions $G(\Sigma) \subset \widetilde \Sigma$ and $G^a(\widetilde \Sigma) \subset \Sigma$ are fulfilled then $G^a G(\Sigma) \subset \Sigma$ and $G G^a(\widetilde \Sigma) \subset \widetilde \Sigma$. The two positive norm eigenvectors $\omega_1^\prime$ and $\omega_2^\prime$ of the selfadjoint operator $G^a G$ thus span the plane $\Sigma$. Now, the complex structure $\Omega_2^\prime=\omega_1^\prime + i \omega_2^\prime$ is fixed up to an overall rescaling by a complex number (which is equivalent to an SO(2) rotation of $\omega_1^\prime$ and $\omega_2^\prime$ and a rescaling by a real number). The same is true for the complex structure $\widetilde \Omega_2$ of $\tK$ and the matrix $GG^a$.

 \item We may now apply this reasoning to concrete examples: Given an explicit flux matrix one can calculate $G^aG$ and thus $\omega_1^\prime$ and $\omega_2^\prime$. The holomorphic two-form is then obtained via $\Omega_2^\prime=\omega_1^\prime + i \omega_2^\prime$. In fact, since $\omega_1^\prime$ and $ \omega_2^\prime$ are vectors computed {\it in an explicit basis}, the $\Omega_2^\prime$ constructed in this way is nothing but the period vector $\Pi'$. One now uses the rescaling freedom to define $\Omega_2 = \alpha \Omega_2^\prime$ such that the first component of the period vector is set to one half: $\Pi_1 = 1/2$. The values at which $S$ and $U$ are stabilized can now be read off by comparing the period vector obtained in this way with \eqref{bigperiodk3}. We will give an example of how to apply this procedure in \secref{flatdirinflaton}.
\item The tadpole cancellation condition following from $\chi (\tK\times K3) = 576$ can be written as
\begin{equation} \label{tadpole}
 \operatorname{tr} \left( G^aG \right)\equiv \operatorname{tr} \left( G^T M G \widetilde M\right) = 48  , \quad \text{where} \quad \widetilde M \equiv M. 
\end{equation}
\end{itemize}

\subsubsection[Flat Directions for the Inflaton on $\tK \times T^2/\mathds{Z}_2$]{Flat Directions for the Inflaton on {\boldmath $\tK \times T^2/\mathds{Z}_2$}} \label{flatdirinflaton}
We will now specify a flux and analyze, along the lines of the previous subsection, how this flux stabilizes moduli. Our flux choice is required to obey the following conditions: The superpotential $W=\int G \wedge \Omega_4 = \tP G \Pi$ should be independent of the D7-brane positions in order not to destroy the shift symmetry along those directions in moduli space. This can be achieved by not turning on any flux on the cycles corresponding to periods which depend on the brane positions. Furthermore, the flux should lead to a non-zero value  $W_0$ for the superpotential in the vacuum and satisfy the tadpole constraint \eqref{tadpole}. We choose the following flux\footnote{We omit the $16\times 16$ entries of the flux matrix corresponding to the brane cycles. All those entries are chosen to vanish.}
\begin{equation}
 G = \begin{pmatrix} 0_{2\times 2} & & \\  & 1 & 1 \\ & 1 & 5 \end{pmatrix}  .
\end{equation}
This implies
\begin{equation} \label{secondblock}
 G^aG \equiv G^T M G \widetilde M = \begin{pmatrix} 0_{2\times 2} & & \\ &24 & 40 \\ &  8 & 24  \end{pmatrix}  ,
\end{equation}
so that $G$ satisfies (\ref{tadpole}). The positive norm eigenvector in direction of the first $2\times 2$ block is arbitrary and thus does not stabilize the expression $SU-C^2$. The positive norm eigenvector in direction of the second block (\ref{secondblock}) is given by $\omega = \sqrt{5} \beta + e_2$. Here, $\beta$ and $e_2$ denote the basis elements corresponding to the third and fourth component of \eqref{bigperiodk3} (see also \appref{k3surface}). Since the overall normalization of the periods is not fixed yet, this flux only stabilizes the ratio $S/U$ at $S/U=\sqrt{5}$. The resulting vacuum is non-supersymmetric ($W_0 \neq 0$), with all brane moduli and one complex structure modulus unfixed.

Quite obviously, our $\tK \times T^2/\mathds{Z}_2$ example is just a toy model. We expect that for compactifications on more general orientifolds it is possible to choose a flux which stabilizes all complex structure moduli. However, even if one or more complex structure moduli were left unfixed in the generic case, loop corrections would eventually stabilize those.

\subsubsection{The Effect of Coordinate Shifts on Flat Directions}
The idea to use a shift symmetry in the D7-brane sector in order to protect the inflaton (which is identified with the D7-brane modulus) from obtaining a large mass due to supergravity corrections was previously discussed in \cite{Hsu:2003cy,Hsu:2004hi}. However, there is an apparent problem with such a mechanism: The K\"ahler potential of the D7-brane moduli space undergoes a K\"ahler transformation under a change of coordinates which needs to be compensated by a corresponding redefinition of the axio-dilaton, whose moduli space is non-trivially fibered over the D7-brane moduli space. Such a redefinition involves the D7-brane coordinate and thus the superpotential, manifestly depending on the axio-dilaton, cannot be independent of the brane positions in all coordinate patches. Since the superpotential is a holormorphic function of the fields, a dependence on the brane moduli naively contradicts the existence of a shift symmetry (cf.\ the discussion in \cite{Kachru:2003sx,McAllister:2005mq,Burgess:2006cb} of 
similar issues in the D3-brane sector). Nevertheless, as already noted in \cite{Hebecker:2012aw}, the shift symmetry will not simply be lost, but is just obscured by the choice of coordinates. We will now elucidate the fate of the shift 
symmetry in 
the D7-brane sector in the $\tK \times K3$ example.

The moduli space of the axio-dilaton $S$ is fibered over the moduli space of the D7-brane positions $C^a$. To appreciate the implications of this statement recall the structure of the K\"ahler potential in the weak coupling limit \cite{Jockers:2004yj,Jockers:2005zy}
\begin{equation}\label{kaehlerd7g}
 \ccK \supset -\log\left(-i(S-\ov S) -k\left(C,\ov C\right)\right) .
\end{equation}
Under coordinate transformations on the D7-brane moduli space the K\"ahler potential $k\left(C,\ov C\right)$ will generically undergo a non-trivial K\"ahler transformation. Since the physics is untouched one has to be able to promote this transformation on the D7-brane moduli space to a change of basis on the combined moduli space of $C$ and $S$, such that the full K\"ahler potential $\ccK$ changes by at most some K\"ahler transformation. It is clear from the structure of \eqref{kaehlerd7g} that the transformation of $k\left(C,\ov C\right)$ will have to be absorbed in a shift of $S$.
Therefore, the superpotential $W$ (which explicitly depends on the axio-dilaton) cannot be independent of the brane moduli in all patches. Since the dependence on $C$ will be holomorphic, the shift symmetry will not be manifest anymore.

Equipped with the explicit example of a $\tK \times K3$ compactification we would like to illustrate this feature. Starting point is the well-known K\"ahler potential of F-theory compactified on $\tK \times K3$ at the orientifold point (cf.\ \eqref{kaehlerk3-2} and \eqref{explicitPeriods})
\begin{equation}\label{eq:KahlerBeforeShift}
 \ccK(S,\ov S; C,\ov C; U,\ov U) = -\log\left((S -\ov{S})(U-\ov{U}) - (C -\ov C)^2 \right).
\end{equation}
For simplicity we only consider one brane with coordinate $C \in B_\epsilon (0)$, where $B_\epsilon (0)$ is some small neighborhood of the origin, and suppress the dependence on all other brane coordinates.
Now imagine moving that D7-brane from its original position by the amount $U$, i.e.\ moving it once around the pillow depicted in figure~\ref{K3Illustration} in the `vertical direction'. The K\"ahler potential which describes the situation after the shift (and which we obtain by analytic continuation of \eqref{eq:KahlerBeforeShift}) reads
\begin{equation}\label{eq:KahlerAfterShift}
\ccK_{\text{Shift}}(S,\ov S; C,\ov C; U,\ov U)  = -\log \left((S -\ov{S})(U-\ov{U}) -  ((C + U) -(\ov{ C + U}))^2 \right) ,
\end{equation}
where again $C \in B_\epsilon (0)$. However, since we moved the D7-brane once around the whole pillow, the physical situation is identical before and after the shift. This means, in particular, that we must be able to obtain \eqref{eq:KahlerAfterShift} from \eqref{eq:KahlerBeforeShift} by a pure redefinition of coordinates. This is indeed possible: Defining $S = S^\prime+U+2{C}^{\prime}$ and $C = C^\prime$ we find
\begin{align}
 \ccK(S^\prime,\ov{ S^\prime}; C^\prime,\ov{ C^\prime}; U,\ov U) &= -\log\left((S^\prime -\ov{S^\prime})(U-\ov{U}) - (C^\prime -\ov{ C^\prime})^2 \right)\nonumber\\
 &= \ccK_{\text{Shift}}(S,\ov S; C,\ov C; U,\ov U).
\end{align}
As expected from our general arguments below \eqref{kaehlerd7g}, the transformation of $S$ explicitly involves the brane coordinate. The need for a redefinition can be interpreted as a monodromy in the moduli space of $K3$.

Consider the flux choice of \secref{flatdirinflaton} which led to a manifestly flat direction in the superpotential before shifting the D7-brane, i.e.\ $W \equiv W(S,U)$. The superpotential, being independent of the brane coordinate, will not be affected by the periodic shift. After the shift and performing the coordinate change as above we therefore find
\begin{equation}
 W^\prime(S^\prime, U,{C}^\prime) \equiv W_{\text{Shift}}(S,U)  = W(S,U)= W(S^\prime+U+2{C}^{\prime},U). 
\end{equation}
Now, generically, the imaginary parts of $S^\prime$ and $C^\prime$ as well as the whole complex variable $U$ are fixed, as they appear explicitly in either the K\"ahler potential or the superpotential. The flat direction is not manifest anymore. However, it still exists and is given by $\Re\left( S^\prime+U+2{C}^\prime \right) = \text{const.}$ Thus, the shift symmetry did not disappear after the coordinate shift, but is only obscured by the redefinition of the coordinates in the new patch. This does not constitute a problem since we only rely on the assertion that there is a coordinate patch in which the brane-deformation scalar enjoys the shift 
symmetry.

\subsection{Open-String Landscape}
In this section we would like to relate our results to the discussion in \cite{Gomis:2005wc}. In this reference it was shown that for a compactification of Type IIB string theory on a toroidal orbifold $T^2_1 \times T^2_2 \times T^2_3/\mathds{Z}_2 \times \mathds{Z}_2$ it is possible to choose a flux that leaves D7-brane positions unfixed. In this model the tree-level superpotential vanishes in the vacuum. Clearly, this is not what we are after eventually. However, we regard it as useful to try to reproduce the results of \cite{Gomis:2005wc} in our framework and, in particular, to understand them from an F-theory point of view, i.e.\ taking into account the backreaction of the D7-branes. As we will see, the latter effect is in general non-negligible.

In accordance with the literature we denote the three kinds of D7-branes in the toroidal orbifold model by D$7_i$, where the index $i=1,2,3$ labels the torus $T^2_i$ in which the brane D$7_i$ is point-like. The flux chosen in \cite{Gomis:2005wc} reads
\begin{align}
F_3 &= 4\pi^2 \alpha' N \, ({\d}x^1\wedge {\d}x^2 \wedge {\d}y^3 + {\d}y^1 \wedge {\d}y^2 \wedge {\d}y^3 ), \label{F3} \\
H_3 &= 4\pi^2 \alpha' N \, ( {\d}x^1\wedge {\d}x^2 \wedge {\d}x^3 + {\d}y^1 \wedge {\d}y^2 \wedge {\d}x^3 ), \label{H3}
\end{align}
where $N \in \mathds{Z}$ and $0 \leq x^i, y^i \leq 1$. Given the complex coordinate $z^i=x^i+ \tau_i y^i$ on each torus and denoting the axio-dilaton by $\tau_0$, this choice of flux gives the following bulk superpotential:
\begin{align}
W_b \sim (1+\tau_1\tau_2)(1+\tau_3\tau_0) \label{wmarchesano} .
\end{align}
The supersymmetric minima of $W$ are then determined by
\begin{align}
1+\tau_1\tau_2 = 0 , \qquad 1+\tau_3\tau_0=0 \label{susycondmarchesano} .
\end{align}

Given the form of $H_3$ in (\ref{H3}) one can now explicitly calculate the pullback of the underlying $B_2$-field to the branes \cite{Gomis:2005wc}. In the absence of brane flux $F$, the supersymmetry conditions on $B_2$ are that $B_2$ is a primitive (1,1)-form. The analysis of \cite{Gomis:2005wc} shows the following: With this explicit choice of flux, the position moduli of D$7_1$-branes and D$7_2$-branes are completely fixed. By contrast, the position moduli of D$7_3$-branes remain unfixed, as the conditions in (\ref{susycondmarchesano}) already imply that the pullback of $B_2$ to D$7_3$ branes is a primitive (1,1)-form and thus no further condition on the brane moduli is imposed. As shown in \cite{Martucci:2006ij}, the analysis of \cite{Gomis:2005wc} is equivalent to the minimization of a Type IIB brane superpotential given in \cite{Gomis:2005wc,Martucci:2006ij}.

\subsubsection{F-Theory Flux Leaving D7-Branes Unfixed}
In the following, we will try to understand the moduli stabilization mechanism of this example from an F-theory point of view. We therefore work with an F-theory compactification on $K3 \times T^2 \times T^2$, where the $K3$ manifold is elliptically fibered. This F-theory compactification is dual to a Type IIB compactification on $T^2/\mathds{Z}_2 \times T^2 \times T^2$ and thus gives a different Type IIB background than the model discussed in \cite{Gomis:2005wc}. In particular, there is only one type of D7-branes and O7-planes present (those which are point-like in  $T^2 / \mathds{Z}_2$) and O3-planes are absent. Therefore, contrary to the previously discussed Type IIB setup where we considered different types of branes (stabilized and unstabilized), we will now consider different types of fluxes in the F-theory model. These fluxes will give rise to the analogs of the stabilized and unstabilized branes of the Type IIB analysis, however, with some crucial differences due to brane backreaction. We start by 
modeling the analog of the unfixed type of branes.

Denoting the complex coordinate of the F-theory fiber by $x^0 + \tau_0 y^0$, the F-theory flux that we choose is given by
\begin{align}
G_4 =& 4\pi^2 \alpha' N \, ({\d}x^1\wedge {\d}x^2 \wedge {\d}y^3 \wedge {\d}y^0 + {\d}y^1 \wedge {\d}y^2 \wedge {\d}y^3 \wedge {\d}y^0 \nonumber \\
 &+ {\d}x^1\wedge {\d}x^2 \wedge {\d}x^3\wedge {\d}x^0 + {\d}y^1 \wedge {\d}y^2 \wedge {\d}x^3 \wedge {\d}x^0)  \label{G4marchesano} ,
\end{align} 
which, in the orientifold limit, reproduces the Type IIB bulk flux given in (\ref{F3}) and (\ref{H3}) by wedging it with the holomorphic one form of the fiber torus and integrating out the fiber. In the orientifold limit, the elliptic fibration becomes trivial and $K3$ reduces to $\left(T^2 \times T^2\right)/\mathds{Z}_2$, where the volume of the first $T^2$ vanishes in the F-theory limit. The holomorphic two-form of $K3$ then splits into a product of the holomorphic one-forms of the two factors in $\left(T^2 \times T^2\right)/\mathds{Z}_2$. By comparison with \eqref{bigperiodk3} we deduce
\begin{equation}\label{eq:OmegaInOrientifoldLimit}
 \Omega_2 = {\d} x^0 \wedge {\d} x^3 - \left( C^2  - \tau^0 \tau^3  \right){\d} y^0 \wedge {\d} y^3 + \tau^0 {\d} y^0 \wedge {\d} x^3 + \tau^3 {\d} x^0 \wedge {\d} y^3  .
\end{equation}
This captures only the bulk cycles. The relative cycles which measure the brane separation are irrelevant in this consideration, as we do not turn on flux along those cycles and they have vanishing intersection with the bulk cycles.
Using \eqref{eq:OmegaInOrientifoldLimit} and \eqref{G4marchesano} we can calculate the full (bulk plus brane) superpotential for the chosen flux:
\begin{align}
W \sim (1+\tau_1\tau_2)(1+\tau_3\tau_0 - C^2) . \label{wfullmarch}
\end{align}
The minimization conditions are given by
\begin{align}
\tau_1 \tau_2 = -1\, , \qquad \tau_3\tau_0 -C^2=-1  . \label{minmarch}
\end{align}
The resulting minimum is supersymmetric ($W=0$). Note that \eqref{minmarch} provides two equations for 20 moduli, thus 18 moduli remain unfixed.

On the other hand, the full superpotential \eqref{wfullmarch} can be split into a bulk and a brane superpotential:
\begin{align}
W_b \sim (1+\tau_1\tau_2)(1+\tau_3\tau_0)  , \qquad W_{\text{D7}} \sim (1+\tau_1\tau_2)C^2  . \label{wsplitmarch}
\end{align}
The stabilization mechanism of \cite{Gomis:2005wc} can now be understood as follows. First, the bulk superpotential $W_b$ is used to stabilize the bulk complex structure moduli at $\tau_1 \tau_2 = -1$ and $\tau_3\tau_0=-1$. This automatically implies $W_{\text{D7}}=0$ in the minimum (analogously, $B_2$ is automatically primitive and of type $(1,1)$ in the minimum), such that there are no further conditions on the D$7$-brane positions.

The difference of both viewpoints is that the F-theory description takes into account the backreaction of the D7-branes on the background geometry. Thus, when the unstabilized branes move, i.e.\ $C^2$ changes, the bulk moduli $\tau_3 $ and $\tau_0$ have to change as well.

To analyze the relevance of this effect, let us solve the second equation in (\ref{minmarch}) for $\tau_3$:
\begin{equation} \label{tau3march}
 \tau_3 = -\frac{1}{\tau_0}(1-C^2)  .
\end{equation}
At the orientifold point $C^2$ will be zero and equation (\ref{tau3march}) tells us that $\tau_3=-1/\tau_0$. In this case, the result is in full accordance with the result of \cite{Gomis:2005wc}. If we demand $\tau_0$ to be large and imaginary (corresponding to a small value for $g_s$), $\tau_3$ will be small and imaginary. When we move the D7-branes off the O7-planes this analysis is still correct as long as $C^2$ is small compared to unity. As soon as $C^2$ is of order one the effect on the stabilization of $\tau_3$ becomes relevant: $\tau_3$ decreases by a relative factor of $1-C^2$ as compared to the case where we neglected backreaction. Thus, by using the full F-theory superpotential, we explicitly see that the influence of the D7-brane moduli on the stabilization conditions for the bulk moduli can be very large, even at small string coupling. In the Type IIB approach of \cite{Gomis:2005wc} this effect has not been considered. It would be interesting to investigate the implications of this effect more 
generally.

\subsubsection{F-Theory Flux Fixing D7-Brane Positions}
In order to model the analogs of the stabilized type of branes in \cite{Gomis:2005wc} we will change the relative position of the flux (\ref{G4marchesano}) in the F-theory approach. This can be done by interchanging the indices 2 and 3 in (\ref{G4marchesano}). The D$7_1$-case can be treated analogously. After switching indices, the F-theory flux is given by
\begin{align}
G_4 =& -4\pi^2 \alpha' N \, ({\d}x^1\wedge {\d}y^2 \wedge {\d}x^3 \wedge {\d}y^0 + {\d}y^1 \wedge {\d}y^2 \wedge {\d}y^3 \wedge {\d}y^0 \nonumber \\
 & + {\d}x^1\wedge {\d}x^2 \wedge {\d}x^3\wedge {\d}x^0 + {\d}y^1 \wedge {\d}x^2 \wedge {\d}y^3 \wedge {\d}x^0). \label{G4marchesano2}
\end{align} 
The corresponding superpotential reads
\begin{align}
W \sim 1 + \tau_1 \tau_3 + \tau_1\tau_2(\tau_0 \tau_3 - C^2) + \tau_0 \tau_2 , \label{wfullmarch2}
\end{align}
and the minimization conditions are:
\begin{align}
0 &= \tau_3 - C^2\tau_2 + \tau_0\tau_2\tau_3 , \nonumber \\
0 &= \tau_0 - C^2\tau_1 + \tau_0\tau_1\tau_3 , \nonumber  \\
0 &= \tau_1 + \tau_0\tau_1\tau_2 , \nonumber  \\
0 &= \tau_2 + \tau_1\tau_2\tau_3 ,  \nonumber \\
0 &= -2 C^b\tau_1\tau_2  , \qquad b=1,\dots,16 . \nonumber 
\end{align}
In contrast to the example given before, all branes are stabilized at $C^b=0$. It is not hard to check that the resulting minimum is supersymmetric ($W=0$). 

In order to understand the stabilization mechanism of \cite{Gomis:2005wc} in our F-theory setting, we again split the full superpotential (\ref{wfullmarch2}) in a bulk and a brane part:
\begin{align}
W_b \sim 1 + \tau_1 \tau_3 + \tau_1\tau_2\tau_0 \tau_3 + \tau_0 \tau_2  , \qquad W_{\text{D7}} \sim \tau_1\tau_2 C^2  . \label{wsplitmarch2}
\end{align}
As the full superpotential is not a product of two terms anymore, the stabilization conditions on $W_b$ do not automatically imply $W_{\text{D7}}=0$. Thus, minimization of $W_{\text{D7}}$ additionally gives 16 conditions on the D$7$-brane moduli and stabilizes them at $C^b=0$. Therefore, in this case the results of the Type IIB analysis are fully reproduced.

The above findings result from the fact that, without brane fluxes, the only possible term in the full superpotential containing the D7-brane moduli is proportional to $C^2$. In this situation, if the branes are stabilized by the minimization conditions for a supersymmetric minimum, stabilization will occur at $C^b=0$ for all $b=1,\dots,16$. The D7-brane moduli then drop out of the stabilization conditions for the bulk moduli. This is in full accordance with \cite{Gomis:2005wc}. Now consider a possible brane-flux-dependence of the superpotential, which will lead to contributions to the superpotential proportional to some $C^b$. It is then possible that the minimization conditions imply $C^b \neq 0$ and thus, generically, $C^2 \neq 0$. Consequently, the stabilization of the brane moduli at non-vanishing $C^b$ induces a non-trivial backreaction on the bulk moduli. This backreaction is neglected in \cite{Gomis:2005wc}, where the bulk superpotential is minimized and the dynamics of D7-branes is considered only 
afterwards.
Our more general analysis shows that the inclusion of brane backreaction is important and can lead to significant changes in the stabilization of bulk moduli, even at small string coupling (i.e.\ large $\Im(\tau_0)$).

The split into bulk and brane superpotential is discussed more generally in the following \secref{ModStabGen}.

\section{Type IIB K\"ahler Potential and Superpotential from F-Theory}
\label{ModStabGen}
In this section we review some facts about compactifications of M-theory to three dimensions with a focus on the dual F-theory description. In particular we discuss the F-theory K\"ahler potential and superpotential and consider the orientifold limit in which the corresponding Type IIB quantities emerge. This section thus generalizes the discussion of explicit examples in \secref{Flat Directions}. Some useful facts about the orientifold limit of F-theory are collected in \appref{senslimit}.

\subsection{Type IIB Orientifold Moduli Space from F/M-Theory}\label{modulispaceftheory}
Given M-theory compactified on an elliptically fibered Calabi-Yau fourfold $Y$, the resulting three dimensional supergravity will have various moduli, amongst them the geometric moduli of $Y$ ($h^{3,1}(Y)$ complex structure and $h^{1,1}(Y)$ K\"ahler moduli). The dual F-theory compactification on $Y$ describes Type IIB theory compactified on the double cover $\wt{X}$ of the base $X$ of the elliptic fibration together with an orientifold action on it. Points in the base where the fiber degenerates correspond to positions of the 7-branes. The corresponding four dimensional supergravity theory comprises $h^{2,1}_-(\wt{X}) + h^{1,1}_+(\wt{X})$ geometrical (closed-string) moduli, the axio-dilaton $\tau$ and the D7-brane (open-string) moduli. We will review how the purely geometric description of D7-brane positions and the axio-dilaton in F-theory translates to the Type IIB language. We will closely follow \cite{Denef:2008wq}.

The complex structure moduli space of $Y$ is encoded in period integrals of the $(4,0)$-form $\Omega_4$ over four-cycles of $Y$. It is convenient to work instead with $\wt{Y}$, which is formally constructed as the elliptic fibration over the double cover $\wt{X}$ of the base. In order to construct a basis of four-cycles on $\wt{Y}$, recall that there are two distinct one-cycles in the fiber of the fourfold, commonly called $A$- and $B$-cycle. We use conventions such that, in the weak coupling limit, $A$ is not subject to any monodromy and collapses to zero size at the D7-brane loci. On the other hand, $B$ will undergo monodromies $B \rightarrow B+A$ when going around D7-brane loci. The holomorphic $(1,0)$-form on the torus is normalized such that 
\begin{equation}
\int_A \Omega_1 = 1 , \qquad \int_B \Omega_1 = \tau  ,
\end{equation}
where $\tau$ is the modular parameter of the torus. In the orientifold limit it can be written as (cf.\ \appref{senslimit})
\begin{equation}
\tau \approx \tau_0 + \frac{i}{2 \pi}\ln\left(\frac{P_{\text{O7}}}{P_{\text{D7}}}\right).\label{dilatonprofile} 
\end{equation}
The quantities $P_{\text{O7}}$ and $P_{\text{D7}}$ are polynomials in the base coordinates which vanish at the loci of the O7-planes and D7-branes, respectively.

Let us denote three-cycles on the double cover of the base with negative parity under the orientifold involution by $\Sigma_i$ ($i=1,\dots,b^3_-(\wt{X})$) (the corresponding periods then have positive parity since $\Omega_3$ of the threefold also has negative parity). Furthermore, let $\Gamma_\alpha$ ($\alpha=1,\dots, h^{2,0}_-(S)$\footnote{For subtleties regarding the definition of this number see \cite{Collinucci:2008pf}. $S$ denotes a divisor of $\wt{X}$ wrapped by the D7-branes.}) denote three-chains on $\wt{X}$ which are swept out by two-cycles of a brane / image-brane pair as they are pulled off the O7-plane. We can now fiber the $A$- and $B$-cycles over those three-cycles and three-chains to define a basis of four-cycles as follows:
\begin{equation} \label{4cyclesftheory}
\Sigma_i \times A  , \qquad \Sigma_i \times B  , \qquad \Gamma_\alpha \times A  .
\end{equation}
Their intersection matrices are
\begin{align} \label{intersectionsftheory}
Q_{\alpha \beta} &:= (\Gamma_\alpha \times A)\cdot (\Gamma_\beta \times A) = -\left.(\partial \Gamma_\alpha)\cdot (\partial \Gamma_\beta)\right|_S ,\\
Q_{ij} &:= (\Sigma_i \times A)\cdot (\Sigma_j \times B)= - \left.\Sigma_i \cdot \Sigma_j \right|_{\wt{X}}  .
\end{align}

Starting from the holomorphic $(4,0)$-form on $\wt{Y}$ we define the holomorphic $(3,0)$-form on the base via\footnote{For more details on how to integrate out the torus-cycles in order to get the $(3,0)$-form see \cite{Denef:2008wq}.}
\begin{equation}
\int_{\Sigma_i \times A} \Omega_4 = \int_{\Sigma_i} \Omega_3 =: \Pi_i(z)  ,
\end{equation}
where we indicated the dependence of the periods $\Pi_i$ on the complex structure moduli $z$ of $\wt{X}$. The remaining periods are then\footnote{Generically, the three-cycle $\Sigma_i$ will intersect the branch cut of the logarithm, the latter being a five-chain of the base, swept out by the D7-branes as they are pulled off the O7-plane. This is just a manifestation of the fact that the $B$-cycle may undergo monodromies when moving along the three-cycle. Integrating, however, does not constitute a problem as the only divergences of the logarithm in \eqref{ftheoryperiods} are at the loci of the D7-branes and O7-planes. Those divergences are very weak, such that the integral is well behaved.\label{IntegrationFootnote}}
\begin{align}
\int_{\Sigma_i \times B} \Omega_4 = \int_{\Sigma_i} \tau \,  \Omega_3 &= \tau_0 \, \Pi_i + \frac{i}{2 \pi}\int_{\Sigma_i} \ln\left(\frac{P_{\text{O7}}}{P_{\text{D7}}}\right)  \Omega_3  \nonumber  \\[0.2cm] 
& =: \tau_0 \, \Pi_i(z) + \chi_i(z,\xi),   \label{ftheoryperiods} \\
\int_{\Gamma_\alpha \times A} \Omega_4 =  \int_{\Gamma_\alpha} \Omega_3 &=: \Pi_\alpha(z,\xi) , \label{ftheoryperiods3}
\end{align}
where $\xi$ indicates the dependence on D7-brane moduli.

The K\"ahler potential on the complex structure moduli space of $\wt{Y}$ is \cite{Haack:2001jz}\footnote{Working on $\wt{Y}$ instead of $Y$ allows us to use the periods of $\wt{X}$ in the next step.}
\begin{equation}
\ccK_{\wt{Y}}= -\ln \left(\frac{1}{2} \int_{\wt{Y}} \Omega_4 \wedge \ov \Omega_4\right) .
\end{equation}
Using (\ref{ftheoryperiods}) and (\ref{ftheoryperiods3}) we can thus write 
\begin{align} 
\ccK_{\wt{Y}} &= -\ln \left( \frac{1}{2} \left[(\tau_0- \ov \tau_0)\Pi_i Q^{ij} \ov \Pi_j + \chi_i Q^{ij}\ov \Pi_j + \ov \chi_i Q^{ij} \Pi_i - \Pi_\alpha Q^{\alpha \beta} \ov \Pi_\beta\right]\right) \nonumber \\
&=: \ccK_{\tau_0} + \ccK_{\wt{X}} + g_s \ccK_{\text{D7}} + \mathcal{O}(g_s^2) ,\label{kahlerpot1} \\
 \ccK_{\tau_0}&:= -\ln \left(-\frac{i}{2}(\tau_0- \ov \tau_0)\right)  , \qquad \ccK_{\wt{X}} :=-\ln \left(i\Pi_i Q^{ij} \ov \Pi_j\right)  ,\\
 \ccK_{\text{D7}} &:= \frac{1}{2} e^{\ccK_{\wt{X}}} \left(\chi_i Q^{ij}\ov \Pi_j + \ov \chi_i Q^{ij} \Pi_i - \Pi_\alpha Q^{\alpha \beta} \ov \Pi_\beta\right).
 \end{align}
This is the generalized version of \eqref{kaehlerk3-2} and \eqref{explicitPeriods}.

\subsection{Superpotential from F-Theory and Type IIB Perspective}\label{fluxesinmtheory}
We now discuss the superpotential. Starting from the F-theory perspective the aim is to recover the well-known brane superpotential for a D7-brane in Type IIB theory \cite{Jockers:2005zy,Martucci:2006ij,Martucci:2007ey}
\begin{equation}\label{WishSuperpot}
 \wt{W}_{\text{D7}} = \int_{\Gamma_5} \ccF_2 \wedge \Omega_3 , \quad \ccF_2 := F_2 - B_2,
\end{equation}
in addition to the standard Type IIB bulk superpotential. Here, $\Gamma_5$ is the five-chain swept out by a pair of D7-branes as they are pulled off the O7-plane. We again follow \cite{Denef:2008wq}.

Starting point is an expansion of the harmonic and quantized flux $G_4$ in a basis of four-forms that are the Poincar\'{e} duals of the basis of four-cycles defined in (\ref{4cyclesftheory}):
\begin{equation}
G_4 =  N^i \ [\Sigma_i \times A] -  M^i \ [\Sigma_i \times B] +  N^\alpha \ [\Gamma_\alpha \times A] . 
\end{equation}
The Gukov-Vafa-Witten superpotential $W=\int_{\wt{Y}} G_4 \wedge \Omega_4 = \int_{[G_4]} \Omega_4$ can then be written as
\begin{equation} 
W =  (N^i - \tau_0 M^i)\Pi_i (z) -   M^i \chi_i(z,\xi) +  N^\alpha \Pi_\alpha (z,\xi) . \label{bulkbranesuperpot} 
\end{equation}
We identify the first term in this expression as the Type IIB bulk superpotential which, in terms of the weak coupling Type IIB bulk fluxes
\begin{equation}
F_3 =  N^i \ [\Sigma_i] , \qquad H_3 =  M^i \ [\Sigma_i], 
\end{equation}
can be written as
\begin{equation}\label{eq:BulkSuperpot}
 W_{\text{b}}= \int_{\wt{X}} (F_3 - \tau_0 H_3) \wedge \Omega_3 =: \int_{\wt{X}} G_3 \wedge \Omega_3 .
\end{equation}
The other two terms contribute to the brane superpotential $W_{\text{D7}}$, such that $W = W_{\text{b}} + W_{\text{D7}}$. Consider the last term in \eqref{bulkbranesuperpot}:
\begin{equation} \label{F2}
 \int_{\Gamma (F_2)} \Omega_3 =  \int_{\Gamma_5} F_2 \wedge \Omega_3  ,
\end{equation}
where $F_2$ is defined to be the Poincar\'{e} dual of $\Gamma (F_2)$ on the five-chain $\Gamma_5$.\footnote{In agreement with most of the literature we use the name $F_2$ to denote both the brane-localized flux $F_2 =  N^\alpha \ [\partial \Gamma_\alpha]$ as well as the brane flux extended to the five-chain as in \eqref{F2}. It should be clear from the context what is meant by $F_2$ in each case.} For $H_3 \equiv 0$ we can choose $B_2 \equiv 0 $ everywhere consistently and the brane superpotential (\ref{F2}) exactly reproduces \eqref{WishSuperpot}.

In order to discuss the term $M^i \chi_i$ consider first the situation where one unit of $H_3$ flux threads the cycle $\Sigma_i$ and integrate $\Omega_3$ over this cycle. This is just by definition $\Pi_i$:
\begin{equation}\label{PiEval}
 \Pi_i \equiv \int_{\Sigma_i} \Omega_3 = \int_{\wt{X}} H_3 \wedge \Omega_3 = \int_{\wt{X}\setminus \Sigma_i} {\d} B_2 \wedge \Omega_3 = \int_{B(\Sigma_i)} B_2 \wedge \Omega_3.
\end{equation}
Here we successively used Poincar\'{e} duality on $\wt{X}$ and the fact that we can write $H_3 = {\d} B_2$ locally. Then we performed a partial integration, using ${\d} \Omega_3 = 0$. The fact that we cannot write $H_3 = {\d}B_2$ globally gives rise to a boundary term which is integrated along the five-cycle $B(\Sigma_i)$ introduced by cutting along $\Sigma_i$. The boundary can be imagined to be $S^2 \times \Sigma_i$, i.e.\ each point on the three-cycle is surrounded by an infinitesimally small two-sphere. $\Omega_3$ is constant on such a sphere and the integral just gives the flux quantum number (which we chose to be one),\footnote{This can be thought of as being the higher-dimensional analog of the integration of the one-form gauge potential over an infinitesimally small $S^1$ winding around the Dirac string.} thus giving back the expression on the LHS of \eqref{PiEval}.
We now want to use the same technique to evaluate the term $M^i \chi_i(z,\xi) = \frac{i}{2\pi} \int_{M^i \Sigma_i} \ln \left(\frac{P_{\text{O7}}}{P_{\text{D7}}}\right)\Omega_3$ in \eqref{bulkbranesuperpot}. To do that it is convenient to consider the term 
\begin{align}\label{MChi}
 \frac{i}{2\pi} \int_{\wt{X}} H_3 \wedge \ln \left(\frac{P_{\text{O7}}}{P_{\text{D7}}}\right)\Omega_3 =& \frac{i}{2\pi}  \int_{B(\Sigma_i)} B_2 \wedge \ln \left(\frac{P_{\text{O7}}}{P_{\text{D7}}}\right)\Omega_3 
 + \frac{i}{2\pi} \int_{B(\Gamma_5)} B_2 \wedge \ln \left(\frac{P_{\text{O7}}}{P_{\text{D7}}}\right)\Omega_3 \nonumber\\
 =& \frac{i}{2\pi} \int_{M^i \Sigma_i} \ln \left(\frac{P_{\text{O7}}}{P_{\text{D7}}}\right)\Omega_3 - \int_{\Gamma_5} B_2 \wedge \Omega_3.
\end{align}
The first term on the RHS of \eqref{MChi} follows in analogy to \eqref{PiEval} and using ${\d} \left[\ln \left(\frac{P_{\text{O7}}}{P_{\text{D7}}}\right)\Omega_3\right]= 0$. The only additional complication in the evaluation of this expression is the fact that, in addition to $H_3$, the logarithm is not globally well-defined either. It jumps when crossing the five-chain $\Gamma_5$ and diverges at the D7-brane and O7-plane loci. Thus, when performing the partial integration, there appears an additional term which comes from an integration over the boundary five-cycle which is introduced by cutting along $\Gamma_5$.
Recall that $\tau$ and therefore also $\frac{i}{2\pi}\ln \left(\frac{P_{\text{O7}}}{P_{\text{D7}}}\right)$ jumps by one when circling around a D7-brane locus. Therefore, the integral over $B(\Gamma_5)$ can be replaced by an integral of $B_2 \wedge \Omega_3$ along the branch cut five-chain $\Gamma_5$. Note that the negative sign of this term in \eqref{MChi} corresponds to an integration which starts at the O7-plane and ends on the D7-brane locus.

\begin{figure}
\centering
\begin{overpic}[width=0.5\textwidth,tics=10]{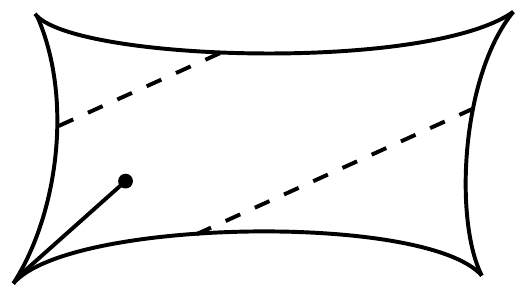}
 \put (12,17) {$\Gamma_5$} \put (67,19) {$M^i \Sigma_i$}
\end{overpic}
\caption[D7-brane superpotential on $T^2/\mathds{Z}_2$.]{The dashed line represents the one-cycle of the flux three-cycle $M^i \Sigma_i$ on $T^2/\mathds{Z}_2$ while the straight line represents the branch cut of the logarithm in (\ref{log-cut}) with a D7-brane at its end.}\label{D7potfig}
\end{figure}

Figure \ref{D7potfig} visualizes the situation for
the simple orientifold $X =\tK \times T^2/\mathds{Z}_2$, where we only look at the complex direction of $X$ in which the D7-branes are point-like (i.e.\ the pillow $T^2/\mathds{Z}_2$ with one O7-plane located at each of its corners). The flux $H_3$ has exactly one leg along $T^2/\mathds{Z}_2$ (there is no one-form and, correspondingly, no three-form on $K3$). Therefore, $M^i \Sigma_i$ can be visualized as a one-cycle in $T^2/\mathds{Z}_2$.

What is special about this example is the possibility to choose the three-cycle and the five-chain to be non-intersecting. Generically this will not be the case. From the perspective of the $\Gamma_5$-integration, intersections with the flux three-cycle correspond to subspaces along which the $B_2$ field behaves non-trivially. However, as we will find presently, the final form of $W_{\text{D7}}$ will depend only on the combination $(F_2-B_2 )$ which is known to be gauge invariant. Thus we expect nothing special to occur at those loci. On the other hand, from the perspective of the integration over the flux three-cycle, the subspaces in $M^i \Sigma_i $ at which the logarithm jumps and diverges appear already in the definition of $\chi_i$. As already mentioned in footnote \ref{IntegrationFootnote}, these divergences are weak enough such that the integral is well behaved.

In summary we find
\begin{equation}
 M^i \chi_i  = \frac{i}{2\pi} \int_{M^i \Sigma_i} \ln \left(\frac{P_{\text{O7}}}{P_{\text{D7}}}\right)\Omega_3 = \frac{i}{2\pi} \int_{\wt{X}} H_3 \wedge \ln \left(\frac{P_{\text{O7}}}{P_{\text{D7}}}\right)\Omega_3 + \int_{\Gamma_5} B_2 \wedge \Omega_3 . \label{log-cut}
\end{equation}
The full brane superpotential is then given by
\begin{equation} \label{D7pot4}
W_{\text{D7}} = -\frac{i}{2\pi} \int_{\wt{X}} H_3 \wedge \ln\left(\frac{P_{\text{O7}}}{P_{\text{D7}}}\right) \Omega_3 + \int_{\Gamma_5} (F_2-B_2) \wedge \Omega_3   .
\end{equation}
which is almost \eqref{WishSuperpot} except for the term $-\frac{i}{2\pi} \int_{\wt{X}} H_3 \wedge \ln\left(\frac{P_{\text{O7}}}{P_{\text{D7}}}\right) \Omega_3$ which compensates the appearance of the non-integral expression $\int_{\Gamma_5} B_2 \wedge \Omega_3$.

Thus, starting from the assumption that the superpotential $W=\int_{\wt{Y}} G_4 \wedge \Omega_4$ gives the correct description of the low energy effective action of F-theory, we have taken the weak coupling limit and derived the corresponding Type IIB quantity. After splitting off the bulk part $W_{\text{b}}$ in \eqref{eq:BulkSuperpot} we identified the brane superpotential $W_{\text{D7}}$ in \eqref{D7pot4}, which differs from \eqref{WishSuperpot}, found in \cite{Jockers:2005zy,Martucci:2006ij,Martucci:2007ey}. We believe that \eqref{D7pot4} is the correct expression for the brane superpotential in Type IIB string theory.

\section{Mirror Symmetry: K\"ahler Potential at Large Complex Structure}
\label{Mirror-Symmetry}
In this section we set out to motivate the existence of shift symmetries in more general examples beyond the toy models discussed in \secref{Flat Directions}. We start from the observation that the K\"ahler potential of the complex structure moduli space of a Type IIB compactification on a Calabi-Yau threefold exhibits a manifestly shift-symmetric form in the vicinity of the point of `large complex structure'. This shift-symmetric structure can be understood via mirror symmetry, which maps the complex structure moduli space of Type IIB string theory to the K\"ahler moduli space of Type IIA string theory. The Type IIA K\"ahler moduli are two-cycle volumes which are complexified by the Kalb-Ramond $B_2$ field, integrated over these two-cycles. The shift symmetry is a remnant of the 10d gauge symmetry $B_2 \to B_2 + {\d}\Lambda_1$. Since mirror symmetry extends to Calabi-Yau manifolds beyond complex dimension three, we are led to the expectation that the complex structure moduli spaces of fourfolds exhibit 
shift-symmetric structures at the point of large complex structure. This is exactly what we are after, as D7-brane positions are encoded in the complex structure of the F-theory fourfold.

\subsection{Type IIA K\"ahler Moduli Space at Large Volume}
Classically, the metric on the K\"ahler moduli space of a Calabi-Yau threefold $W$ is derived from the K\"ahler potential \cite{Candelas:1990pi,Grimm:2004ua}
\begin{equation}\label{IIBKahlerPot}
e^{-\mathcal K_{\ccV}} =  \frac{8}{3!}\int_W J \wedge J \wedge J =\frac{i}{3!}\kappa_{abc} (\bft^a -\ov \bft^a) (\bft^b-\ov \bft^b)(\bft^c -  \ov \bft^c)= \frac{8}{3!}\kappa_{abc}t^a t^b t^c  = 8\ccV^s ,
\end{equation}
where $t^a$ are volumes of two-cycles appearing as the expansion coefficients of the K\"ahler form $J$ in a basis of two-forms $\omega_a$, $a=1,\ldots,h^{1,1}(W)$, and $\ccV^s$ is the volume of $W$, measured in units of $\ell_s$ in the ten-dimensional string frame. The $\kappa_{abc}$ are triple intersection numbers of four-cycles of the threefold. In the context of Type IIA string compactifications this should be read as a function of complex variables $\bft^a$ which are the complexified two-cycle volumes defined via
\begin{equation}
 B_2+iJ=: \bft^a \omega_a.
\end{equation}

The K\"ahler potential can be expressed in $\mathcal N = 2$ language by defining a holomorphic prepotential
\begin{equation}
\label{prepotk11}
\mathcal{F}(\bft):=-\frac{1}{3!} \kappa_{abc} \frac{T^a T^b T^c}{T^0}.
\end{equation}
Here we have introduced a new set of coordinates $T^A$, $A=0,...,h^{1,1}(W)$, related to the $\bft^a$ via $\bft^a :=\frac{T^a}{T^0}$. In terms of these projective coordinates the K\"ahler potential $\mathcal K_{\ccV}$ for the K\"ahler moduli $\bft^a$ can be written as \cite{Candelas:1990pi}
\begin{equation}
\label{kpotk11}
e^{-\mathcal K_{\ccV}} = \left.-i(T^A \ov\partial_A \ov{\mathcal{F}} - \ov T^A \partial_A \mathcal{F})\right\vert_{T^0=1} .
\end{equation}

The prepotential \eqref{prepotk11} receives quantum corrections at the non-perturbative level which are exponentially suppressed by two-cycle volumes.

\subsection{Type IIB Complex Structure Moduli Space at Large Complex Structure}\label{IIBAtLCS}
Via mirror symmetry \cite{Hosono:1994av} the Type IIA K\"ahler moduli space is mapped to the complex structure moduli space of Type IIB string theory compactified on the mirror Calabi-Yau $M$.
We thus expect to be able, in a certain limit and under an appropriate identification of variables, to describe the complex structure moduli space of a Type IIB compactification by a K\"ahler potential which takes the shift-symmetric form \eqref{IIBKahlerPot}. This is indeed the case: There exists a set of projective coordinates $X^i$ in which the prepotential of the Type IIB complex structure moduli space, expanded around the point of large complex structure\footnote{The point of `large complex structure' (LCS) \cite{Morrison:1991cd,Hosono:1994av} is defined as follows: It is a singular point in the complex structure moduli space, where the divergence structure of the periods is characterized by certain monodromies. Let $u^i$ be a suitable set of local coordinates on the complex structure moduli space in which the LCS point is at $u^i =0$, $\forall i$. Then, for a certain set of three-cycles $\Sigma_A$ there is one invariant period. This period is scaled to one in the end. For the periods associated with 
the 
special coordinates $z^i$ one finds $z^i \sim \log u^i$ in the vicinity of the LCS point. Furthermore, at leading order the remaining periods then have the simple structure implied by the prepotential \eqref{lcsl}.} \cite{Morrison:1991cd} (which is the Type IIB equivalent of the large volume limit on the Type IIA side), reads
\begin{equation}
\label{lcsl}
\mathcal{G} (z) =\frac{1}{3!}\kappa_{ijk} \frac{X^i X^j X^k}{X^0} + \ldots .
\end{equation}
Consequently, the K\"ahler potential, expressed in affine coordinates $z^i = X^i / X^0$, takes the shift-symmetric form
\begin{equation}\label{lokp}
 e^{-\mathcal K_{\text{CS}}}=i\int_M \Omega_3 \wedge \ov \Omega_3 = i(\ov X^I \partial_I \mathcal{G}-X^I \ov{\partial}_I \ov{\mathcal{G}})=\frac{i}{3!}\kappa_{ijk} (z^i -\ov z^i) (z^j-\ov z^j)(z^k -  \ov z^k) +\ldots .
\end{equation}

To understand the meaning of the quantities involved and, in particular, to appreciate the limit in which these expressions are valid, recall the standard description of the complex structure moduli space of a Calabi-Yau threefold: The period vector $\Pi_A := \int_{\Sigma_A}\Omega_3$ is conveniently expressed in terms of a symplectic basis of three-cycles $\Sigma_A=(A_0,A_i,B_i,B_0)$, $A=0,\ldots,2h^{2,1}(M)+1$, $i = 1,\ldots,h^{2,1}(M)$, where $A_I \cdot B^J=\delta_I^{\phantom{I}J}$, $I,J = 0,\ldots,h^{2,1}(M)$, such that
\begin{equation}
 \Pi=\begin{pmatrix} X^0 \\ X^i \\ \mathcal{G}_i \\ \mathcal{G}_0 \end{pmatrix} ,\quad X^I = \int_{A_I}\Omega_3 , \quad \mathcal{G}_I = \int_{B^I}\Omega_3 .
\end{equation}
A convenient way to parametrize the complex structure moduli space is to take half of the periods (e.g.\ the $X^I$) as $h^{2,1}(M)+1$ projective coordinates. The scaling redundancy of $\Omega_3$ can be `gauge fixed' by setting one of those coordinates, say $X^0$, to unity. The other $h^{2,1}(M)$ coordinates $z^i=X^i/X^0$ are called \textit{special coordinates} and completely determine the complex structure of $M$. Therefore, the remaining $h^{2,1}+1$ periods $\mathcal{G}_I$ are not independent, but functions of those special coordinates: $\mathcal{G}_I=\mathcal{G}_I(z)$. In fact it turns out that the complex structure moduli space is fully described by a holomorphic prepotential homogeneous of degree two in the projective variables $X^I$, such that $\mathcal{G}_I = \partial_I \mathcal{G}$.

The general form of the prepotential expanded around the point of large complex structure, including lower order corrections, reads \cite{Hosono:1994av,Grimm:2009ef}
\begin{equation}
\label{prepot}
 \mathcal{G}(X)=\frac{1}{3!}\kappa_{ijk}\frac{X^iX^jX^k}{X^0}+\frac{1}{2}a_{ij}X^iX^j+b_iX^iX^0+\frac{1}{2}c(X^0)^2+\mathcal{G}_{\text{inst}}(e^{2\pi i z}),
\end{equation}
where $\kappa_{ijk}$ are the triple intersection numbers of the mirror manifold $W$, the quantities $a_{ij}$, $b_i$ are real numbers, $c$ is purely imaginary and $\mathcal{G}_{\text{inst}}$ is an infinite sum over exponential terms $\sim e^{2\pi i z^k}$. At leading order this has precisely the form \eqref{prepotk11} (up to a sign which is purely a choice of convention).
At the level of the periods one finds (after setting $X^0=1$)
\begin{equation}
\label{lcsperiods}
 \Pi=\begin{pmatrix} 1 \bigstrut\\ z^i  \bigstrut\\ \frac{1}{2}\kappa_{ijk}z^jz^k + a_{ij}z^j + b_i +\mathcal{O}(e^{2\pi i z})\bigstrut\\ -\frac{1}{3!}\kappa_{jkl}z^j z^k z^l + b_j z^j + c + \mathcal{O}(e^{2\pi i z})\bigstrut \end{pmatrix}.
\end{equation}

The K\"ahler potential derived from the general prepotential \eqref{prepot} has the shift-symmetric structure \eqref{lokp} (i.e.\ it is invariant under $z \rightarrow z+ \delta$, $\delta \in \mathds{R}$) at the perturbative level. It coincides with the Type IIA K\"ahler potential for the K\"ahler moduli at large volume upon the identification
\begin{equation}
\label{mirrormap}
 \mathbf{t}^i=z^i .
\end{equation}
The expression \eqref{lokp} is corrected only by a constant term proportional to $c$ and exponentially suppressed contributions which break the continuous symmetry to a discrete one (see \cite{Wen:1985jz} and footnote 33 in \cite{Hosono:1994av}). 
However, the continuous symmetry remains intact approximately in the large complex structure limit, since the corrections are negligible for $\Im (z^i) \gg  1 \ \forall i$.

We can understand this fact from a more fundamental perspective on the Type IIA side: The complex partners of the K\"ahler deformations $t^i$ are the zero modes $b^i$ of the two-form field $B_2$. The shift symmetry $\bft^i \rightarrow \bft^i + \delta$, $\delta \in \mathds{R}$ has its origin in the gauge symmetry $B_2 \rightarrow B_2 + {\d}\Lambda_1$ of the two-form $B_2$ in the ten-dimensional theory. In the effective field theory it is respected to all orders in perturbation theory but it gets broken to a discrete shift symmetry by worldsheet instantons on two-cycles \cite{Wen:1985jz}. The correction to the prepotential is thus exponentially suppressed by the volumes of the wrapped two-cycles. This explains the presence of an approximate shift symmetry in the limit where the two-cycle volumes on the IIA side become large, which corresponds to large complex structure on the Type IIB side. 

In summary, mirror symmetry relates the $\mathcal N =2$ Type IIA K\"ahler moduli space at large volume to the $\mathcal N =2$ Type IIB complex structure moduli space at large complex structure. The identifications can be summarized as
\begin{equation}\label{mm}
\begin{array}[h]{ccc}
 \text{IIB on }M && \text{IIA on }W \bigstrut[b] \\
 z^k &\longleftrightarrow& \bft^k \bigstrut \\
 \mathcal{G}(z) &\longleftrightarrow& \mathcal{F}(\bft) \bigstrut \\
i\int_M \Omega_3 \wedge \ov \Omega_3 &\longleftrightarrow& \int_W J \wedge J \wedge J. \bigstrut[t] \\
\end{array}
\end{equation}

\subsection{Mirror Symmetry for Orientifolds}
For a mirror pair of Calabi-Yau orientifolds $(M/\sigma_B,W/\sigma_A)$ the same story holds. This has been analyzed in \cite{Grimm:2004ua} and it turns out that one finds essentially just a truncated version of $\mathcal{N}=2$ mirror symmetry discussed in the previous section: The map from the Type IIB complex structure moduli space to the Type IIA K\"ahler moduli space is exactly the same as in the $\mathcal{N}=2$ case with the only difference that on either side $h^{2,1}_+(M)=h^{1,1}_+(W)$ fields are projected out by the orientifold action. One crucial difference is in the structure of loop corrections: In the Calabi-Yau case they are very restricted and do not mix K\"ahler and complex structure moduli. In particular, the one-loop correction in Type II compactifications is explicitly known and higher order corrections are argued to be absent \cite{RoblesLlana:2006ez}. On the other hand, orientifold compactifications have a much richer structure of loop corrections (see e.g.\ \cite{Berg:2005ja}). They exist 
to all orders and intertwine K\"ahler and complex structure moduli spaces. We will discuss loop corrections in \secref{String Loop Corrections}.

\subsection{The Strominger-Zaslow-Yau Conjecture}
\label{syzformulation}
\begin{figure}[t]
	\centering
\begin{overpic}[width=\textwidth]{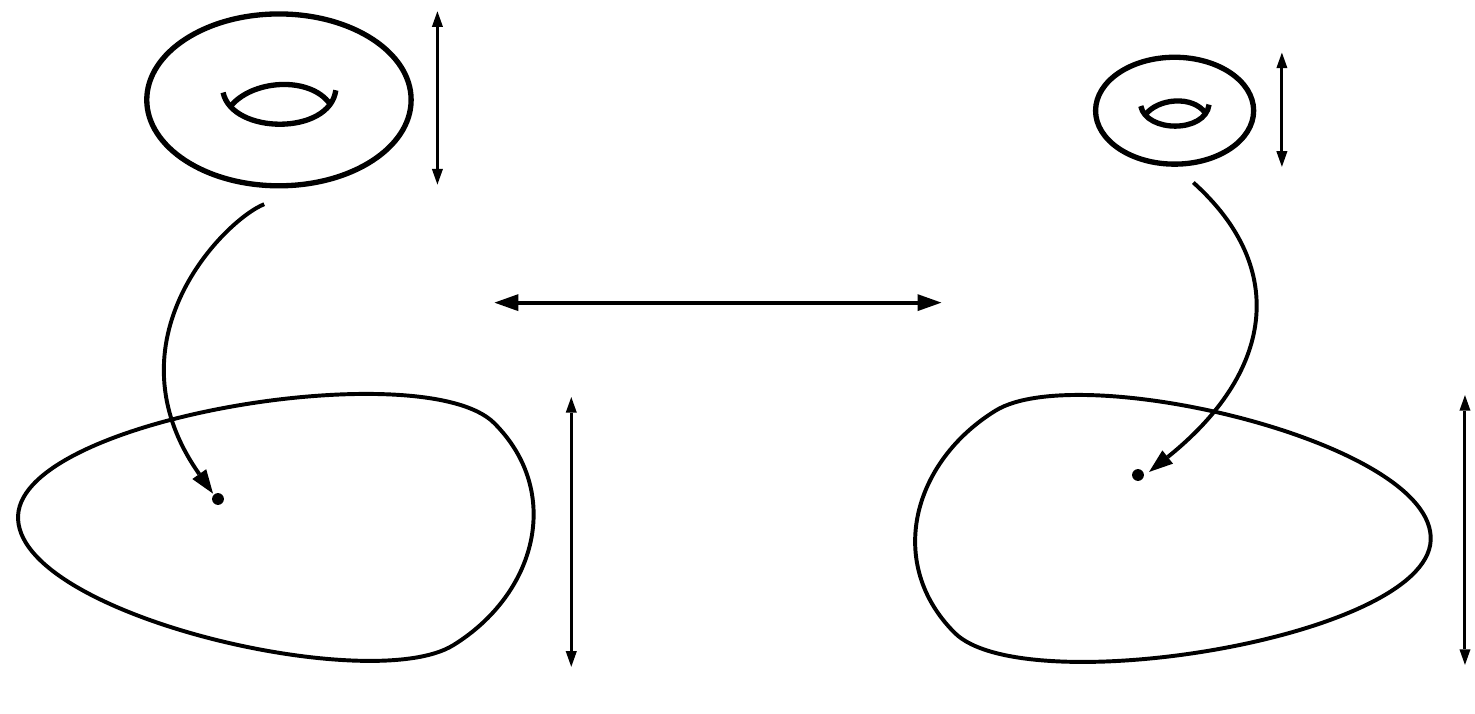}
 \put (1,18) {$B_3$}\put (7,44) {$T^3$}\put (31,41) {$R^s$} \put (40,11) {$L^s$}  \put (40,29) {mirror symmetry}\put (87,40) {$\frac{1}{R^s}$}\put (100,11) {$L^s$}
\end{overpic}
\caption{Mirror symmetry in the SYZ-picture. A Calabi-Yau is described as a $T^3$-fibration over a three dimensional base $B_3$. Mirror symmetry is realized as a chain of three T-duality transformations along the torus fiber, mapping a cycle of length $R^s$ to a cycle of length $1/R^s$.}
	\label{fig:syz}
\end{figure}
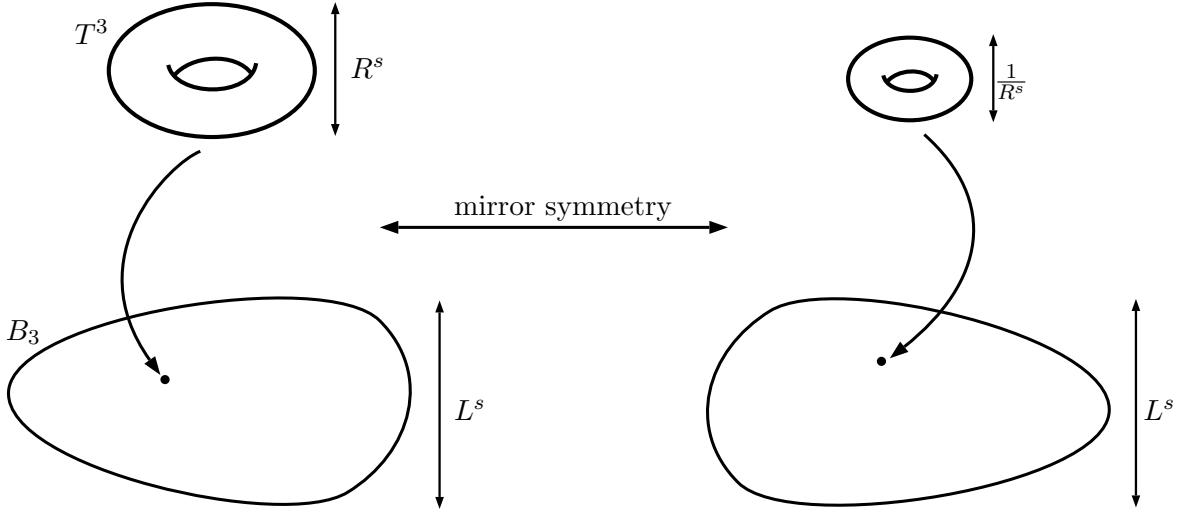

It is widely believed, that mirror symmetry holds for the full quantum string theory. In reference \cite{Strominger:1996it} Strominger, Zaslow and Yau derive implications of this statement for the geometry of Calabi-Yau manifolds. In particular these authors conjecture that every Calabi-Yau can be described as a $T^3$-fibration and mirror symmetry is a chain of three T-duality transformations along the fibers. This is depicted schematically in figure~\ref{fig:syz}.

Using this picture, we can reproduce the essential properties of the mirror map: Consider a model with only one complex structure modulus and one K\"ahler modulus. In the SYZ-picture this is realized in the most simple way by assuming a $T^3$-fiber of typical string frame length scale $R^s$ and a base of typical string frame length scale $L^s$ (cf.\ figure~\ref{fig:syz}). The two-cycle volume $\Im (\bft)$ and the complex structure modulus $z$ then scale as 
\begin{eqnarray*}
\Im (z) &\sim & \frac{L^s}{R^s}, \\
\Im (\bft) &\sim & L^sR^s.
\end{eqnarray*}
In this picture, large volume means large fiber and base size $L^sR^s\gg 1$, whereas large complex structure means that the fiber size is small compared to the base $\frac{L^s}{R^s} \gg 1$. Note that both limits can be taken simultaneously, provided that $L^s \gg R^s$. To figure out the mirror map, let us start with a Type IIB compactification with string coupling $g_s^B$ and define $\Im (z_B)=\frac{L^s}{R^s_B}$. Now perform three T-duality transformations along the fiber directions. One such transformation acts as (see e.g.\ \cite{Polchinski:1998rq})
\begin{equation*}
 \begin{array}[h]{lll}
 R^s &\longrightarrow& \frac{(2\pi)^2\alpha'}{R^s}, \bigstrut \\
 g_s &\longrightarrow& \frac{2\pi\sqrt{\alpha'}}{R^s} g_s. \bigstrut \\
\end{array}
\end{equation*}
All together we find the following correspondence (setting $\ell_s \equiv 2\pi\sqrt{\alpha'}=1$ for simplicity):
\begin{equation}
\label{mirrormapsyz}
\begin{array}[h]{lll}
 \Im (z_B) &=& \frac{L^s}{R^s_B} = L^sR^s_A = \Im (\bft_A), \bigstrut \\
 g_s^B &=& (R^s_B)^3g_s^A= \sqrt{\frac{\ccV_B^s}{\ccV_A^s}}\; g_s^A, \bigstrut \\ 
\end{array}
\end{equation}
in agreement with equation \eqref{mm}. The mirror map between the K\"ahler variables of the Type IIB K\"ahler moduli space and the Type IIA complex structure moduli space is more involved and will not be of importance in our discussion. For a detailed analysis see e.g.~\cite{Grimm:2004ua}.

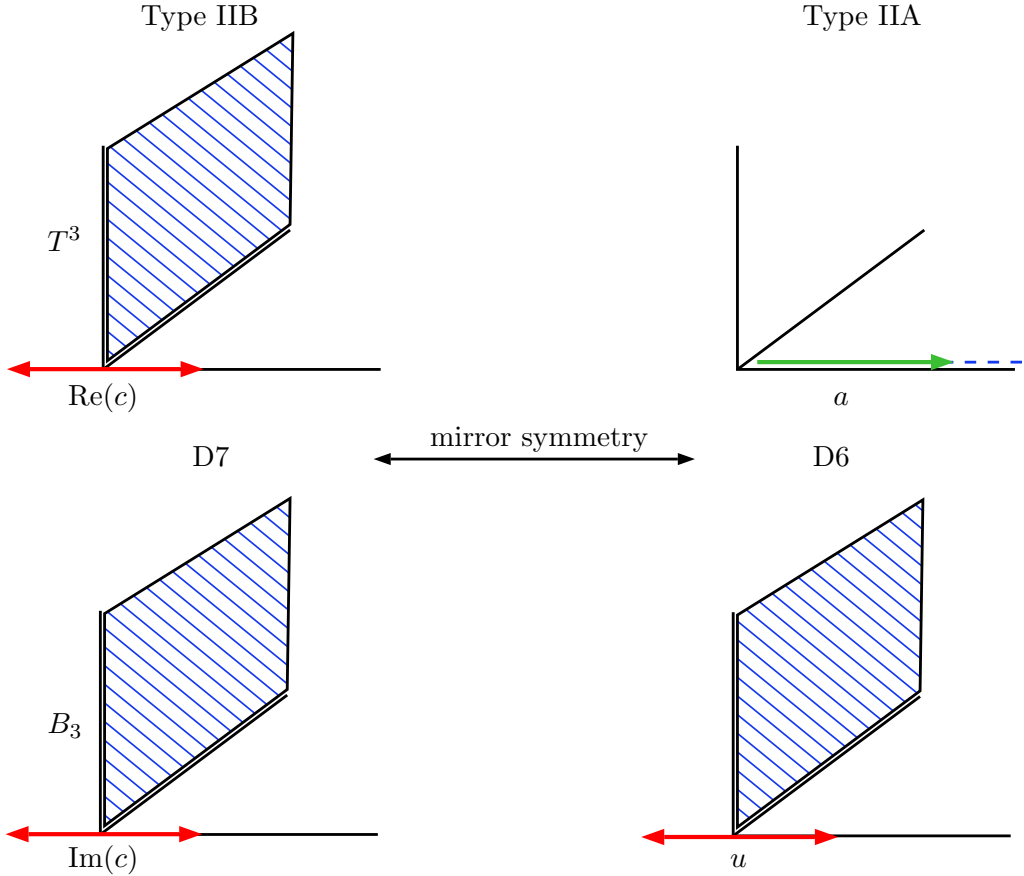
\begin{figure}[t]
	\centering
  \begin{overpic}[width=0.9\textwidth]{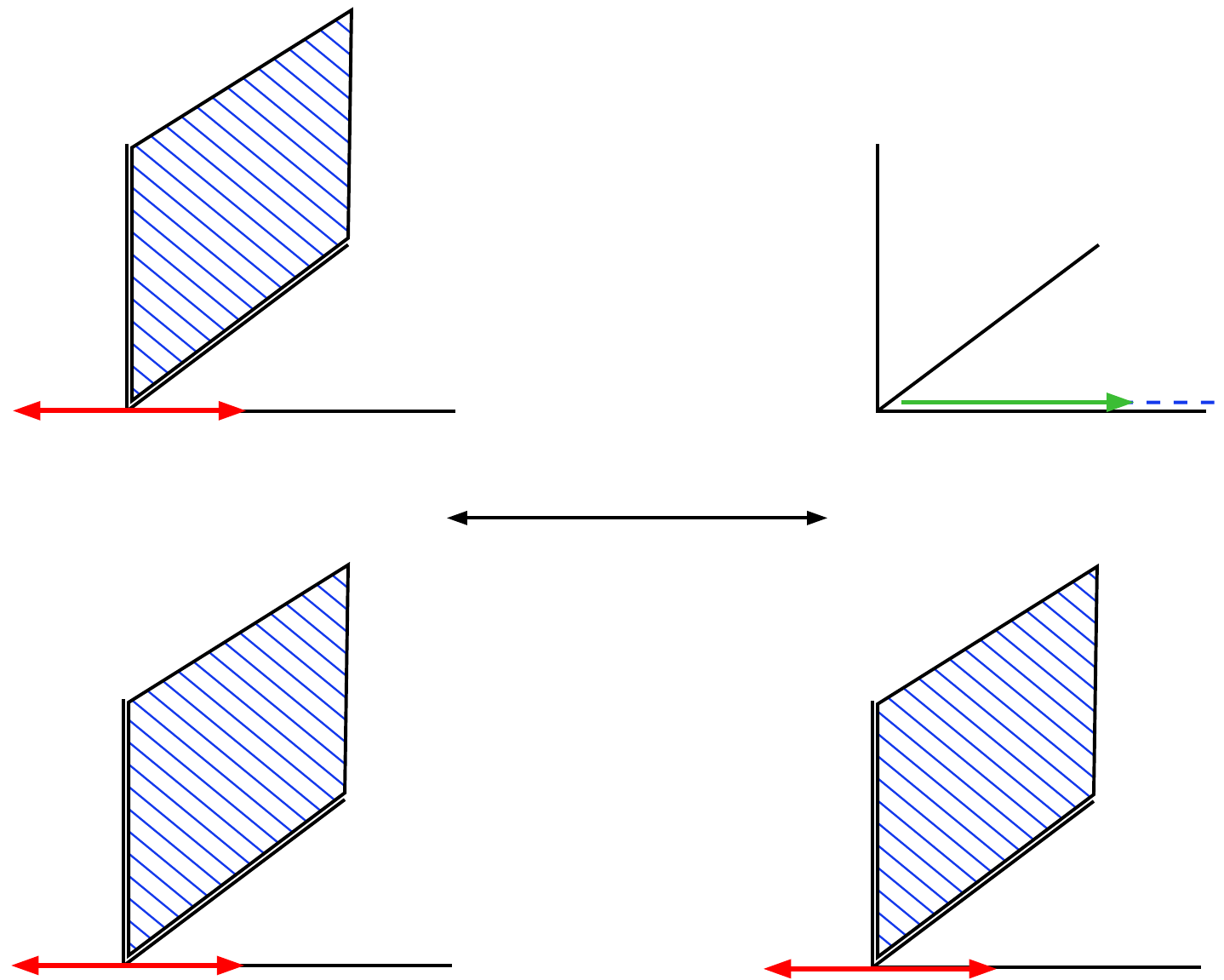}
 \put (5,11) {$B_3$}\put (5,58) {$T^3$}\put (19,37) {D7} \put (14,80) {Type IIB}\put (78,80) {Type IIA} \put (7,-2) {$\Im (c)$}  \put (42,39) {mirror symmetry}\put (79,37) {D6}\put (71,-2) {$u$}\put (81,43) {$a$}\put (7,43) {$\Re(c)$}
\end{overpic}
	\caption{A mirror brane configuration. The upper triads indicate the three real directions of the $T^3$-fiber while the lower ones correspond to the three directions in the base $B_3$. The D7-brane position $\Re (c)$ in the SYZ-fiber in Type IIB corresponds to a Wilson line $a$ on the Type IIA side, while the D7-brane position $\Im (c)$ in the base is related to the D6-brane position $u$ in the base of the mirror manifold.}
	\label{fig:syzbranedef}
\end{figure}

\subsection{Approximately Flat Directions at Large Complex Structure}
\label{flatdirectionslcs}
We have seen in \secref{IIBAtLCS} that the complex structure moduli space of a Calabi-Yau three-fold has an approximate shift symmetry in a certain corner of the complex structure moduli space -- the large complex structure limit. In this section we give an argument why this should also be true for D7-brane deformations. This approximate shift symmetry in the D7-brane sector has been explored in the context of Higgs phenomenology in \cite{Hebecker:2012qp,Hebecker:2013lha}. Our arguments closely follow their analysis.

In the SYZ-picture discussed in \secref{syzformulation}, consider a D7-brane wrapped on a holomorphic four-cycle with two legs along the fiber and two legs along the base. Deformations of that D7-brane in the normal directions are described by a complex scalar $c$.
The components $\Re (c)$ and $\Im (c)$ measure brane deformations along the fiber and base direction, respectively. Under mirror symmetry this brane configuration is mapped to a D6-brane in Type IIA string theory, wrapping a special Lagrangian three-cycle with the same two legs in the base but extending along the transverse cycle in the fiber. This is depicted in figure~\ref{fig:syzbranedef}.

Moving the D7-brane along the fiber corresponds to turning on a Wilson line $a=\int_{\gamma}A$ along the D6-brane direction in the fiber on the IIA side. Here, $\gamma$ is the cycle in the fiber which is wrapped by the D6-brane. Together with the real brane deformation modulus $u$ in the base, the Wilson line makes up a complex scalar $a + iu$ which, under mirror symmetry, is related to the D7-brane deformation $c$ \cite{Grimm:2011dx}.

\begin{figure}[t]
	\centering
  \begin{overpic}[width=0.5\textwidth]{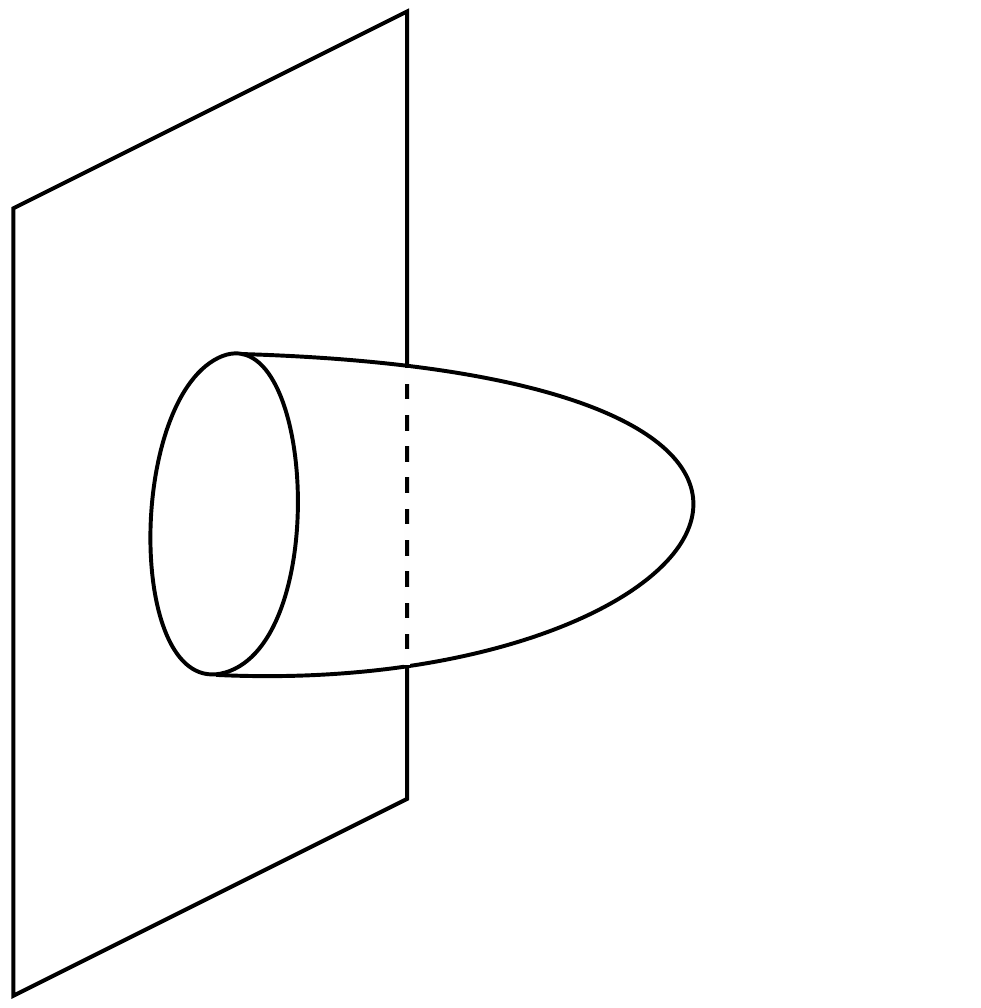}
 \put (18,3) {D6}\put (12,62) {$\gamma$}\put (69,58) {$w$}
\end{overpic}
	\caption{A disk instanton. The boundary of the disk is the one-cycle $\gamma$ of the Wilson line whose K\"ahler potential gets corrected non-perturbatively by terms $\sim e^{-w}$. Here, $w$ is the area of a minimal disk with boundary $\gamma$.}
	\label{fig:disk}
\end{figure}

The crucial point is that, to all orders in $\alpha'$, the effective action of a Type IIA compactification is invariant under shifts of the Wilson line scalar $a$ by a real constant. This comes about because $a$ couples only derivatively in the effective action \cite{Kerstan:2011dy,Grimm:2011dx,Hebecker:2012qp,Hebecker:2013lha}. The origin of this fact lies in the 10d gauge symmetry for the gauge potential $A$ in the worldvolume theory on the brane. The corresponding K\"ahler potential in Type IIB language reads
\begin{equation}\label{eq:shiftsymmetricKahler}
  \ccK \supset - \ln \left(-i \left(\tau_0 - \ov \tau_0\right) - k_{\text{D7}}\left(z, \ov z; c - \ov c\right)\right).
\end{equation}

The shift symmetry gets broken by gauge theory loops (to be discussed in \secref{String Loop Corrections}) and non-perturbative effects from disk instantons, i.e.\ holomorphic maps from the open-string worldsheet into the Calabi-Yau with boundary on the D-brane \cite{Kachru:2000ih}.
For a disk ending on a topologically non-trivial one-cycle $\gamma$ the corrections to the superpotential and the K\"ahler potential are proportional to $e^{-w}$. Here, $w=\int_{D}J$ is the area of a minimal disk $D$ whose boundary $\partial D=\gamma$ is the cycle on which the Wilson line lives (see figure~\ref{fig:disk}).

In our toy model the Wilson line cycle $\gamma$ would be the one-cycle in the $T^3$-fiber wrapped by the D6-brane. It is plausible to assume that, in the large volume limit where the two-cycle volume $t_A$ becomes large, the area of the disk becomes large as well. Intuitively, if the two-cycle volume scales as $t_A \sim \lambda$, the length of the Wilson line cycle scales as $r(\gamma) \sim \sqrt{\lambda}$. We conjecture that $w = \beta t_A$, where $\beta$ is some constant.
Then all  corrections depending on the Wilson line scalar $a$ to the effective theory are exponentially suppressed by $e^{-\beta t_A}$ and hence the real Wilson line scalar exhibits a shift symmetry $a \rightarrow a + \delta$, $\delta \in \mathds{R}$ in the large volume limit. On the Type IIB side the relevant limit is the large complex structure limit. According to these considerations we expect an approximate shift symmetry
\begin{equation}
c \rightarrow c + \delta, \quad \delta \in \mathds{R} 
\end{equation}
of the D7-brane deformations along the fiber direction in the limit of large complex structure. Furthermore, according to this reasoning, all shift-symmetry-breaking $\alpha'$-corrections to the K\"ahler potential and superpotential of the effective theory are exponentially suppressed by a factor $e^{-\beta \Im (z_B)}$.

\subsection{Shift Symmetry from the Fourfold Perspective}
Mirror manifolds do exist in complex dimensions larger than three \cite{Greene:1993vm}. For a recent analysis of fourfold K\"ahler potentials see e.g.\ \cite{Honma:2013hma}. There it was found that the K\"ahler potential for the complex structure moduli of M-theory compactified on a Calabi-Yau fourfold at large complex structure takes the form
\begin{equation}\label{eq:CSofFourfold}
 \ccK_{\text{CS}}^{\text{LCS}} = -\ln\left(\frac{1}{4!}\kappa_{ijkl}(z^i - \ov z^i)(z^j - \ov z^j)(z^k - \ov z^k)(z^l - \ov z^l) + \ldots\right).
\end{equation}
This suggests to conjecture that the 7-brane coordinate, being a complex structure modulus of the F-theory fourfold, has a shift symmetry at large complex structure of the fourfold. We now give some more details on how we think the shift symmetry comes to the fore in the F-theory formalism. This view complements the arguments presented in \secref{flatdirectionslcs}. 

Recall from \secref{ModStabGen} that the K\"ahler potential of the $\mathcal{N}=1$ effective action of F-theory on an elliptically fibered Calabi-Yau fourfold $Y$ reads
\begin{equation}
 \mathcal K_{\text{CS}} = -\log \Pi_I Q^{IJ} \ov \Pi_J,
\end{equation}
where $\Pi_I$, $I=1,...,b^4(Y)$ are the periods\footnote{A method for finding an integral monodromy basis for the periods of the fourfold, in analogy to \eqref{lcsperiods}, has been developed in \cite{Bizet:2014uua}.} of the holomorphic four-form $\Omega_4$ integrated over a basis of four-cycles $\Sigma_I$ with intersection form $Q_{IJ}=\Sigma_I \cdot \Sigma_J$. We have seen that in the weak coupling limit it is possible to choose a basis of four-cycles that allows us to identify the brane and bulk moduli dependent parts of the K\"ahler potential:
\begin{equation}\label{K_CS in WKL}
 \mathcal K_{\text{CS}}^{g_s \to 0} = -\log\left((\tau_0 - \ov\tau_0)\Pi_i(z) Q^{ij}\ov\Pi_j(\ov z)+ f(z,\ov z;c,\ov c)\right),
\end{equation}
where $\Pi_i$ are the periods of the orientifold $\wt X/\sigma$ and 
\begin{equation}
 f(z,\ov z;c,\ov c)=\chi_i(z,c) Q^{ij}\ov \Pi_j(\ov z) + \ov\chi_i (\ov z, \ov c)Q^{ij}\Pi_j(z) - \Pi_{\alpha}(z,c) Q^{\alpha \beta}\ov\Pi_{\beta} (\ov z,\ov c)
\end{equation}
is the brane and bulk moduli dependent correction to the K\"ahler potential.
Rewriting \eqref{K_CS in WKL}, one recovers the familiar structure of the Type IIB K\"ahler potential upon dimensional reduction \cite{Jockers:2004yj,Jockers:2005zy}
\begin{equation}
 \mathcal K_{\text{CS}}^{g_s \to 0} = -\log\left(-i(\tau_0 - \ov\tau_0)+ \frac{f(z,\ov z;c,\ov c)}{i\Pi_i Q^{ij}\ov\Pi_j}\right) - \log \left(i\Pi_i Q^{ij}\ov\Pi_j\right).
\end{equation}

In view of \eqref{eq:CSofFourfold} we expect that at large complex structure of $ Y$ the correction $f(z,\ov z;c,\ov c)$ takes a form in which there is a shift symmetry in the brane moduli space, i.e.\ $c\to c + \delta$, $\delta \in \mathds{R}$. Thus, in this limit we precisely reproduce \eqref{eq:shiftsymmetricKahler}. The above statement is actually weaker than our conjecture from \secref{flatdirectionslcs} which states that the shift symmetry is present at weak coupling $\Im (\tau_0) \gg 1$ and large complex structure of the base $\wt X/\sigma$. From the fourfold perspective it is not clear why there should not be shift-symmetry-violating terms that are suppressed by a factor $\sim e^{-\Im (c)}$ only. However, even if these terms are present, we can, in principle, suppress them by making $\Im (c)$ large (recall that inflation occurs in the direction of $\Re (c)$). Motivated by the considerations in \secref{flatdirectionslcs} we do, nonetheless, believe that the shift symmetry indeed exists at weak 
coupling and large complex structure of the base~$\wt X/\sigma$.

An instructive explicit example is again F-theory on the fourfold $\tK \times K3$ as discussed in \secref{modulispacek3}. The brane-position- (i.e.\ $C^a$-) dependent part of the K\"ahler potential in this case reads
\begin{equation}
 \mathcal K_{\text{CS}}^{K3} = -\log \left[-\left( (S - \ov S)(U - \ov U) - \sum\limits_{a}(C^a - \ov C^a)^2 \right)\right] + \ldots ,
\end{equation}
where $S \equiv \tau_0$, $U$ is the complex structure of $K3$ in the orientifold limit, and the $C^a$ are 16 brane positions.\footnote{This K\"ahler potential is exact, i.e.\ there are no instanton-type corrections (this is due to certain integrability conditions of the Picard-Fuchs equations, see e.g.\ \cite{Jockers:2009ti}).}

What changes if one considers a pair of D7-branes (as in fluxbrane inflation) instead of one isolated brane (as in D7-brane chaotic inflation)? As indicated in \eqref{K_CS in WKL} the $\tau_0$-dependence of the K\"ahler potential in the weak-coupling limit is very simple. In particular, $\tau_0$ does not show up in $f(z,\ov z;c,\ov c)$. Corrections of the K\"ahler potential in the weak coupling limit due to brane-brane interactions (which are higher order in $\Im (\tau_0)^{-1}$) are thus exponentially suppressed in $\tau_0$ \cite{Denef:2008wq} and therefore not part of $f(z,\ov z;c,\ov c)$ (the branes don't `see' each other at this order). Consequently, in analogy to the $K3$-example, we expect the function $f(z,\ov z;c,\ov c)$ to be additive in a suitable parametrization of the brane positions. Terms which break this structure are suppressed as $\sim e^{-\Im (\tau_0)}$.\footnote{In addition there are of course the usual gauge-theory loops which add corrections to the scalar potential. But this is a 
different issue and will be discussed in \secref{String Loop 
Corrections}.}

In this context it would also be interesting to exploit the existing literature on $\mathcal{N}=1$ mirror symmetry \cite{Lerche:2002yw} which deals with the calculation of disk instanton corrections to a mirror Type IIA compactification with branes. The idea is roughly that the period vector of the closed-string sector can be extended by so-called relative periods which encode the geometry of the open-string sector and that this period vector satisfies differential equations which are similar to the Picard-Fuchs equations. Schematically,
\begin{equation}
 \Pi(z,\hat z) = (1,z,\hat z,\partial_z\mathcal{G}(z), \mathcal{W}(z,\hat z),...),
\end{equation}
where $\mathcal{G}(z)$ is the prepotential of the bulk moduli space, $\hat z$ is a relative period related to brane deformations, and $\mathcal{W}(z,\hat z)$ encodes the open- and closed-string instanton contribution to the superpotential of the mirror Type IIA brane configuration.\footnote{It is interesting that the periods displayed on p.\ 48 in \cite{Jockers:2009ti}, which were obtained using the open-closed duality \cite{Mayr:2001xk} (which determines the geometry of a toric open-closed background in terms of a toric Calabi-Yau fourfold), give rise to exactly the same type of K\"ahler potential as in equation \eqref{kahlerpot1}.} It is conjectured that $\mathcal{W}(z,\hat z)$ admits an expansion of the form \cite{Ooguri:1999bv}
\begin{equation}
 \mathcal{W}(z,\hat z) = \sum\limits_{k=0}^{\infty} \sum\limits_{\alpha, \, \beta}n_{\alpha, \beta}q^{k\alpha}p^{k \beta},
\end{equation}
where $q=e^{2\pi i z}$, $p=e^{2 \pi i \hat z}$ and $n_{\alpha, \beta}$ are integers, referred to as Ooguri-Vafa invariants. One attempt to find suitable models where D7-brane inflation at large complex structure can be realized could be to look for open-closed backgrounds where $n_{0, \beta}=0\ \forall \beta$, such that all $\hat z$-dependent corrections are exponentially suppressed by the bulk complex structure $z$.

\section{Loop Corrections to the K\"ahler Potential}
\label{String Loop Corrections}
In this section we will discuss string loop corrections to the K\"ahler potential. The superpotential will not be affected by perturbative effects (cf.\ \cite{Witten:1985bz})
due to the standard non-renormalization theorem.\footnote{Because of its holomorphicity, volume moduli can enter the superpotential only in their complexified version, i.e.\ paired up with an axion. Such an axion, however, enjoys a shift symmetry at the perturbative level, which thus forbids an appearance of the corresponding complex field in the tree-level superpotential. Consequently, the superpotential does not depend on the K\"ahler moduli at this order. The axionic shift symmetry can be broken by
non-perturbative effects, which then induce a K\"ahler moduli dependence in the superpotential at the non-perturbative level.}
On one hand, the K\"ahler potential corrections play an essential role for K\"ahler moduli stabilization. On the other hand, they will generically
depend on the open-string moduli, in particular on the D7-brane moduli. A proper discussion of the induced terms in the scalar potential is thus crucial for inflation phenomenology.

Regarding string loop corrections in Type IIB orientifolds without any dependence on open-string or complex structure moduli (i.e.\ those moduli are assumed to be fixed at some higher scale and the only light fields are K\"ahler moduli), it was shown \cite{vonGersdorff:2005bf,Berg:2007wt,Cicoli:2007xp} that the leading order contributions to the scalar potential induced by those corrections cancel due to the `extended no-scale' structure. This structure renders $g_s$-corrections less important in the limit of large volume than, for example, $\alpha'$-corrections \cite{Becker:2002nn}. We will demonstrate that the extended no-scale structure holds even when including branes, at least in an exemplifying toy model.

\subsection{Tree-Level Masses in D7-Brane Inflation}
To set the stage we calculate some tree-level masses for open-string moduli in D7-brane inflation. More precisely, we consider the open-string sector of a pair of D7-branes and compute masses for the components of the 4d scalar SU(2) multiplet which contains the relative deformation modulus of the two branes. In the fluxbrane inflation model, this multiplet describes, amongst others, the inflaton field. As we will see, the structure of the K\"ahler and superpotential which is forced upon us by string theory is crucial for both D7-brane inflation models to be viable: Assuming a shift-symmetric K\"ahler potential for the brane deformation modulus, the only term which violates the shift symmetry in the $F$-term potential corresponds to the SUSY mass term for the zero modes which couple to the deformation modulus (the waterfall fields in fluxbrane inflation). By this mass term, these modes are stabilized at zero vev during inflation.

To obtain this result, recall that the open-string sector of a pair of D7-branes can be described in terms of a higher dimensional SU(2) multiplet, which is a $\ccN = 1 $ vector multiplet in 8d, consisting of a vector, a complex scalar, and fermions. These components can be thought of as arising from the reduction of a 10d vector to 8d. From the 4d perspective we can construct the 8d multiplet in terms of several 4d $\ccN =1$ multiplets, namely a vector multiplet and three chiral multiplets (see e.g.\ \cite{ArkaniHamed:2001tb}).

The internal components of the 10d gauge field will be the lowest components of the chiral superfields $\phi_i$,
\begin{equation}
 {\phi_j}_{\mid \theta = \ov \theta = 0} = \frac{1}{\sqrt{2}} \left(A_{4+2j}+iA_{3+2j} \right), \quad j \in \{1,2,3\}.
\end{equation}
With this given field content, three scalars and one vector, one can now go on and build an action. This has been done in \cite{ArkaniHamed:2001tb}:
\begin{align}
\begin{split}
S_{10} &= \int d^{10}x \int d^2 \theta \textnormal{Tr} \left( \frac{1}{4kg^2}W^\alpha W_\alpha
+ \frac{1}{2kg^2}\epsilon^{ijk}\phi_i\left(\partial_j\phi_k + \frac{1}{\sqrt{2}}[\phi_j, \phi_k] \right) \right) \\
&+ \int d^{10} x \int d^4 \theta \frac{1}{kg^2} \textnormal{Tr} \left( (\sqrt{2}\ov \partial^i + \ov \phi^i)
e^{-V}(-\sqrt{2}\partial_i + \phi_i)e^V + \ov \partial^ie^{-V}\partial_i e^V\right) \\
&+ \text{WZW term},
\end{split}
\end{align}
with $\phi \equiv \phi^a T^a$, $T^a$ being the generators of SU(2), $\phi^a := \{c, \chi^1, \chi^2 \}$, $\operatorname{Tr} T^aT^b = k\delta^{ab}$,
$W_\alpha = -\frac{1}{4} \ov D\,  \ov D e^{-V}D_\alpha e^V$ \cite{Bagger:1990qh}.

After compactification to 4d (and using the same symbol for the 10d fields and their zero-modes in 4d) one can read off the superpotential
\begin{equation}
W \sim \textnormal{Tr} \left( \epsilon^{ijk}\phi_i [\phi_j, \phi_k] \right)
\end{equation}
and the K\"ahler potential
\begin{equation}
K\sim \textnormal{Tr} \left(\ov \phi^i \phi_i\right).
\end{equation}
Using the structure constants of the SU(2) algebra together with $\operatorname{Tr} T^aT^b = k\delta^{ab}$ one finds that the only non-vanishing terms in the superpotential are the
ones $\sim \phi_i^a \phi_j^b \phi_k^c$, where $a \neq b \neq c$ and $i \neq j \neq k$. We now recall the parametrization of $\phi$ in terms of $c$, $\chi^1$, $\chi^2$ to obtain the superpotential
\begin{equation}
W \sim \lambda c_i \chi_j^1 \chi_k^2, \quad i\neq j\neq k.
\end{equation}
Furthermore, the K\"ahler potential is given by
\begin{equation}\label{eq:CanonicalKahlerWaterfall}
K \sim \ov \chi^1_i \chi^1_i + \ov \chi^2_i \chi^2_i + \ov c_i c_i. 
\end{equation}

The inflaton in D7-brane inflation will be associated with the diagonal (neutral) component of the 8d complex SU(2) scalar. Let's call this component $c_3$ for definiteness. Due to the completely antisymmetric structure of the superpotential, it is multiplied by components of the 4d chiral SU(2) multiplets which arise from dimensionally reducing the 8d vector. Those components are $\chi^1_1$, $\chi^2_2$ as well as $\chi^2_1$, $\chi^1_2$. For simplicity we focus only on $\chi^1_1 \equiv \chi_1$ and $\chi^2_2 \equiv \chi_2$. The origin of these fields is important for extracting the K\"ahler moduli dependence of their {\it supergravity} K\"ahler potential. While the K\"ahler potential for the component transverse to the brane is given by $K\sim c\ov c$, the one for the components parallel to the brane reads $K\sim \chi_i \ov \chi_i / (\bT + \ov \bT)$ \cite{Aparicio:2008wh},\footnote{Note that this expression is related to \eqref{eq:CanonicalKahlerWaterfall} via a field redefinition.} where $\bT$ is a K\"ahler 
modulus whose real part $\Re(\bT) = T$ measures a four-cycle volume. Additionally, from our previous considerations we expect the K\"ahler potential to have a shift-symmetric structure, such that it is independent of $\Re (c)$. We therefore work 
with
\begin{equation}
 K = -3\ln \left(\bT + \ov \bT\right) - \frac{\left(c - \ov c \right)^2}{2} + \frac{1}{\bT + \ov \bT} \left(\chi_1 \ov \chi_1 + \chi_2 \ov \chi_2\right)
\end{equation}
and
\begin{equation}
 W = W_0 + \lambda c \chi_1 \chi_2.
\end{equation}

The scalar $F$-term potential computed from these quantities reads
\begin{align}\label{eq:F-termPotForSU2}
 V_F =& e^{-\frac{1}{2}(c - \ov c)^2} \left\{\frac{1}{(\bT + \ov \bT)^2} |\lambda|^2 c\ov c (\chi_1 \ov \chi_1 + \chi_2 \ov \chi_2 ) \right.\nonumber\\
 & -\frac{1}{(\bT + \ov \bT)^3}\left[(c - \ov c )^2 |W_0|^2 +\left\{ (-(c-\ov c) + c(c-\ov c)^2)  \ov W_0 \lambda\chi_1 \chi_2 + \text{h.c.}\right\}\right]\nonumber\\
 & \left.-\frac{1}{(\bT + \ov \bT)^4}(c-\ov c)^2(\chi_1 \ov \chi_1 + \chi_2 \ov \chi_2 ) |W_0|^2\right\}\nonumber\\
 &+ \text{terms higher order in }\chi_i \text{ and } \frac{1}{(\bT + \ov \bT)}.
\end{align}
There are several observations to be made:
\begin{itemize}
 \item After canonically normalizing $\chi_i \to \chi_i\sqrt{\bT + \ov \bT} $, the first term (being the SUSY mass for the fields which couple to the inflaton in the superpotential) scales with $\gym \sim (\bT + \ov \bT)^{-1}$, exactly as expected from the analysis of \cite{Hebecker:2011hk}.
 \item The second term is the only one without $\chi_i$-dependence. It fixes the imaginary part of $c$ at the origin.
 \item All terms except for the first one are proportional to $(c-\ov c)$. Since this difference is stabilized at zero, no SUSY-breaking mass term for the charged fields $\chi_i$ is obtained at tree level. This is similar in spirit to the fact that for a K\"ahler potential of the form $K = -3\ln\left(\bT + \ov \bT - \chi \ov \chi\right)$ known from no-scale supergravity \cite{Cremmer:1983bf,Ellis:1984bm} (see also \cite{Brignole:1997dp}), the potential is exactly flat, i.e.\ no SUSY-breaking $\chi$-mass is induced by non-vanishing $F$-terms of the modulus $\bT$.
 \item A SUSY-breaking mass for the waterfall fields is thus introduced only by subleading effects, such as loop corrections.
\end{itemize}

\subsection[General Form of $g_s$-Corrections and Extended No-Scale Structure]{General Form of {\boldmath $g_s$}-Corrections and Extended No-Scale Structure}
We now consider loop corrections to the above setting.
For general Calabi-Yau manifolds the explicit form of the string loop corrections to the K\"ahler potential is not known.
In the toroidal case the corrections can be computed \cite{Berg:2005ja,Berg:2005yu,Berg:2007wt,Cicoli:2007xp}, however, the complex structure and brane moduli dependence is rather complicated.
It seems thus hard to make an educated guess for this dependence in the general Calabi-Yau orientifold case.
What is somewhat simpler, however, is the K\"ahler moduli dependence in the large volume limit. It was conjectured that the string loop corrections to the K\"ahler potential for general Calabi-Yau orientifolds read \cite{Berg:2007wt}
\begin{equation}\label{eq:GenLoopCorr}
\delta \ccK_{(g_s)} \sim \sum_{i=1}^{h_{1,1}} g_s \frac{C_i (U,c)a_{ij}t^j}{\mathcal{V}}
 + \sum_{i=1}^{h_{1,1}}\frac{D_i(U,c)}{b_{ij}t^j\mathcal{V}}. 
\end{equation}
Here, $U$ and $c$ denote complex structure and open-string moduli. The scaling with the inverse overall volume $\ccV$ was argued to arise from the Weyl rescaling in order to go to the 4d Einstein frame.
The masses of the particles running in the loops (Kaluza-Klein and winding states) scale with some power of two-cycle volumes $t^i$, which results in a corresponding dependence of $\delta \ccK_{(g_s)}$ on those moduli.

Regarding the volume scaling, the corrections \eqref{eq:GenLoopCorr} seem to dominate over $\alpha'$-corrections which are of the general form \cite{Becker:2002nn}
\begin{equation}
\ccK_0 + \delta \ccK_{(\alpha')} = -2 \log \left( \mathcal{V} + \frac{\hat\xi}{2 (2\pi)^3 g_s^{3/2}} \right) 
= -2 \log(\mathcal{V}) - \frac{\hat\xi}{ (2\pi)^3 g_s^{3/2} \mathcal{V}} + \mathcal{O}(1/\mathcal {V}^2).
\end{equation}
Here, $\hat \xi = -\frac{\zeta(3)\chi (M)}{2}$ and $\chi(M)$ is the Euler characteristic of the compactification manifold $M$.
However, due to an intricate cancellation at the level of the scalar potential, the `extended no-scale' structure \cite{vonGersdorff:2005bf,Berg:2007wt,Cicoli:2007xp}, this is not the case, at least assuming $U$ and $c$ to be stabilized at some higher scale. More precisely, if one considers a toy-model with only one K\"ahler modulus $\bT$ (where $\Re(\bT)$ measures a four-cycle volume\footnote{Recall that in the low energy effective action of Type IIB string theory the proper K\"ahler variables are complexified versions of four-cycle volumes.}) one finds
\begin{equation}
\delta V_{(g_s)} \sim \frac{W_0^2}{(\bT + \ov \bT)^5} \ll \frac{W_0^2}{(\bT + \ov \bT)^{9/2}} \sim \delta V_{(\alpha')} \quad \text{for} \quad \Re(\bT) \gg 1.
\end{equation}
In the following we demonstrate that this conclusion remains true even if one includes a further light degree of freedom which, in our case, is the inflaton.

\subsection{Extended No-Scale Structure with Dynamical Branes}
\label{Extended_no-scale_with_branes}
Consider a toy model of Type IIB string theory compactified to 4d whose low energy spectrum contains only one K\"ahler modulus $\bT = T + ic_4$ and dynamical branes. Here, $T$ is the (Einstein frame) volume of the four-cycle $\Sigma$ and 
$c_4$ is the RR four-form $C_4$ reduced along $\Sigma$, i.e.\ $c_4 = \int_\Sigma C_4 $.
As in the previous subsection, the brane deformation modulus whose real part is associated to the inflaton is denoted by $c$. All complex structure moduli $U$ and the
axio-dilaton $S$ are assumed to be fixed through bulk fluxes at a higher scale. Additionally, we will not consider the $\chi_i$ to be dynamical fields, as they are stabilized supersymmetrically by the leading term in the potential \eqref{eq:F-termPotForSU2}.
The leading order K\"ahler potential for the dynamical fields reads
\begin{equation}\label{eq:LeadingKahlerPot}
\ccK_0 = -3 \ln \left(  \bT + \ov \bT \right) + \ccK_{\text{D7}}(c,\ov c),
\end{equation}
where $\ccK_{\text{D7}}(c, \ov c)$ denotes the K\"ahler potential for the open-string moduli $c$.
Motivated by the discussion in the previous sections one can think of
$\ccK_{\text{D7}} \sim (c - \ov c)^2$. 
Let us now consider a correction to the K\"ahler potential of the type \eqref{eq:GenLoopCorr},\footnote{From a Type IIB perspective the term $\ccK_{\text{D7}}(c, \ov c)$ is already a $g_s$-correction, cf.\ \secref{modulispaceftheory} and \cite{Denef:2008wq}:
The K\"ahler potential reads \mbox{$\ccK \supset \ccK_{\tau_0} + \ccK_{\text{CS}} + g_s \ccK_{\text{D7}} + \ccO(g_s^2)$}, where $\ccK_{\tau_0}$ and $\ccK_{\text{CS}}$ denote the
axio-dilaton and complex structure moduli K\"ahler potentials of the Type IIB compactification and $\ccK_{\text{D7}}$ is the brane moduli K\"ahler
potential. The string loop corrections that we are considering here are,
however, of different nature, as they involve volume moduli.} i.e.
\begin{equation}\label{GenCorrKP}
\delta \ccK  =  \frac {\beta (c, \ov c)}{\bT + \ov \bT} ,
\end{equation}
and compute the effect of such a term in the scalar potential
\begin{equation}
V = e^\ccK \left( \ccK^{i \ov \jmath}  D_i W D_{\ov \jmath} \ov W - 3|W|^2 \right).
\end{equation}

In performing this calculation we follow the methods used for similar purposes in \cite{Cicoli:2007xp}: Suppose one would like to calculate the inverse of a matrix $\ccK^0_{\ov \imath j} + \delta \ccK_{\ov \imath j}$, where $\delta \ccK_{\ov \imath j}$ is thought of as a correction to the leading order expression $\ccK^0_{\ov \imath j}$. Then we rewrite $\ccK^0_{\ov \imath j} + \delta \ccK_{\ov \imath j} = \ccK^0_{\ov \imath k} (\delta^k_{\phantom{k}j} + \ccK_0^{k \ov l} \delta \ccK_{\ov l j})$ and thus, using the Neumann series $(1 - B)^{-1} = \sum_{i = 0}^{\infty} B^i$, we find
\begin{equation}
\ccK^{i \ov \jmath} \equiv \ccK_0^{i \ov \jmath} + \delta \ccK^{i \ov \jmath} + \ldots = \ccK_0^{i \ov \jmath}  - \ccK_0^{i \ov l} \delta \ccK_{\ov l k} \ccK_0^{k \ov \jmath}  + \ldots.
\end{equation}

Using the explicit expressions \eqref{eq:LeadingKahlerPot} and \eqref{GenCorrKP} we obtain, for $i,j\in\{\bT,c\}$,
\begin{align}
\renewcommand{\arraystretch}{2.0}
\ccK^0_{\ov \imath j}  &= \begin{pmatrix}
 \frac{3}{(\bT + \ov \bT)^2} & 0 \\
0 & \ccK^{\text{D7}}_{\ov c c}
\end{pmatrix}, \\
\ccK_0^{i \ov \jmath} &= \begin{pmatrix}
  \frac{(\bT + \ov \bT)^2}{3} & 0    \\                   
0 & \left( \ccK^{\text{D7}}_{\ov c c} \right)^{-1}  
\end{pmatrix}, \\
\delta \ccK_{\ov \imath j}  &= \begin{pmatrix}
 \frac{2\beta}{(\bT + \ov \bT)^3} & - \frac{\beta_c}{(\bT + \ov \bT)^2} \\
-\frac{\beta_{\ov c}}{(\bT + \ov \bT)^2} & \frac{\beta_{\ov c c}}{(\bT + \ov \bT)}
\end{pmatrix}, \\
\delta \ccK^{i \ov \jmath} &= \begin{pmatrix}
  - \frac{2\beta}{9(\bT + \ov \bT)^{-1}} &
  \frac{\beta_c \left( \ccK^{\text{D7}}_{\ov c c}\right)^{-1}}{3} \\                   
  \frac{\beta_{\ov c} \left( \ccK^{\text{D7}}_{\ov c c}\right)^{-1}}{3} &
- \frac{\left( \ccK^{\text{D7}}_{\ov c c} \right)^{-2} \beta_{\ov c c}} {(\bT + \ov \bT)}  
\end{pmatrix},
\end{align}
in terms of which the first order correction to the scalar potential reads
\begin{align}
\begin{split}
\delta V_1 = e^{\ccK_0} \Big( &\ccK_0^{\bT \ov \bT} \ccK^0_{\bT} \delta \ccK_{\ov \bT}
+ \ccK_0^{\bT \ov \bT} \delta \ccK_{\bT}  \ccK^0_{\ov \bT} 
+ \delta \ccK^{\bT \ov \bT} \ccK^0_{\bT} \ccK^0_{\ov \bT}  \\
+\ &\ccK_0^{c \ov c} \ccK^0_{c} \delta \ccK_{\ov c}
+\ \ccK_0^{c \ov c} \delta \ccK_{c}  \ccK^0_{\ov c} 
+\ \delta \ccK^{c \ov c} \ccK^0_{c} \ccK^0_{\ov c} \\
+\ &\delta \ccK^{\bT \ov c} \ccK^0_{\bT} \ccK^0_{\ov c}
+\ \delta \ccK^{c \ov \bT} \ccK^0_{c} \ccK^0_{\ov \bT} \\
+\ &\delta \ccK \ccK_0^{c \ov c} \ccK^0_c \ccK^0_{\ov c} \Big)|W|^2.
\end{split}
\end{align}
The first line vanishes due to the extended no-scale structure. The last line is present due to the expansion of the prefactor
$e^{\ccK_0 + \delta \ccK}$ (which for $\ccK_{\text{D7}}(c, \ov c) = 0$ was absent due to the no-scale structure of the potential). Plugging
in the expressions above we see that almost all terms cancel except the last one of the second line and the one in the last line, leaving
\begin{equation}
\delta V_1 = e^{\ccK_0} \left( \beta -  \left( \ccK_{\ov c c}^{\text{D7}} \right)^{-1} \beta_{\ov c c}\right) \frac{\left( \ccK_{\ov c c}^{\text{D7}} \right)^{-1}  \ccK^{\text{D7}}_c \ccK^{\text{D7}}_{\ov c}}
{(\bT + \ov \bT)}|W|^2.
\end{equation}
As $e^{\ccK_0} \sim 1/(\bT + \ov \bT)^3$, at first glance it looks like the presence of branes destroys the extended no-scale
structure and leads to a term $\sim W_0^2 / (\bT + \ov \bT)^4$ in the scalar potential. However, fixing the imaginary part
of the brane position modulus by the leading order term
\begin{equation}
 V_0 = e^{\ccK_0}\left(\left( \ccK_{\ov c c}^{\text{D7}} \right)^{-1}  \ccK^{\text{D7}}_c \ccK^{\text{D7}}_{\ov c}\right)|W|^2
\end{equation}
implies
\begin{equation}
\ccK^{\text{D7}}_c = \ccK^{\text{D7}}_{\ov c} = 0. 
\end{equation}
Hence, $\delta V_1 = 0$. We conclude that the extended no-scale structure, e.g.\ the 
cancellation of terms $\mathcal{O}\left(\frac{1}{T^4} \right)$ in the scalar potential, holds even in the presence of branes which are not fully stabilized at leading order.

\subsection{Relevance of Loop Corrections}
In the previous subsection we found that we expect the inflaton-dependent loop corrections to appear at $\ccO\left(|W_0|^2 \ccV^{-10/3} \right)$ in the scalar potential. On one hand, this is good news as the inflaton, being an additional light degree of freedom entering the loop corrections, does not spoil the extended no-scale structure. On the other hand, these loop corrections are often used to stabilize K\"ahler moduli which are `transverse' to the overall volume, as for example in the Large Volume Scenario \cite{Balasubramanian:2005zx,Conlon:2005ki,Berg:2007wt,Cicoli:2007xp,Cicoli:2008va}. While this need not necessarily be worrisome for the D7-brane chaotic inflation model \cite{Hebecker:2014eua}, it generically presents a problem for the fluxbrane inflation model: The stabilization mechanism used in \cite{Hebecker:2012aw} balances loop corrections against the $D$-term, thereby stabilizing the latter at a small value which is phenomenologically required. But since the loop corrections are generically 
inflaton-dependent, this leads to an $\eta$-problem, i.e.\ the $D$-term vacuum energy density will be of the same size as the loop-induced mass term for the inflaton.

One can ask whether the loop corrections involving the inflaton are suppressed by additional small numbers. After all, we found in \cite{Hebecker:2011hk} that the loop-induced $D$-term potential for the inflaton features an additional suppression by the quantity $\gym^2 \left(\int J \wedge \ccF\right)^2$ which needs to be small in order to satisfy the cosmic string bound. However, expecting this would indeed be too optimistic. Let us consider the $D$-term potential, including the Coleman-Weinberg-type loop corrections, which was calculated in \cite{Hebecker:2011hk}:
\begin{align}\label{CW-D-term-Potential}
 V_D &=  V_0 \left(1 + \alpha_{\ln}\ln \left(\frac{\varphi}{\varphi_0}\right) \right), \\
 \alpha_{\ln} &= \frac{\gym^2}{16\pi^2}\left(-2 \int \ccF\wedge \ccF + \frac{\gym^2}{2\pi} \left(\int J\wedge \ccF\right)^2\right).
\end{align}
In \cite{Hebecker:2011hk,Hebecker:2012aw} the first term in the expression for $\alpha_{\ln}$ was turned off (by an appropriate flux choice) for phenomenological purposes. However, assume for a moment that this term is there and, instead, neglect the (potentially small) second term $\sim \left(\int J\wedge \ccF\right)^2$. In this case, $\alpha_{\ln}$ is proportional to the number of chiral multiplets running in the loop (or, equivalently, the induced D3-brane charge) and the potential is the usual one of $D$-term hybrid inflation, found in field-theoretic approaches \cite{Binetruy:1996xj,Halyo:1996pp} as well as, for example, in D3/D7-inflation \cite{Dasgupta:2002ew,Haack:2008yb}.
Let us now try to rephrase the loop correction in \eqref{CW-D-term-Potential} in terms of a correction to the K\"ahler potential. To this end we make the following educated guess\footnote{A similar type of correction was considered already in the Wilson line inflation papers \cite{Kaplan:2003aj,ArkaniHamed:2003mz}.}
\begin{equation}
 \mathcal K \supset  -3 \ln (\bT + \ov \bT) - \frac{(c - \ov c)^2}{2} + \frac{3 \gym^2 }{16 \pi^2}\ln \sqrt{c \ov c} , \quad \gym^2  = \frac{4\pi}{\bT + \ov \bT}, \quad \varphi = \Re (c).
\end{equation}
Using the general form of the $D$-term potential in supergravity (following e.g.\ the conventions of \cite{Haack:2006cy})
\begin{equation}
 V_D = \frac{\gym^2}{2}\left(\frac{Q}{(2\pi)^2}\partial_{\bT} \mathcal K \right)^2
\end{equation}
we now precisely reproduce \eqref{CW-D-term-Potential} up to the factor $2 \int \ccF\wedge \ccF$ which, as stated above, only counts the number of fields running in the loop and could easily be included in the above ansatz. Here, $Q$ is the charge of the superfield $\bT$ which shifts under the U(1) symmetry and $\gym^{-2}$ is the real part of the gauge kinetic function.

Now the naive hope might be that turning off $\int \ccF \wedge \ccF$ could also turn off the above correction. What remains would be suppressed by a further power of $\gym^2 \left(\int J \wedge \ccF\right)^2$ ({\it in addition} to the $\gym^2$-suppression) which can be stabilized at some very small value as discussed in \cite{Hebecker:2012aw}. This is, however, indeed too naive generically: While the supergravity calculation performed in section 4 of \cite{Hebecker:2011hk} actually admits the cycle which is wrapped by the D7-brane to be compact (and therefore captures the corrections $\sim \gym^2 \int \ccF \wedge \ccF$), curvature-induced D3-brane charge \cite{Douglas:1995bn,Green:1996dd,Cheung:1997az,Minasian:1997mm,Morales:1998ux,Stefanski:1998he,Scrucca:1999uz,Blumenhagen:2006ci,Collinucci:2008pf} was neglected. As long as this charge is not canceled {\it locally} there is no reason to assume the BHK-type corrections \cite{Berg:2005ja,Berg:2005yu,Berg:2007wt,Cicoli:2007xp} to be absent.

Here, we therefore follow a different strategy and stabilize the `transverse' directions by leading order $D$-terms, rather than loop corrections. More details on this are contained in the following section.

Still, gaining a better understanding of the structure of loop corrections to the K\"ahler potential (in particular concerning their behavior at large complex structure) is desirable. This includes an analysis of known corrections on toroidal orbifolds \cite{Berg:2005ja} and on $\tK\times T^2/\mathds{Z}_2$. This is work in progress.
Besides being compactifications in which the loop corrections are actually computable, in these examples the D7-branes can be parallel-displaced, i.e.\ there is no self-intersection curve of the D7-brane divisor. Other such examples may be provided by $K3$-fibrations. Having in mind our fluxbrane inflation model, such settings are rather attractive.

\section{Phenomenology of Fluxbrane Inflation}\label{sec:pheno}
Having discussed the general prerequisites for D7-brane inflation to work, we now turn to a more detailed discussion of the fluxbrane inflation scenario. In particular we aim at quantifying the required size of the stringy model parameters in order for the model to be viable.

\subsection{Moduli Stabilization}\label{sec:modstab}
From the analysis of the previous sections we now know that the generic size of the loop corrections calculated by \cite{Berg:2007wt} is\footnote{Regarding our normalization conventions, we follow \cite{Hebecker:2012aw}.}
\begin{equation}\label{eq:GenInflatonLoop}
 \delta V_{\text{loop}}(\varphi) \sim \frac{g_s W_0^2}{\ccV^{10/3}}\beta(\varphi),
\end{equation}
where $\beta(\varphi)$ is some function which involves the brane deformation modulus, i.e.\ the inflaton, and which we assume to have no specific structure except for its periodicity.\footnote{Note that $\beta(\varphi)$ may contain additional factors of $g_s$. We analyze this in the course of this section.} What matters now is the relative size of \eqref{eq:GenInflatonLoop} with respect to the constant energy density during inflation. To quantify this we have to specify the vacuum of our theory, i.e.\ we have to discuss moduli stabilization.

We start with the assumption that the axio-dilaton as well as all complex structure moduli are stabilized by fluxes at some high scale, such that we are left with an effective theory of the K\"ahler and D7-brane moduli. Recall from \cite{Hebecker:2012aw} that one needs more than two K\"ahler moduli in order to implement the fluxbrane inflation scenario. This comes about as follows: The constant energy density during inflation in fluxbrane inflation is due to supersymmetry breaking flux on the D7-branes which annihilates during reheating. In the effective theory this flux gives rise to a non-vanishing contribution to the $D$-term potential which is given by
\begin{equation}\label{eq:D-termGeneric}
 V_D = \frac{g_{\text{YM}}^2 \xi^2}{2}, \quad\text{where}\quad g_{\text{YM}}^2 = \frac{2\pi}{\ccV_{\text{D7}}}, \quad   \xi = \frac{1}{4\pi}\frac{\int J\wedge \ccF }{\ccV} .
\end{equation}
The cosmic string bound forces
\begin{equation}\label{eq:CSBound}
 \xi = \frac{1}{4\pi}\frac{\int J \wedge \ccF}{\ccV}\lesssim 4\cdot 1.3 \cdot 10^{-7}.
\end{equation}
One can try to satisfy this bound by either making the overall volume large, or by having a small $\int J \wedge \ccF$.
Making the overall volume large is problematic generically: Recall, that in order to avoid a runaway in the K\"ahler moduli space to infinite volume one needs the $F$-terms and the $D$-term to be of comparable size. However, due to the no-scale structure a non-zero K\"ahler moduli $F$-term potential is only induced via higher-order ($\alpha'$ and non-perturbative) corrections and thus scales as $V_F \sim |W_0|^2 / \ccV^3$. This implies a tuning $W_0^2 \sim \ccV$ in order to enhance the size of the $F$-term potential, which is suitable for stabilizing the overall volume $\ccV$, up to a level where it is comparable to the $D$-term potential \eqref{eq:D-termGeneric}.
On the other hand, in order to trust the validity of the supergravity approximation, the Kaluza-Klein scale should be larger than the gravitino mass, the ratio of the two being $m_{3/2} / \mkk \sim W_0 / \ccV^{1/3} \sim \ccV^{1/6}$. This can work for small volumes due to the appearance of $\pi$-factors (which we have neglected here), but goes the wrong way parametrically for large $\ccV$ \cite{Hebecker:2012aw,Cicoli:2013swa}.

This forces us to stabilize $x = \int J \wedge \ccF / \sqrt{\ccV_{\text{D7}}}$ at a small value in order to satisfy the cosmic string bound. Thus, we need to consider models with more than two K\"ahler moduli. Reference \cite{Hebecker:2012aw} discusses a situation in which the flux $\ccF$ is dual\footnote{`Dual' in this case refers to Poincar\'{e} duality on the worldvolume of the 7-brane.} to an effective curve (i.e.\ a curve inside the Mori cone) on the brane. Therefore, given an overall volume, there was a minimal value of $x$ below which the volume of this two-cycle becomes sub-stringy. If this is the case, one cannot trust the supergravity approximation anymore. Consequently, in the discussion of \cite{Hebecker:2012aw} there was a lower bound on the size of $x$. In more generic situations, however, this lower bound will not be present due to the fact that the dual two-cycle need not be an effective two-cycle of the brane (in particular it can be a linear combination of effective two-cycles with 
coefficients of 
either 
sign). In this case, nothing prevents $x$ from being extremely small during inflation.

Stabilization of the directions in K\"ahler moduli space which are `transverse' to the overall volume can be achieved in different ways:  The strategy pursued in \cite{Hebecker:2012aw} was to stabilize those moduli via loop corrections of the form discussed in \secref{String Loop Corrections}. This thus forces a balance of those loop corrections and the $D$-term which is undesirable because one would then have $V_0 \sim \delta V_{\text{loop}}(\varphi)$, implying $\eta \sim 1$.

One can, however, also follow a different strategy and stabilize those `transverse' directions via $D$-terms as in \cite{Cicoli:2011qg}. That means one turns on worldvolume fluxes on D7-branes other than the two branes which are responsible for inflation. These fluxes induce parametrically dominant $D$-terms in the scalar potential which are strictly positive and stabilized at zero value, thereby fixing relative sizes of two-cycle volumes. The remaining two flat directions are then stabilized within the standard Large Volume Scenario, giving rise to a vacuum energy density $V_{\text{AdS}} \sim - |W_0|^2 / \ccV^3$. This vacuum is uplifted first to a Minkowski minimum and then, subsequently, to dS via two different $D$-terms.\footnote{Think of the corresponding two U(1) theories as linear combinations of the gauge theories living on the two fluxbranes, as in section 2 of \cite{Hebecker:2012aw}.} In order not to introduce a runaway potential for the overall volume $\ccV$, these $D$-terms have to be 
parametrically smaller than their generic value $\sim 1/\ccV^2$, 
which can be achieved by fine-tuning the relative sizes of two-cycle volumes. The maximum value of the energy density during inflation is thus roughly given by $|V_{\text{AdS}}|$, and we will parametrize $V_0 = \gamma |V_{\text{AdS}}|$ with $\gamma \lesssim 1$ in the following.\footnote{Values for $\gamma$ in the range $\gamma = 10^{-1}\ldots 10^{-2}$ are 
required quite generally if the flux responsible for the uplift to Minkowski and the flux responsible for the inflationary de Sitter uplift live on the same two-cycle \cite{Hebecker:2012aw}. In this case, stability of the uplifted vacuum requires the flux quantum number for the annihilating flux to be a fraction of the total flux quantum number, leading to a hierarchy between $|V_{\text{AdS}}|$ and $V_0$.\label{foot:SizeGamma}} From a model building point of view we consider the tuning in the $D$-term superior to a tuning of loop coefficients, as the ability to compute $D$-terms exceeds by far the ability to compute loop corrections in general Calabi-Yau compactifications.
It would be interesting and instructive to construct an example for this uplifting proposal.

\subsection{The Relative Size of Loop Corrections from a Microscopic Point of View}
\label{sec:ComputingAlpha}
What is the generic size of $V_0$ and $\delta V_{\text{loop}}$, including the relevant parameters and factors of $\pi$? The AdS minimum in the Large Volume Scenario is at \cite{Hebecker:2012aw}
\begin{equation}\label{eq:AdSMin}
 V_{\text{AdS}} = -\frac{3}{8}\frac{(2\gamma_s)^{1/3}}{(4\pi)^3\sqrt{2\pi}} \frac{\hat \xi^{2/3}}{\sqrt{\ln\left(\frac{8\pi A_s}{3\gamma_s}\frac{\ccV}{W_0}\right)}} \frac{|W_0|^2}{\ccV^3},
\end{equation}
where again $\hat \xi = -\chi(M)\zeta(3)/2 $, $A_s$ is the prefactor of the instanton correction to the superpotential involving the small four-cycle of the Large Volume Scenario, $\delta W = A_s e^{-2\pi T_s}$, and $\gamma_s = \frac{2^{3/2}}{3! \sqrt{\kappa_{sss}}}$, where $\kappa_{sss}$ is the triple self-intersection number of the small four-cycle. An uplift of this potential to dS allows for a maximum energy density of $V_0 \simeq |V_{\text{AdS}}|$ during inflation, without destabilizing the minimum. In that minimum, the small four-cycle is stabilized such that
\begin{equation}
 2\pi T_s = \ln\left(\frac{8\pi A_s}{3\gamma_s}\frac{\ccV}{W_0}\right) = \frac{1}{2\pi g_s}\left(\frac{\hat \xi}{2\gamma_s}\right)^{2/3}.
\end{equation}

The loop corrections are known explicitly only for certain orbifolds/orientifolds of the factorisable torus $T^2_1 \times T^2_2 \times T^2_3$, where they take the form \cite{Berg:2005ja}
\begin{align}\label{eq:BHKWindingCorrKahler}
 \delta K_{\text{BHK}} &=\delta K_{\text{BHK}}^{\text{KK}} + \delta K_{\text{BHK}}^{\text{W}}\nonumber\\
 & = -\frac{2g_s}{(4\pi)^4} \sum_{I=1}^3 \frac{\mathcal{E}_I^{\text{KK}}(\varphi, U^I)}{T_{T^4_I}}-\frac{2}{(4\pi)^4} \left. \sum_{I=1}^3 \frac{\mathcal{E}_I^{\text{W}}(\varphi, U^I)}{T_{T^4_J}T_{T^4_K}}\right|_{I \neq J \neq K} .
\end{align}
Here, $U^I$ is the complex structure of the two-torus $T^2_I$ and $T_{T^4_I}$ is the volume of the four-torus $T^2_J \times T^2_K$ (where $I \neq J \neq K$). The superscripts KK and W indicate whether the corrections are due to Kaluza-Klein or winding modes of the strings. The KK-corrections arise from the exchange of Kaluza-Klein modes between D7-branes (and, potentially, D3-branes and the respective O-planes). The W-corrections are due to the exchange of strings which wind around one-cycles along the intersection locus of two D7-branes \cite{Berg:2007wt}. Therefore, the presence of those winding corrections in a given model depends on the topology of the compact space and, in particular, of the intersection locus of the two D7-branes. For example, in the $\tK \times K3$ model discussed in \secref{modulispacek3} the D7-branes do not intersect at all. Thus we expect the corrections associated to winding modes to be absent in this model \cite{Neuenfeld}.

The KK-corrections are precisely of the form \eqref{GenCorrKP} and are thus subject to the extended no-scale structure investigated in \secref{Extended_no-scale_with_branes}. On the other hand, the W-corrections enter the K\"ahler potential as a homogeneous function of degree $-2$ in the four-cycle volumina and are therefore expected to appear in the scalar potential at linear order in $\delta K_{\text{BHK}}^{\text{W}}$. Counting $\pi$-factors in the toroidal computation \cite{Berg:2005ja} and factors of $g_s$, the loop corrections due to winding modes are generically the largest in the scalar $F$-term potential.

The functions $\mathcal{E}_I^{\text{KK,W}}(0, U^I)$ in \eqref{eq:BHKWindingCorrKahler} are proportional to a particular non-holomorphic Eisenstein series, 
\begin{equation}
 E_2 (U) = \sum_{(n,m)\neq (0,0)}\frac{\operatorname{Im}(U)^2}{|n + m U |^{4}} .
\end{equation}
For a square torus one has $E_2 (i) \simeq 6$. The factors of proportionality (which we will call $N^{\rm KK}$, $N^{\rm W}$) depend on the particular orbifold/orientifold model and its brane content. They are essentially traces over matrices which specify the action of the orbifold and orientifold group on the CP labels of an open-string state. We view them as integral, topological data. While they can be of the order of thousands (see e.g.\ \cite{Berg:2005yu}), we do not expect such large factors to be a generic feature. The generic form of the correction to the scalar potential, induced by \eqref{eq:BHKWindingCorrKahler} is thus
\begin{equation}\label{eq:GenLVSLoopCorr}
 \delta V_{\text{loop}} \simeq \frac{g_s}{16\pi}\frac{|W_0|^2}{\ccV^2}\frac{1}{(4\pi)^4 \ccV^{4/3}}\left\{\frac{g_s^2\left(N^{\text{KK}}C^{\text{KK}}\right)^2}{(4\pi)^4} \beta^{\text{KK}} (\varphi) + N^{\rm W}C^{\text{W}} \beta^{\text{W}}(\varphi)\right\} ,
\end{equation}
expecting that this also applies in the more general Calabi-Yau context, as long as there are no big hierarchies of four-cycle volumes (except for potential blow-up modes). The quantities $C^{\text{KK,W}}$ account for the complex structure dependence of this expression and are expected to be $\ccO(1)$ generically. As expected, the KK-corrections are suppressed with respect to the W-corrections, as the corresponding corrections to the K\"ahler potential feature a common $(4\pi)^{-4}$-factor, but the KK-corrections are subject to the extended no-scale structure. Therefore, for the KK-corrections the $(4\pi)^{-4}$-factor is squared in the scalar potential.

Now let us compute $\alpha$, defined in \eqref{cosPot}, microscopically. Recall that $\alpha$ quantifies the relative size of the loop corrections \eqref{eq:GenLVSLoopCorr} with respect to the constant \eqref{eq:AdSMin} of the potential. From \eqref{eq:GenLVSLoopCorr} it is clear that, generically, $\alpha_{\text{micro}}$ will be dominated by the loop corrections due to winding modes. However, as noted above, those are only present as long as there are non-trivial one-cycles along the intersection locus of D7-branes. This need not be the case. Therefore, we will distinguish $\alpha_{\text{micro}}^{\text{KK}}$ and $\alpha_{\text{micro}}^{\text{W}}$. Regarding the inflaton dependence of the loop corrections, we assume that the $\beta^{\text{KK,W}}(\varphi)$ in \eqref{eq:GenLVSLoopCorr} vary by $\ccO(1)$ as $\varphi$ moves the maximal distance in field space. Thus, assuming $N^{\text{KK,W}} = C^{\text{KK,W}} = 2\gamma_s= 1$, $\alpha_{\text{micro}}^{\text{KK,W}}$ is directly computed as
\begin{align}
 \alpha_{\text{micro}}^{\text{KK}}&= \frac{2}{3}\frac{g_s^{5/2}}{\gamma(4\pi)^6  \hat \xi^{1/3}\ccV^{1/3}},\\
 \alpha_{\text{micro}}^{\text{W}} &= \frac{2}{3}\frac{\sqrt{g_s}}{\gamma(4\pi)^2  \hat \xi^{1/3}\ccV^{1/3}}. 
\end{align}
Here we have used the fact that, as detailed at the end of \secref{sec:modstab}, the uplift is such that $V_0 = \gamma |V_{\text{AdS}}|$ with $\gamma \lesssim 1$.

\subsection{Consequences of Experimental Constraints}
\begin{figure}
\centering
 \includegraphics[width = 0.6\textwidth]{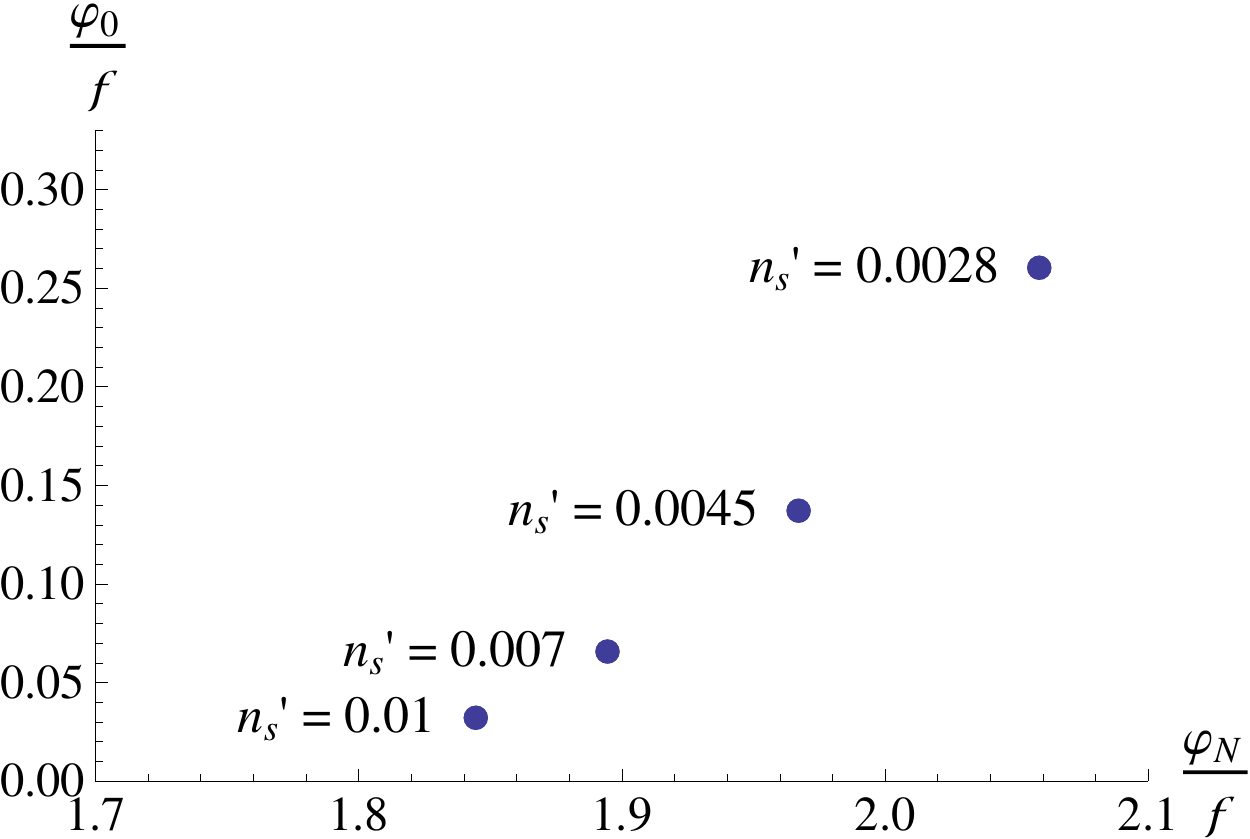}
\caption{Field values of the inflaton at the beginning of the last 60 $e$-foldings and the end of inflation for different values of $n_s'$.}\label{fig:fieldvalues}
\end{figure}

In order to analyze the experimental constraints for the relative size of $V_0$ and $\delta V_{\text{loop}}(\varphi)$ (i.e.\ the magnitude of $\alpha$ in \eqref{cosPot}), let us now return to the phenomenological analysis of \secref{PhenoConstraints}. For exploring the parameter space of the fluxbrane inflation model let us parametrize it in terms of the quantity $\varphi_0 /f$ which describes the point of tachyon condensation in units of the total field space (up to a factor of $2\pi$). Using the expression for $\eta$ \eqref{eq:slow-roll-parameters} and $N$ \eqref{eq:e-folds} together with the experimental constraints \eqref{phenoConstr} and assuming $\epsilon \ll |\eta|$ we find
\begin{equation}\label{eq:NinPhif}
 N = \frac{2 \cos\left(\frac{\varphi_N}{f}\right)}{n_s -1}\ln\left(\frac{\tan\left(\frac{\varphi_N}{2f}\right)}{\tan\left(\frac{\varphi_0}{2f}\right)}\right).
\end{equation}
For $N=60$ this implicitly defines $\varphi_N / f $ in terms of $\varphi_0 /f$. Furthermore, the running of the spectral index is easily computed as
\begin{equation}\label{eq:ns'inPhi0}
 n_s' = \frac{(n_s -1)^2}{2}\tan^2 \left(\frac{\varphi_N}{f}\right).
\end{equation}
Combining \eqref{eq:NinPhif} and \eqref{eq:ns'inPhi0} we obtain a relation between $\varphi_0 /f$ and $\varphi_N /f$ which is monotonic. The requirement $n_s' \lesssim 0.01$ thus puts a lower bound on $\varphi_0 /f$ which is given by
\begin{equation}
 \frac{\varphi_0}{f} \gtrsim 0.032.
\end{equation}
Figure \ref{fig:fieldvalues} shows $\varphi_N / f$ and $\varphi_0 / f$ for several different values of $n_s'$.

From \eqref{eq:D-termFieldTheory} and \eqref{phenoConstr} we now compute
\begin{equation}\label{eq:epsFromGandXi}
 \epsilon = \frac{1}{4}\left(\frac{ g_{\text{YM}}\xi}{\tilde \zeta}\right)^2 .
\end{equation}
We thus have introduced the additional parameter $(g_{\text{YM}}\xi)^2$ which, together with $\varphi_0 /f$, parametrizes the model. It is bounded from above due to $\xi \lesssim \xi_{\text{max.}} \equiv 5.2 \cdot 10^{-7}$ and $g_{\text{YM}}^2 \lesssim 2\pi$.\footnote{In order to trust the supergravity approximation we require $\vd \gtrsim 1$ and thus, in view of \eqref{D-termPot}, we find $g_{\text{YM}}^2 = 2\pi / \vd \lesssim 2\pi$.}

The quantity $\alpha$ is now expressed as
\begin{equation}\label{eq:AlphaGen}
 \alpha = \frac{2 \epsilon}{\sqrt{n_s'}}\sqrt{2+\frac{\left(1-n_s\right)^2 }{n_s'}},
\end{equation}
and is thus completely determined by $\varphi_0 /f$ and $(g_{\text{YM}}\xi)^2$. For $g_{\text{YM}}^2 = 2\pi$, $\xi=\xi_{\text{max}}$ and $\varphi_0 /f = 0.032$, such that $n_s'=0.01$, one finds $\alpha = 4.8\times 10^{-5}$. This corresponds to the situation where $\epsilon$ and therefore the tensor to scalar ratio $r = 16\epsilon$ is maximal and given by
\begin{equation}
 r = 4\left(\frac{ g_{\text{YM}}\xi}{\tilde \zeta}\right)^2 \simeq 2.6 \cdot 10^{-5}.
\end{equation}
This is smaller than the value $r \simeq 7.6\cdot 10^{-4}$ found in the related work \cite{Hebecker:2013zda}. The reason for this is the cosmic string bound which was not taken into account in this work. Without the cosmic string bound, the limiting factors are the bounds on $n_s'$ and $f$, from which one then determines a maximal value for $\epsilon$ which lies above the value computed via \eqref{eq:epsFromGandXi} with $g_{\text{YM}}^2 = 2\pi$, $\xi=\xi_{\text{max}}$. This $\epsilon$ then leads to the larger value for $r$.

One can ask what the maximal possible value for $\alpha$ is. In order to determine that value, observe that the axion decay constant is given in terms of $\epsilon$ and $n_s'$ as
\begin{equation}\label{eq:running}
 n_s' = \frac{4\epsilon}{f^2}.
\end{equation}
In view of \eqref{eq:AlphaGen} it is clear that $\alpha$ is maximized for large $\epsilon$ and small $n_s'$, i.e.\ large $f$. However, recall that, in a regime where one controls the effective theory of a string theory compactification, the size of the field space of the axion is constrained as $f \lesssim 1/4\pi$ \cite{Banks:2003sx} (see also \cite{Conlon:2006tq,Goodsell:2009xc,Cicoli:2011yh,Cicoli:2012sz} for an explicit discussion in the case of K\"ahler axions).\footnote{Proposals of realizing inflation in string theory with larger $f$ include \cite{Blumenhagen:2014gta,Grimm:2014vva}.} Also in field theory there are arguments that the quantity $f$ should take only sub-planckian values \cite{Conlon:2012tz}. Using the fiducial value $f = 1/4\pi$ one finds
\begin{equation}
 \alpha^{(4\pi)} = \frac{\sqrt{\epsilon}}{4\pi} \sqrt{2 + \frac{\left(1-n_s\right)^2}{(4\pi)^2 4\epsilon}}.
\end{equation}
This is a monotonically increasing function and maximized for large $\epsilon$. Therefore, setting again $g_{\text{YM}}^2 = 2\pi$ and $\xi=\xi_{\text{max}}$ we find $\alpha^{(4\pi)}_{\text{max}} = 1.9\times 10^{-4}$. Values larger than this one are not compatible with the data. This is a rather stringent constraint which needs to be satisfied by the string theory model.

\subsection{Translation to Parameters of the String Theory Model}
We saw that the field theory model of hybrid natural inflation can be parametrized by $\varphi_0 / f $ (or, equivalently, $n_s'$) and $g_{\text{YM}} \xi$. We now analyze how these two parameters map to corresponding quantities constructed from $\ccV$, $\vd$, $g_s$, $x$ and $z$ of the stringy embedding. Let us start by combining \eqref{phimax} with \eqref{eq:epsFromGandXi} and \eqref{eq:running} to obtain
\begin{equation}
 \frac{g_s}{z} = \frac{(4\pi)^2}{\tilde \zeta^2}\frac{(\gym \xi)^2}{n_s'}.
\end{equation}
Furthermore, from \eqref{D-termPot} it follows that
\begin{equation}
 \frac{x}{\ccV}= \sqrt{8\pi}\;\gym \xi.
\end{equation}
On the other hand we can combine \eqref{D-termPot} with \eqref{eq:phi0} to get
\begin{equation}
 \sqrt{\vd} = \frac{\sqrt{4\pi}}{\tilde \zeta^2}\frac{\gym\xi}{n_s'}\left(\frac{\varphi_0}{f}\right)^2.
\end{equation}
We have thus determined $\vd$, $x/\ccV$ and $g_s/z$ in terms of $\varphi_0 / f $ and $g_{\text{YM}} \xi$. The quantity $\alpha$ is then determined via \eqref{eq:AlphaGen}.

In order to determine the absolute values of $z$, $g_s$, $\ccV$ and $x$ we have to implement the constraints which come from moduli stabilization. As we have discussed in \secref{sec:modstab} the maximum uplift in our model, and thus the maximum energy density during inflation, is given by
\begin{equation}
 V_0 = \gamma\left|V_{\text{AdS}}\right|, \quad \gamma \lesssim 1.
\end{equation}
Furthermore, since the positive energy density is provided by a $D$-term \eqref{eq:D-termFieldTheory} which annihilates at the end of inflation, we find
\begin{equation}\label{eq:VolComputation}
 \frac{(\gym \xi)^2}{2} = \frac{3}{8}\frac{1}{(4\pi)^3 \sqrt{2\pi}}\frac{\gamma\hat \xi^{2/3}}{\sqrt{\ln\left(\frac{16\pi }{3}\frac{\ccV}{W_0}\right)}}\frac{W_0^2}{\ccV^3}.
\end{equation}
Here we have assumed that $ A_s = 1$. Given some $W_0$, $\hat \xi$ and $\gamma$, this equation determines $\ccV$. Finally, the string coupling $g_s$ is given by
\begin{equation}\label{eq:gsComputation}
 g_s = \frac{\hat \xi^{2/3}}{2\pi\ln\left(\frac{16\pi }{3}\frac{\ccV}{W_0}\right)}.
\end{equation}

The following table shows the model parameters for a couple of different values of $n_s'$ and $\xi$. The constant value of the tree-level superpotential is chosen to be $W_0 = 1$ and, furthermore, $\hat \xi = \gamma = 1$. We obtain
\begin{center}
\begin{tabular}{|c|c||c|c|c|c|c|c|c|c|c|}\hline
 $n_s'$& $\xi$&$\ccV$ & $\vd$ &$g_s$ &   $z$ &  $x$ & $R^2$& $T_s$ &$\alpha$  \\\hhline{|=|=||=|=|=|=|=|=|=|=|}
 0.01 & $5.2\cdot 10^{-7}$&380.7 &1.84 &0.018 &   0.324 &  $1.8 \cdot 10^{-3}$ & 639&1.39 & $2.6 \cdot 10^{-5}$ \\\hline
  0.007 & $5.2\cdot 10^{-7}$&683.0 &11.0 &0.017 &   1.27 &  $1.3 \cdot 10^{-3}$ & 49.1&1.49 & $5.3 \cdot 10^{-6}$ \\\hline
  0.007 & $ 10^{-7}$&1172 &2.11 &0.016 &   6.23 &  $1.0 \cdot 10^{-3}$ &89.1&1.57 & $1.0 \cdot 10^{-6}$ \\\hline
\end{tabular} 
\end{center}
Here, $R^2$ is the length squared of the $S^1$ in the fiber which is transverse to the D7-brane, cf.\ \eqref{phimax}. Decreasing $n_s'$ decreases $\alpha$. Lowering $\xi$ lowers the energy scale of inflation and therefore also the value of $\epsilon$ and $\alpha$.
Note that for $n_s' = 0.01 $ and  $\xi = 5.2\cdot 10^{-7}$ the value of $z$ is rather low, such that one might worry about the relevance of non-perturbative corrections to the K\"ahler potential. However, slightly decreasing $n_s'$ and $\xi$ increases $z$, which thus helps in this respect. Furthermore, $n_s' = 0.01$ and $\xi = 10^{-7}$ would have given $\vd <1$ which might pose a problem for the control of the effective theory. This is why we chose $n_s' = 0.007$ and $\xi = 10^{-7}$ in the last row of the above table.

Having a compact space which contains hierarchically different length scales, one faces the danger of having light KK-modes in the spectrum which might spoil the validity of the supergravity approximation. In particular, the ratio $m_{3/2} / \mkk$ might become of order one. Let us briefly estimate this ratio. In the SYZ-picture, the space transverse to the D7-branes, whose size is given by $\ccV/\vd$, consists of one direction with length $R$ in the $T^3$-fiber and one direction with length $L$ in the base. Schematically,
\begin{equation}
 z = \frac{\ccV}{\vd}\frac{1}{ R^2} = LR\frac{1}{R^2} = \frac{L}{R}.
\end{equation}
Thus, recalling our discussion in \secref{syzformulation}, $z$ is the (imaginary part of the) complex structure modulus in the SYZ-picture.
The mass of the KK-modes along the direction with length $L$ is given by $\mkk = \frac{2\pi}{\ell_s L^s}$, where the superscript $s$ denotes that $L^s$ is measured in units of $\ell_s$ in the ten-dimensional string frame. It is related to $L$ via $L g_s^{1/4}  = L^s$. The gravitino mass is $m_{3/2} = \frac{\sqrt{g_s}}{\sqrt{16\pi}}\frac{W_0}{\ccV}$. Furthermore, recall that $ \ell_s^{-1} = \frac{g_s}{\sqrt{4\pi\ccV^s}}M_p$. Putting everything together we have, for $n_s' = 0.01 $,  $\xi = 5.2\cdot 10^{-7}$, $W_0 = 1$ and $\hat \xi =\gamma = 1$,
\begin{equation}\label{eq:m32mkk}
 \frac{m_{3/2} }{\mkk} = \frac{\sqrt{g_s}}{4\pi}\frac{W_0}{\sqrt{\ccV}}zR =\frac{\sqrt{g_s} W_0}{4\pi}\sqrt{\frac{z}{\vd}}  \simeq 4.5\cdot 10^{-3}.
\end{equation}
Therefore, the supergravity approximation is under control in this example.

Let us now compute $\alpha_{\text{micro}}$. One finds that it varies only very weakly with $n_s'$ and $\xi$. In particular
\begin{center}
\begin{tabular}{|c|c||c|c|}\hline
 $n_s'$&$\xi$& $\alpha_{\text{micro}}^{\text{KK}}$& $\alpha_{\text{micro}}^{\text{W}}$   \\\hhline{|=|=||=|=|}
  0.01&$ 5.2\cdot 10^{-7}$ &$1.0\cdot 10^{-12}$& $7.9 \cdot 10^{-5}$\\\hline
  0.007&$ 5.2\cdot 10^{-7}$ &$7.3\cdot 10^{-13}$& $6.3 \cdot 10^{-5}$\\\hline
  0.007&$  10^{-7}$ &$5.3\cdot 10^{-13}$& $5.1 \cdot 10^{-5}$\\\hline
\end{tabular} 
\end{center}

Now let us analyze the scaling of the above quantities with $W_0$, $\hat \xi$, and $\gamma$. Treating the logarithm in \eqref{eq:VolComputation} as constant we find
\begin{equation}\label{eq:Scaling}
 \ccV \sim W_0^{2/3} \hat \xi^{2/9} \gamma^{1/3}, \quad g_s \sim \hat \xi^{2/3},\quad  z\sim \hat \xi^{2/3}, \quad x \sim W_0^{2/3} \hat \xi^{2/9}\gamma^{1/3}, \quad R^2 \sim W_0^{2/3} \hat \xi^{-4/9}\gamma^{1/3}.
\end{equation}
Most importantly,
\begin{equation}
 \alpha \sim \text{const.},\quad \alpha_{\text{micro}}^{\text{KK}} \sim \frac{\hat \xi^{34/27}}{W_0^{2/9}\gamma^{10/9}},\quad \alpha_{\text{micro}}^{\text{W}} \sim \frac{1}{W_0^{2/9}\hat \xi^{2/27}\gamma^{10/9}}.
\end{equation}
It thus seems that we should increase $\hat \xi$ and $W_0 $ as much as possible. However, a natural upper bound on $\hat \xi$ is given by the requirement $g_s \lesssim 1$. Additionally, in view of \eqref{eq:m32mkkScaling} we find that
\begin{equation}\label{eq:m32mkkScaling}
 \frac{m_{3/2} }{\mkk}\sim W_0 \hat \xi^{2/3}.
\end{equation}
As $\hat \xi \gtrsim 1$, for fixed $n_s'$ and $\xi$ this scaling puts an upper bound on $W_0$. A large $\hat \xi$ in \eqref{eq:m32mkkScaling} can in principle be compensated by a small $W_0$. On the other hand, since in the above example both $m_{3/2}/\mkk\simeq 10^{-2} $ and $g_s \simeq 10^{-2}$ and both scale with the same power of $\hat \xi$, this `compensation' by a small $W_0 $ is actually not constructive. Still, as $z$ scales with a positive power of $\hat \xi$, increasing the latter helps suppressing non-perturbative corrections to the K\"ahler potential. Therefore, if one insists on having $z>1$ it is most efficient to increase both, $W_0$ and $\hat \xi$ at the same time, but keeping $m_{3/2}/\mkk \ll 1$. Let us thus choose $n_s' = 0.01$, $\xi = 5.2\cdot 10^{-7}$, $\gamma  = 1$, and $W_0= \hat \xi =10$. We then find
\begin{center}
\begin{tabular}{|c|c|c|c|c|c|c|}\hline
 $\ccV$ &$\vd$ &$g_s$ &   $z$  & $x$ &  $R^2$& $T_s$ \\\hhline{|=|=|=|=|=|=|=|}
2962 &1.84 &0.09 &   1.55  & $1.4 \cdot 10^{-2}$ &  1040& 1.35 \\\hline
\end{tabular}

\begin{tabular}{|c|c|c|c|}\hline
 $\alpha$ &  $\alpha_{\text{micro}}^{\text{KK}}$ & $\alpha_{\text{micro}}^{\text{W}}$ & $m_{3/2} / \mkk$\\\hhline{|=|=|=|=|}
 $2.6 \cdot 10^{-5}$ &  $1.2\cdot 10^{-11}$& $4.0\cdot 10^{-5}$ & 0.22\\\hline
\end{tabular} 
\end{center}
Clearly, the quantity $\alpha_{\text{micro}}^{\text{W}}$ is still slightly larger than $\alpha$. In particular, as discussed at the end of \secref{sec:modstab}, it is generally not possible to realize $\gamma =1$, which further deteriorates the situation. However, if we assume the presence of winding mode corrections, insist on $z>1$, and ignore that typically $\gamma <1$, the above table summarizes the best we can do.

In the analysis presented so far we have neglected the Coleman-Weinberg-type loop corrections which were computed and analyzed for the fluxbrane model in \cite{Hebecker:2011hk,Hebecker:2012aw}. In fact, this term was found to correct the tree-level $D$-term potential as
 \begin{equation}
  V(\varphi) = V_0 \left(1 + \alpha_{\ln} \ln \left(\frac{\varphi}{\varphi_0}\right) + \ldots\right), \quad \alpha_{\ln} =  \frac{\gym^2}{(4\pi)^2}x^2 = \frac{1}{8\pi\vd}x^2.
 \end{equation}
 This gives, for $n_s' = 0.01$, $\xi = 5.2\cdot 10^{-7}$, and $\gamma= 1$,
 \begin{center}
\begin{tabular}{|c|c||c|}\hline
$W_0$ & $\hat \xi$ & $\alpha_{\ln}$   \\\hhline{|=|=||=|}
1 &1 & $7.3 \cdot 10^{-8}$\\\hline
10 & 10 &  $4.4 \cdot 10^{-6}$\\\hline
\end{tabular} 
\end{center}
The second value of $\alpha_{\ln}$ in this table is rather large, due to the relatively large value of $x$. On the other hand, it is still almost one order of magnitude smaller than the corresponding $\alpha$, which means that the corrections to the $F$-term potential remain dominant. Nevertheless, the constraint $\alpha, \alpha_{\text{micro}}^{\text{KK,W}} \gg \alpha_{\ln}$, needed for being able to consistently neglect the Coleman-Weinberg-type loop term, is generically non-trivial and should be taken into account carefully.
 
Up to now we assumed the worst case scenario, i.e.\ we assumed that the winding mode corrections which, regarding the accompanying $\pi$-factors, is the largest exists and depends on the inflaton. However, as detailed in \secref{sec:ComputingAlpha}, these corrections are absent in models in which there are no one-cycles along the intersection loci of two D7-branes \cite{Berg:2007wt}. In particular, on $\tK \times K3$ these corrections are expected to be absent. Let us therefore concentrate on models in which the intersection curves of D7-branes (at least of the ones on which we realize our fluxbrane inflation model) do not have non-trivial one-cycles. Then, the phenomenological quantity $\alpha$ is determined microscopically by $\alpha_{\text{micro}}^{\text{KK}}$ which, for the above values for $n_s'$ and $\xi$ tends to be too small. For $n_s' = 0.007$, $\xi = 10^{-7}$, $\hat \xi = 100$, $\gamma = 10^{-2}$ and $W_0 = 1$ we find
\begin{center}
\begin{tabular}{|c|c|c|c|c|c|c|}\hline
 $\ccV$ &$\vd$ &$g_s$ &   $z$  & $x$  & $R^2$& $T_s$\\\hhline{|=|=|=|=|=|=|=|}
708.9 &2.11 &0.37 &   142  &$6.1 \cdot 10^{-4}$  & 2.37& 1.49  \\\hline
\end{tabular}

\begin{tabular}{|c|c|c|c|}\hline
  $\alpha$ &  $\alpha_{\text{micro}}^{\text{KK}}$  &$\alpha_{\ln}$ & $m_{3/2} / \mkk$\\\hhline{|=|=|=|=|}
 $1.0 \cdot 10^{-6}$ &  $3.3\cdot 10^{-8}$ & $7.1\cdot 10^{-9}$& 0.39\\\hline
\end{tabular}
\end{center}
Larger values for $n_s'$ and $\xi$ increase the ratio $\alpha / \alpha_{\text{micro}}^{\text{KK}}$. A lower value for $n_s'$ leads to $R^2 <1$, whereas a lower $\xi$ leads to an increase in $m_{3/2} / \mkk$, both of which is undesirable. This leaves us with a considerable difference in size between $\alpha$ and $\alpha_{\text{micro}}^{\text{KK}}$ in the above table. However, this discrepancy can easily be resolved by having a larger number of $N^{\text{KK}}$ or $C^{\text{KK}}$ in \eqref{eq:GenLVSLoopCorr}. For example, the function $E_2 (U)$ gives $E_2 (i) \simeq 6$ and grows for larger $\Im (U)$.

\begin{figure}[t]
	\centering
  \begin{overpic}[width=0.6\textwidth]{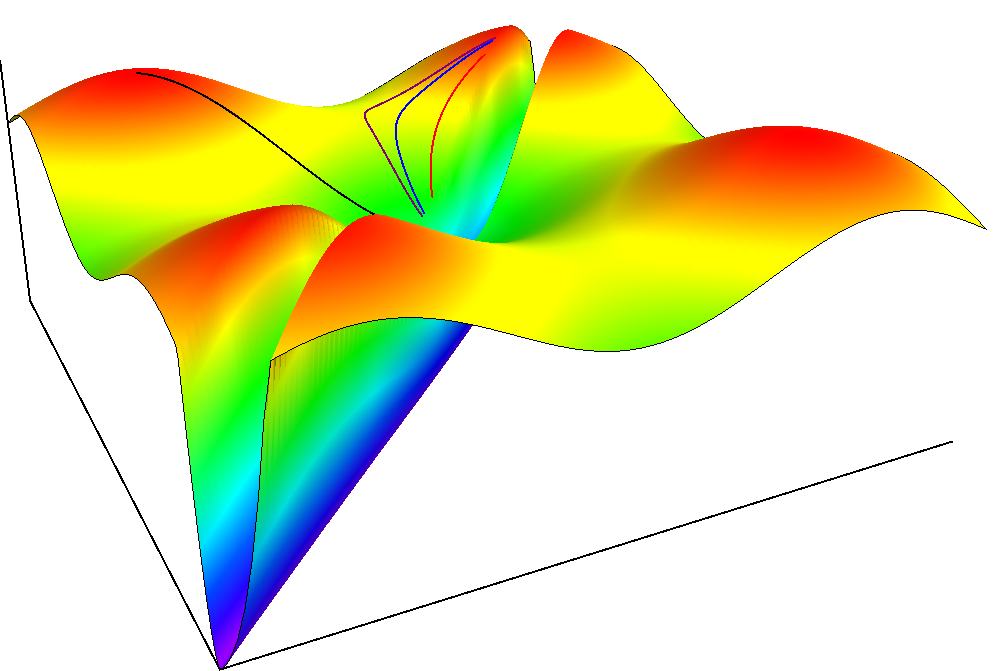}
 \put (5,18) {$\varphi_2$} \put (62,8) {$\varphi_1$} \put (-19,50) {$V\left(\varphi_1 , \varphi_2\right)$}
\end{overpic}
	\caption{Various trajectories in the field space of the two brane coordinates.}
	\label{fig:Multifield}
\end{figure}

Let us make some additional remarks:
\begin{itemize}
 \item As detailed in \secref{sec:modstab}, $x$ has to be tuned to a small value in order to have $F$- and $D$-term of comparable size. The sum of $D$-term and (negative) $F$-term energy density in the minimum of the scalar potential after inflation determines the cosmological constant in our $D$-term uplifting scenario. Thus, the tuning of $x$ in our model constitutes part of the tuning of the cosmological constant to the famous value $10^{-120}$. As mentioned already at the end of \secref{sec:modstab}, $D$-terms can be computed very easily in a given model and thus the tuning of $x$ can in principle be analyzed very explicitly. If one furthermore requires $\gamma \ll 1 $ for stability reasons as in \cite{Hebecker:2012aw} (see also footnote \ref{foot:SizeGamma}), this amounts to additional fine tuning. In the context of a $D$-term inflation model with an unwarped $D$-term uplift and moduli stabilization in terms of the Large Volume Scenario we do not see any way to circumvent this tuning.
  
 \item If one could relax the cosmic string bound, one has more freedom. Using $\xi = 10^{-6}$, $n_s' = 0.01$, $\hat \xi = 30$, $W_0 = 10$, and $\gamma =1$, one finds
\begin{center}
\begin{tabular}{|c|c|c|c|c|c|c|}\hline
 $\ccV$ &$\vd$ &$g_s$ &   $z$ & $x$ &  $R^2$ & $T_s$   \\\hhline{|=|=|=|=|=|=|=|}
3039 &3.54 & 0.18&   1.67  & $2.0 \cdot 10^{-2}$  & 514& 1.36 \\\hline
\end{tabular}

\begin{tabular}{|c|c|c|c|}\hline
 $\alpha$ &  $\alpha_{\text{micro}}^{\text{W}}$ & $\alpha_{\ln}$  & $m_{3/2} / \mkk$ \\\hhline{|=|=|=|=|}
 $5.0 \cdot 10^{-5}$  & $4.0\cdot 10^{-5}$& $4.6\cdot 10^{-6}$ & $2.3\cdot 10^{-1}$\\\hline
\end{tabular} 
\end{center}

\item Alternatively, if one drops the assumption that the inflaton is responsible for the generation of CMB perturbations, for $\tilde \zeta = 5.1\cdot 10^{-5}$, $n_s' = 0.014$, $W_0 = \gamma= 1$, $\hat \xi = 100$ and $\xi = 10^{-7}$ one finds that
\begin{center}
\begin{tabular}{|c|c|c|c|c|c|c|}\hline
   $\ccV$ &$\vd$ &$g_s$ & $z$ &  $x$ & $R^2$ &$T_s$  \\\hhline{|=|=|=|=|=|=|=|}
  4304 &5.16 &0.31 & 5.80 & $2.4 \cdot 10^{-3}$ &  144& 1.78 \\\hline
\end{tabular}

\begin{tabular}{|c|c|c|c|}\hline
  $\alpha$ &  $\alpha_{\text{micro}}^{\text{W}}$ & $\alpha_{\ln}$  & $m_{3/2} / \mkk$ \\\hhline{|=|=|=|=|}
$2.9 \cdot 10^{-5}$ &  $3.1\cdot 10^{-5}$& $4.4\cdot 10^{-8}$ & $4.7\cdot 10^{-2}$\\\hline
\end{tabular}
\end{center}
Here, we have assumed for simplicity that still $n_s=0.9603$.
\end{itemize}

\subsection{Alternative Trajectories}

\begin{figure}
\centering
 \begin{overpic}[width=0.5\textwidth,tics=10]{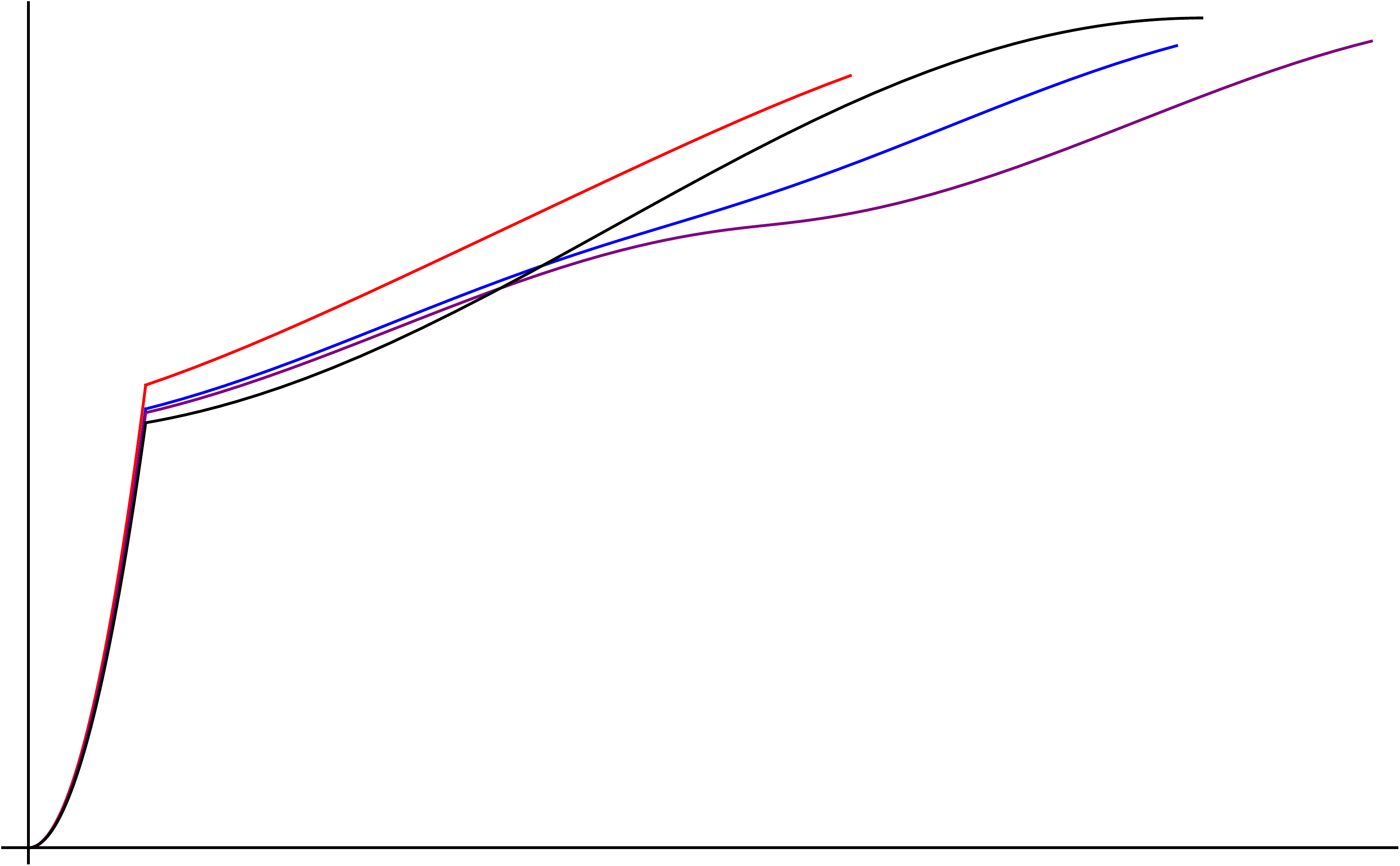}
 \put (102,1) {$\varphi$} \put (-1,65) {$V\left(\varphi\right)$}
\end{overpic}
\caption{One-dimensional plot of the potential along the trajectories drawn in figure~\ref{fig:Multifield}.}\label{fig:Multifield1D}
\end{figure}

We would like to emphasize that the phenomenological analysis performed in the previous subsections was assuming a cosine-shaped potential. In view of figure~\ref{genPot} it is obvious that fluxbrane inflation really is a multifield inflation model and various types of trajectories with different shapes are conceivable. This is illustrated in figure~\ref{fig:Multifield}. The projection of different trajectories onto an effectively one-dimensional inflaton potential is shown in figure~\ref{fig:Multifield1D}. We focused on a `generic' case among those possibilities. However, other realizations are likely to lead to a rather different phenomenological discussion. We will not pursue these ideas any further in the present paper.

Since the field space is two-dimensional, isocurvature modes can be important \cite{Linde:1985yf}. An analysis of those modes is beyond the scope of this paper.

\section{Statistics of Flux Vacua at Large Complex Structure}
\label{statistics}
Given a certain region in the moduli space of complex structures on a Calabi-Yau orientifold $X$, how many supersymmetric flux vacua are there in the region of large complex structure, consistent with the D3-tadpole condition
\begin{equation}
\label{tadpole1}
L:= \int F_3 \wedge H_3 = -\frac{1}{\tau - \ov \tau} \int \ov G_3 \wedge G_3 \leq L_* ?
\end{equation}
Here, $L_* = \chi(Y) / 24$ is related to the Euler characteristic of the corresponding F-theory fourfold $Y$.
In \cite{Denef:2004ze} a counting formula for the number of supersymmetric vacua was derived. In a Calabi-Yau orientifold $X$ with $b_3$ three-cycles and $n=\frac{b_3}{2}-1$ complex structure moduli the number of vacua with flux contribution obeying \eqref{tadpole1} is estimated by
\begin{equation}
\label{countingformula}
\mathcal{N}_{\text{SUSY}}(L\leq L_*)=\frac{(2\pi L_*)^{b_3}}{b_3!}|\det{\eta}|^{-1/2}\int\limits_{\mathcal{R}\subset\mathcal{M}}{\d}^2\tau {\d}^{2n}z (\det{g}) \rho(z), 
\end{equation}
where $\eta$ is the four-cycle intersection form of the fourfold $Y$, $\mathcal{R}$ is a region in the complex structure and dilaton moduli space $\mathcal{M}$, $g$ denotes the metric on the moduli space and $\rho(z)$ is the density of supersymmetric vacua per unit volume in that moduli space. In particular, the density is independent of the axio-dilaton \cite{Denef:2004ze}. In the above counting the quantization of fluxes is neglected which is valid, as the authors of \cite{Denef:2004ze} argue, as long the radius $r$ of the region $\mathcal{R}$ (measured with the metric on the moduli space) fulfills $r^2 > \frac{2b_3}{L_*}$ (recall that the number of distinct fluxes is $2b_3$). An explicit formula for $\rho(z)$ is given in \cite{Denef:2004ze}. Whilst it is hard to evaluate analytically for $n>1$, it in principle allows to study the statistics of string vacua numerically.

An instructive example which admits an explicit computation \cite{Denef:2004ze} is the mirror quintic \cite{Candelas:1990qd}. It is defined as a $\mathds{Z}_5^3$-quotient of the hypersurface $x_1^2+x_2^2+x_3^2+x_4^2+x_5^2=5 \psi x_1x_1x_3x_4x_5$ in $\mathds{CP}^4$ and has one complex structure parameter $\psi$ whose fundamental domain is the wedge $-\frac{\pi}{5} < \arg \psi < \frac{\pi}{5}$. The boundaries of the wedge are identified. The parameter $\psi$ is related to the usual $z$ modulus via $5\psi=e^{-2\pi i z/5}$ and hence the fundamental domain of $z$ is $-\frac{1}{2} < \operatorname{Re}z < \frac{1}{2}$. There are two special loci in the moduli space where the mirror quintic becomes singular: the conifold point at $\psi=1$ and the large complex structure point at $\psi = \infty$. In terms of $z$ the conifold point is located at $z=i 5 \ln{5}/2\pi$ while the large complex structure point is at $\operatorname{Im}z = \infty$. At both points the metric on the moduli space degenerates. At large complex 
structure, $\operatorname{Im}z \gg 1$, the metric on the 
moduli space is given by $g_{z \ov z}= -\frac{3}{(z - \ov z)^2}$ \cite{Candelas:1990qd}. Written in terms of real coordinates $x=\operatorname{Re}z$ and $y=\operatorname{Im}z$ the line element reads ${\d}s^2=\frac{3}{4 y^2}({\d}x^2 +{\d}y^2)$. Thus, the region around $\operatorname{Im}z = \infty$ is a throat: while the large complex structure point is at infinite distance, the volume of a region around this point is finite. For $n=1$ the density function $\rho(z)$ universally becomes constant at large complex structure, $\rho(z)=\frac{2}{\pi^2}$ \cite{Denef:2004ze}.

We now calculate the integral over the moduli space of $\tau$ and $z$, cutting off the $z$-integral at $\operatorname{Im} z = y_*$. The integral over the fundamental domain of $\tau$ gives a factor of $\frac{\pi}{12}$. The result is 
\begin{equation}
 \int_{\mathcal{M}}{\d}^2 \tau {\d}^2z \det{g} \rho(z) = \frac{\pi}{12} \frac{2}{\pi^2}  \! \! \! \! \! \! \! \! \! \int \limits_{\substack{\operatorname{Im}z > y_*\\ -1/2 \, < \, \operatorname{Re}z \, < \, 1/2}} \! \! \! \! \! \!\!\!\!  \frac{3}{4 (\operatorname{Im} z)^2}{\d}^2z= \frac{1}{8\pi y_*}.
\end{equation}
Inserted into the counting formula \eqref{countingformula} we find 
\begin{equation}
 \mathcal{N}_{\text{LCS}}=\frac{(2\pi L_*)^4}{4!}|\det{\eta}|^{-1}\frac{1}{8\pi y_*} = \frac{2 \pi^3}{4!} |\det{\eta}|^{-1} \frac{L_*^4}{y_*} \simeq 2.58 |\det{\eta}|^{-1} \frac{L_*^4}{y_*}.
\end{equation}
Comparing this to the total number of vacua, computed numerically in \cite{Denef:2004ze}
\begin{equation}
\mathcal{N}_{\text{total}}\simeq \frac{(2\pi L^*)^4}{4!} |\det{\eta}|^{-1} 5.46 \cdot 10^{-2} \simeq 3.55 |\det{\eta}|^{-1}  L_*^4, 
\end{equation}
leads to the conclusion that a fraction of $0.73 \frac{1}{y_*}$ vacua are located at $\operatorname{Im}z > y_*$ \cite{Denef:2004ze}. It seems that for orientifolds of the mirror quintic a relatively large fraction of all flux vacua are at large complex structure.

Generalizing this to higher dimensional complex structure moduli spaces we expect to encounter the scaling
\begin{equation}
\int \limits_{\operatorname{Im}z_i > y_*} {\d}^{2n}z \det{g} \rho \sim \frac{1}{y_*^n}.
\end{equation}
Indeed, expressed in terms of the volume of the mirror Type IIA compactification $\ccV^A = \kappa_{ijk}y^i y^j y^k$ one can approximate \cite{Douglas:2004zg}
\begin{equation}
\label{countingestimate}
 \mathcal{N}_{\text{LCS}} \sim \frac{(2\pi L_*)^{b_3}}{b_3!} (\ccV^A_*)^{-n/3} \sim  \left(\frac{2\pi L_*}{b_3}\right)^{b_3} (\ccV^A_*)^{-b_3/6},
\end{equation}
where we have used that $n \approx \frac{b_3}{2}$ for large $n$ and Stirling's approximation $b_3! \approx b_3^{b_3}$. Furthermore, $\ccV^A_* := \ccV^A(y^i = y_*)$. This makes sense intuitively: the volume of a ball of fixed radius $\epsilon < 1$ in $\mathds{R}^n$ decreases with the number of dimensions $n$ as $\epsilon^n$. Accordingly, the largest value of $\ccV^A_*$ where we would still expect vacua to be present is of the order \cite{Douglas:2004zg}
\begin{equation}\label{eq:MaxVol}
 \ccV^A_* \sim \left(\frac{2 \pi L_*}{b_3}\right)^6 .
\end{equation}

Consider the Calabi-Yau orientifold on $\mathds{CP}^3$ which has $n=149$ and $L_*=972$ as discussed e.g.\ in \cite{Denef:2008wq}. Let us define for the mirror manifold $\ccV^A(1) := \ccV^A(y^i = 1)$. Then, $\ccV^A_* = y_*^3 \ccV^A(1)$. Thus, from \eqref{eq:MaxVol} we expect to find vacua at $\Im(z^i) \geq y_*$ for
\begin{equation}\label{eq:yStarMax}
 y_* \lesssim \left(\frac{2 \pi L_*}{K}\right)^2\ccV^A(1)^{-1/3} \simeq \frac{4\cdot 10^2}{\ccV^A(1)^{1/3}} .
\end{equation}
Provided the typical values of the intersection numbers are not too high it seems that in this model $\operatorname{Im}z_i$ can be made rather large. On the other hand, since $\ccV^A(1) = \frac{1}{3!}\sum_{i,j,k=1}^n \kappa_{ijk}$ the expression for $\ccV^A(1)$ contains $n^3 \simeq 3.3\cdot 10^6 $ terms, each of which is an integer (up to the overall factor of $1/3!$). Therefore, the size of $\ccV^A(1)$ is likely to exceed $10^6$, giving $y_* \lesssim 1$ from \eqref{eq:yStarMax}. This makes it unlikely to find vacua at large complex structure in this model. A similar conclusion was drawn in \cite{Denef:2004dm} for the $\mathcal{F}_{18}$-example with $b_3=190$ and $L_*=552$.

In summary, it seems difficult to find flux vacua at large complex structure $\Im (z^i) > y_* \gg 1$ in models with a large number of complex structure moduli. In view of \eqref{eq:yStarMax}, models with a smaller number of complex structure moduli are more promising in this respect, though it is not clear how large $L_*$ can be for those models. For the mirror quintic, which represents the extreme case with $n=1$, we found that a large fraction of flux vacua will be at $\Im (z^i) > y_* > 1$. It would be interesting to know $L_*$ in an orientifold version of this example.

However, ensuring the existence of flux vacua at large complex structure is not enough: in order to be able to tune the cosmological constant with a precision of $10^{-120}$ along the lines of \cite{Bousso:2000xa}, a large number of string vacua (typically $\mathcal{N}>10^{120}$) is needed. A way to realize such a number in our model could be not to go into the `complete' large complex structure limit $\operatorname{Im}z^i \gg 1\ \forall i$, but only demand this for one or a few $z^i$. In the intuitive SYZ-picture discussed in \secref{syzformulation} one can see that this has a chance to work: Loosely speaking, for every Wilson line associated with a one-cycle $\gamma$ in the $T^3$-fiber on the Type IIA side there are corresponding K\"ahler moduli $t_A^j$ controlling the size of the two-cycles which suppress the instanton corrections. Accordingly, for every D7-brane position $c^i$ 
on the Type IIB side there are complex structure moduli $z^j$ which ensure the flatness of $\operatorname{Re}c^i$ if $\operatorname{Im}z^j$ is made large. For our purposes it would suffice to make only those $\operatorname{Im}z^j$ large which suppress the corrections to the D7-brane pair whose distance is associated to the inflaton. Obviously there should be many more vacua of this `partial large complex structure' type than vacua which have $\operatorname{Im}z^i \gg 1\ \forall i$. 

Finally, we should mention that we have neglected the huge `open-string landscape' \cite{Gomis:2005wc} descending from the large number of different supersymmetric brane configurations and brane fluxes (e.g.\ two-form fluxes on D7-branes) in our discussion. A generalization of the results of \cite{Denef:2004ze} to F-theory vacua can be found in \cite{Denef:2008wq}. The prefactor in front of the density integral in the F-theory case is
\begin{equation*}
 \frac{(2\pi L_*)^{b/2}}{\left(\frac{b}{2} \right) !},
\end{equation*}
where $b$ is the number of four-form fluxes with one leg along the fiber (these are the admissible four-form fluxes in F-theory that do not break Lorentz-invariance of the four dimensional effective theory). For the F-theory uplift of the orientifold on $\mathds{CP}^3$ discussed before we have $L_*=972$ and $b=23320$ \cite{Denef:2008wq}, so this gives a number of the order  
\begin{equation*}
 \frac{(2\pi L_*)^{b/2}}{\left(\frac{b}{2} \right) !} \sim 10^{1787}.
\end{equation*}
This has to be contrasted to $10^{521}$ in the orientifold limit, neglecting brane fluxes.

\section{Conclusions}
\label{Conclusions}
The purpose of this paper was twofold: First, we analyzed in general several features of the D7-brane moduli space, such as the shift symmetry at large complex structure, the appearance of the D7-brane position modulus in the superpotential, and the persistence of the extended no-scale structure in case one component of a complex scalar field remains massless. These features are  crucial ingredients in two otherwise very different realizations of slow-roll inflation in string theory, using D7-brane position moduli as the inflaton. Second, we set out to develop a consistent overall picture of the parameter regime in which one of these models, namely fluxbrane inflation \cite{Hebecker:2011hk,Hebecker:2012aw}, is viable.
A similar parametric analysis of the large-field realization, dubbed `D7-brane chaotic inflation', was performed already in \cite{Hebecker:2014eua}, referring to some results obtained in the present paper.

In the model of D7-brane chaotic inflation the inflaton potential is generated at tree-level via an explicit appearance of the brane modulus in the superpotential due to a proper flux choice. This flux leads to a monodromy and inflation occurs as the D7-brane circles around a certain direction in moduli space which, in the absence of flux, would be periodic. The coefficients of the brane modulus in the superpotential are tuned small, such that the corresponding $F$-terms, which break supersymmetry spontaneously during inflation, do not interfere with moduli stabilization.

On the other hand, fluxbrane inflation is the attempt to realize non-fine-tuned hybrid natural inflation in string theory. The inflationary energy density in this model is due to a non-vanishing relative flux between two D7-branes, whose relative deformation is associated to the inflaton. As such, fluxbrane inflation combines several appealing features: Similarly to the large-field chaotic inflation model discussed above, a shift symmetry is used to protect the inflaton from the supergravity $\eta$-problem. The appearance of the brane moduli in the superpotential is avoided by an appropriate flux choice. Thus, at leading order, the potential for the inflaton is exactly flat. The shift symmetry is broken by non-perturbative corrections to the K\"ahler potential of the Calabi-Yau (equivalently, corrections to the periods of the F-theory fourfold) and loop corrections from an open-string exchange between the two branes. These corrections thus induce a potential for the inflaton, which is the starting point for 
a 
phenomenological analysis. The non-perturbative corrections to the K\"ahler 
potential are exponentially suppressed at `large complex structure' and therefore expected to be small.

Both models require a detailed understanding of the D7-brane position moduli space, which was therefore the main theme of the present paper. Before we started to analyze this moduli space in generality, in \secref{Flat Directions} we considered the familiar example of a compactification of F-theory on $\tK \times K3$ which, in the orientifold limit, reduces to Type IIB string theory compactified on $\tK\times T^2/\mathds{Z}_2$. In this example the K\"ahler potential is explicitly known and has a shift-symmetric structure. For a specific flux choice we were able to show that all brane moduli are left as flat directions, and the ratio of the axio-dilaton and the complex structure of $T^2/\mathds{Z}_2$ is stabilized. Furthermore, we re-analyzed the `open-string landscape' \cite{Gomis:2005wc} and included `brane backreaction', i.e.\ the effect that the periods of the fourfold which, in the orientifold limit reduce to 
bulk periods, depend on the open-string moduli. This led to the interesting conclusion that, even at small string coupling, brane backreaction cannot be neglected in bulk moduli stabilization if the branes are deformed away from the O-plane.

In the following \secref{ModStabGen} we reviewed how, in general, the Type IIB K\"ahler and superpotential descend from the corresponding F-theory quantities. We saw explicitly how, even if the brane flux $\ccF$ is of Hodge-type $(1,1)$, the superpotential can develop a brane-moduli-dependence. This has to be taken into account when specifying the non-generic flux choice alluded to above.

The general story of the appearance of a shift symmetry in the K\"ahler potential was discussed in \secref{Mirror-Symmetry}. The issue can be approached either via mirror symmetry for the F-theory fourfold, or via the mirror-dual IIA description, in which the D7-brane position becomes a Wilson line on a D6-brane. Starting with the latter, the basic idea was that in Type IIA string theory, Wilson lines on D6-branes are perturbatively protected and enter the potential only non-perturbatively. The non-perturbative corrections are suppressed by volumes of holomorphic discs, which become large in the large volume limit on the IIA side. Thus under mirror symmetry we expect that the K\"ahler potential is corrected by instanton effects which are suppressed in the large complex structure limit on the Type IIB side, the mirror-dual equivalent to the large volume limit on the IIA side. On the other hand we recalled that, also for fourfolds, the K\"ahler potential for the complex structure moduli 
takes a shift-symmetric structure in the vicinity of the point of large complex structure. The Type IIB theory which arises in the weak coupling limit contains D7-brane position moduli which, in the F-theory description, are encoded in the complex structure of the fourfold. Therefore, we inferred that the shift-symmetric form of the F-theory complex structure K\"ahler potential is inherited by the D7-brane moduli K\"ahler potential. In summary, we found that the shift-symmetric inflaton potential is a good approximation in the limit of large complex structure, neglecting loop corrections.

However, loop corrections are certainly present and will lead to a non-vanishing potential for the D7-brane modulus. The relative importance of the induced corrections to the scalar potential with respect to the leading terms depends on the way in which moduli are stabilized. 
Working in Type IIB string theory, we were able to make use of the vast amount of literature dealing with moduli stabilization proposals developed for these compactifications over the past years. In particular, in \secref{String Loop Corrections} we parametrically estimated the size of the loop-induced corrections to the leading order flat potential in the context of the Large Volume Scenario. 
Interestingly, the `extended no-scale' structure analyzed in \cite{vonGersdorff:2005bf,Berg:2007wt,Cicoli:2007xp} survives the inclusion of an additional light degree of freedom in the effective theory, namely the inflaton. As a crucial consequence, the shift-symmetry-breaking loop corrections are subleading relative to the $\alpha'^3$-corrections used to stabilize the overall volume. This fact makes our D7-brane inflation scenarios viable.

The features of the D7-brane moduli space discussed so far are important for both realizations of slow-roll inflation, D7-brane chaotic inflation and fluxbrane inflation. However, in the phenomenological parts of the paper (\secref{d7d7} and \secref{sec:pheno}) we focused on the fluxbrane inflation scenario. In particular, we discussed how the model can be consistent with moduli stabilization.\footnote{A phenomenological discussion of the inflaton potential in D7-brane chaotic inflation, including effects induced by moduli stabilization, is already contained in \cite{Hebecker:2014eua}.}
This is achieved as follows: The axio-dilaton and complex structure moduli are stabilized at a high scale due to bulk fluxes which are chosen such that the superpotential does not depend on the brane moduli (at least not on the one which is associated to the inflaton). Owing to the shift-symmetric structure of the K\"ahler potential at large complex structure, the model does not suffer from the supergravity $\eta$-problem. $D$-terms are used to stabilize the relative sizes of four-cycles, such that only the overall volume and some small exceptional four-cycle is left unfixed. The latter two moduli are fixed in terms of the usual Large Volume Scenario. The resulting AdS minimum is uplifted by means of a SUSY breaking $D$-term which is tuned to a small value by an appropriate stabilization of the relative four-cycle volumes. Flux on the same cycle (but for a different U(1) gauge field) is responsible for the inflationary uplift to de Sitter. This is precisely 
along the 
lines of 
section 2 of \cite{Hebecker:2012aw}, only that the flux in the present paper is generically not along the Poincar\'{e} dual of an effective curve, which leaves more room for tuning the $D$-term to small values without leaving the geometric regime.

Lacking an explicit knowledge of the shift-symmetry-breaking loop corrections to the K\"ahler potential, we quantified their parametric size in the scalar potential, keeping track of $2\pi$-factors wherever it was possible. The resulting potential was then analyzed phenomenologically in \secref{sec:pheno}. We translated the phenomenological constraints imposed by the recent Planck measurement \cite{Ade:2013uln,Ade:2013xla} into constraints on generic parameters which appear in the fluxbrane inflation model. These are the overall volume of the compact manifold, the volume of the divisor on which we wrap the D7-branes, the string coupling constant, the size of the transverse direction along which the branes are separated during inflation, the $D$-term, the tree-level superpotential $W_0$ after complex structure and axio-dilaton stabilization, and the Euler characteristic of the threefold.

The relative size $\alpha$ of the loop-induced corrections to the potential with respect to the constant was then computed purely phenomenologically and in terms of the parameters of the stringy embedding. We found that, by a suitable choice of model parameters, the two can be brought in good agreement, in particular in the case where no corrections due to winding modes are present. The absence of those corrections can be achieved in cases where the self intersection of the D7-brane divisor is either empty or contains no non-contractible one-cycle. Furthermore, even in the presence of corrections due to winding modes the discrepancy between the phenomenologically computed $\alpha$ and the one obtained microscopically is rather small. Thus, a given model which features such winding-mode corrections may well reproduce the correct size of $\alpha$ due to the appearance of $\ccO(1)$-factors which were neglected in our investigation. We were able to fit the correct value of the spectral index, the amplitude of 
curvature 
perturbations, and the number of $e$-foldings. Furthermore, the fluxbrane inflation model satisfies the current cosmic string bound and the running of the spectral index is small, $n_s' \lesssim  10^{-2}$. However, being a small-field inflation model, the tensor-to-scalar ratio is tiny, typically $r \lesssim 2.6\cdot 10^{-5}$.

Finally, in \secref{statistics} we commented on the probability of finding flux vacua with the complex structure moduli stabilized at large values, i.e.\ vacua in which the non-perturbative corrections to the K\"ahler potential are suppressed. While the number of vacua in the large complex structure region is suppressed with an inverse power (proportional to the number of complex structure moduli) of the mirror-dual IIA volume, we found that there are expected to be enough vacua in the desired region of the moduli space, leaving room to also tune the cosmological constant. This is particularly true in a `partial large complex structure' limit, in which only those moduli are stabilized at large values which are needed in order to suppress the inflaton dependence in the K\"ahler potential.

To summarize, we have discussed the quantum-corrected moduli space of D7-brane positions, having in mind two rather distinct realizations of slow-roll inflation using these moduli. Furthermore, we arrived at a consistent overall picture of fluxbrane inflation, with a detailed parametric understanding of the various terms in the scalar potential. The next logical step towards finding explicit models would be to search for Calabi-Yau manifolds on which the phenomenologically required features, such as 
the tuning of the coefficients multiplying the inflaton in the superpotential or the detailed moduli stabilization program, outlined in \secref{sec:pheno}, can be attained. For the fluxbrane inflation scenario, this includes the uplift to de Sitter via $D$-terms. Furthermore, it would be instructive to see the exponential suppression of the inflaton-dependent non-perturbative corrections to the K\"ahler potential in explicit solutions to Picard-Fuchs equations and to carry out the complex structure moduli 
stabilization procedure in detail for such an example. Finally, in the long run, a better understanding of loop corrections to the K\"ahler potential for compactifications beyond orientifolds of tori would be desirable and would allow for a more detailed understanding of the inflaton potential. However, given the techniques currently available, the 
latter seems beyond reach.

If the recent measurement of B-modes by the BICEP2 collaboration \cite{Ade:2014xna} and the attribution of the effect to primordial gravitational waves, generated during an epoch of slow-roll inflation, were correct, then small-field inflation in general and the fluxbrane inflation model in particular would be ruled out. However, as stated in the introduction, we think that it is too early to be certain that the effects measured by the BICEP2 team are due to inflaton perturbations. On the other hand, confirmation of the BICEP2 results would certainly ignite further investigation of large-field inflation models in string theory. We think that the D7-brane chaotic inflation model \cite{Hebecker:2014eua} is a promising candidate in this arena, as it has the prospect of being realized in an explicit and controlled string setting.

\section*{Acknowledgments}
In the course of this work we profited from discussions with Alexander Westphal, Lukas Witkowski, Stefan Sj\"ors, Hans Jockers, Thomas Grimm, and Dominik Neuenfeld. This work was supported by the
DFG Transregional Collaborative Research Centre TRR 33 ``The Dark Universe'', the DFG cluster of excellence ``Origin and Structure of the Universe'', and by the ERC Advanced Grant ``Strings and Gravity'' (Grant.No.\ 32004). SK acknowledges support by the Studienstiftung des deutschen Volkes.

\appendix

\section*{Appendix}

\section{The {\boldmath $K3$} Surface}\label{k3surface}
In this appendix we collect several facts about the $K3$ surface. For a comprehensive review on $K3$ and for references to the mathematical literature on it we refer to \cite{Aspinwall:1996mn}. We closely follow the notation of \cite{Braun:2008pz,Braun:2008ua}.
\subsection{Generalities}
Apart from $T^4$, the $K3$ surface is the only compact Calabi-Yau twofold. The Hodge diamond of a $K3$ manifold is given by:
\begin{equation}
  \begin{array}{ccccc}
    & & 1& & \\ & 0& & 0& \\  1& & 20 & & 1 \\
    &0& &0& \\ & & 1& &
  \end{array}
\end{equation}
The holonomy group of $K3$ is SU(2) = Sp(1), additionally giving it the structure of a hyperk\"ahler manifold. 

Contrary to the familiar case of moduli spaces of Calabi-Yau threefolds, the variation of the holomorphic two-form $\Omega_2$ and the variation of the real K\"ahler two-form $J_{K3}$ lie in the same Dolbeault cohomology group $H^{1,1}(K3)$. There is, nevertheless, a nice geometrical description of the moduli space and we will review the most important facts for our needs here: As usual, coordinates on the moduli space are given by period integrals of $\Omega_2$ and $J_{K3}$ over integral two-cycles or, equivalently, via Poincar\'{e}-duality, by a decomposition of $\Omega_2$ and $J_{K3}$ into integral two-forms:
\begin{equation}
\Omega_2 = \pi^i \eta_i , \qquad J_{K3} = t^i \eta_i , \qquad \eta_i \in H^2(K3,\mathds{Z}) , \qquad i=1,\dots,b^2(K3) ,
\end{equation}
where $\pi^i = \int_{\gamma_i} \Omega_2$, $t^i = \int_{\gamma_i} J_{K3}$ and $\gamma_i$ are two-cycles which are the Poincar\'{e} duals of $\eta_i$. 
The intersection numbers of those two-cycles are given by
\begin{equation}
q_{ij}=\int_{K3} \eta_i \wedge \eta_j .
\end{equation}

The integral two-cycles on $K3$ span a self-dual lattice $\Gamma_{3,19}$ of signature $(3,19)$, where $q$ is a $22\times 22$ matrix of the form
\begin{equation} \label{metricK3}
U \oplus U \oplus U \oplus - E_8 \oplus -E_8 .
\end{equation}
$E_8$ denotes the positive definite Cartan matrix of $E_8$ and
\begin{equation}
U =  \left( \begin{array}{cc} 0 & 1 \\ 1 & 0  \\ \end{array} \right).
\end{equation}
The expansion of an arbitrary vector in the corresponding basis of two-forms can be written as
\begin{equation}
D = p^i e^i + p_j e_j + q_I E_I  ,
\end{equation}
where $i,j = 1,2,3$ and $I,J = 1, \dots, 16$. The $E_I$ can be chosen such that the only non-vanishing inner products among the basis vectors are
\begin{equation}
\label{scapr}
E_I \cdot E_J = - \delta_{IJ} , \qquad e^i \cdot e_j = \delta^i_j  .
\end{equation}

The two-forms $\Omega_2$ and $J_{K3}$ are subject to the following conditions:
\begin{equation} \label{orthK3}
\Omega_2 \cdot \ov{ \Omega}_2 > 0  ,\qquad J_{K3} \cdot J_{K3} > 0  ,\qquad J_{K3} \cdot \Omega_2 = 0  ,\qquad \Omega_2 \cdot \Omega_2 = 0  .
\end{equation}
The first two inequalities ensure that the volume of $K3$ is positive, the other two follow from Hodge decomposition. If we parametrize $\Omega_2$ and $ J_{K3}$ by three real two-forms $\omega_a$, $a=1,2,3$, such that $\Omega_2 =\omega_1 + i \omega_2$ and $ J_{K3}=\sqrt{2v} \, \omega_3$ (where $v$ is the overall volume of $K3$), the conditions given in (\ref{orthK3}) are equivalent to 
\begin{equation}
\omega_a \cdot \omega_b = 0 \ \ \ (a\neq b)  , \qquad \omega_a^2 >0 \ \ \ (a=1,2,3). 
\end{equation}
Thus, the $\omega_a$ are spacelike and span a spacelike three-plane $\Sigma \subset \Gamma_{3,19}$. They can be normalized according to $\omega_a \cdot \omega_b = \delta_{ab}$. If we picture the lattice $\Gamma_{3,19}$ being embedded into $\mathds{R}^{3,19}$, variation of the complex and K\"ahler structure can be described by a rotation of $\Sigma$ within $\mathds{R}^{3,19}$.
We refer to \cite{Aspinwall:1996mn,Braun:2010ff} for further details.

\subsection{From Cycles to Brane Positions}
Two-cycles on an elliptically fibered $K3$ with vanishing volume and their intersection properties classify ADE singularities \cite{Braun:2010ff}. In \cite{Braun:2008ua} it is shown how one can explicitly construct two-cycles on an elliptically fibered $K3$, whose lengths measure the positions of D7-branes relative to the positions of the O7-planes. Since we will be interested in $K3$ at the SO(8)$^4$ point, we will review the corresponding choice of cycles in \cite{Braun:2008ua} in the following table. Each block of the table (denoted by $\mathcal{A},\ldots,\mathcal{D}$) corresponds to one of the four O7-planes. Each block consists of four different cycles (denoted by $1,\ldots,4$) corresponding to a combination of the positions of the four D7-branes attached to each O7-plane.
\vspace{.5cm}

\begin{center}
\begin{tabular}[h]{c|c|c|c|c}
& $\ccA$&$\ccB$&$\ccC$&$\ccD$\\\hline
1&$E_{7}-E_{8}$&$-E_{15}+E_{16}$&$-e_2-E_{1}+E_{2}$&$e_2+E_{9}-E_{10}$\\
2&$E_{6}-E_{7}$&$-E_{14}+E_{15}$ &$-E_{2}+E_3$&$E_{10}-E_{11}$\\
3&$-e_1-E_{5}-E_{6}$&$e_1+E_{13}+E_{14}$&$-E_{3}+E_{4}$&$E_{11}-E_{12}$\\
4&$E_{5}-E_{6}$&$-E_{13}+E_{14}$&$-E_{3}-E_{4}$&$E_{11}+E_{12}$       
\end{tabular} 
\vspace{-1.7cm}
\begin{equation}
\label{AtoD}
\end{equation}
\end{center}
\vspace{1.5cm}
Using (\ref{scapr}) one can easily see that the intersection matrix of the four two-cycles of each block is the $D_4$ matrix, thus the choice of cycles indeed describes the SO(8)$^4$ point.

Let us now focus on one block, say, block $\mathcal{C}$. In \cite{Braun:2008ua} it is shown that one can choose a basis of two-cycles such that the size of those basis cycles measures the position of the D7-branes as they are pulled off the O7-plane corresponding to block $\mathcal{C}$ (the other blocks can be treated analogously). The corresponding change of basis is specified by
\begin{center}
\begin{tabular}[h]{c|c}
D7-brane position & cycle \\ \hline
$C_1$& $\ccC_1+\ccC_2+\frac{1}{2}(\ccC_3+\ccC_4)$ \\ 
$C_2$& $\ccC_2+\frac{1}{2}(\ccC_3+\ccC_4)$ \\ 
$C_3$& $\frac{1}{2}(\ccC_3+\ccC_4)$ \\ 
$C_4$& $\frac{1}{2}(\ccC_4-\ccC_3)$
\end{tabular}
\vspace{-1.6cm}
\begin{equation}
\label{branepos}
\end{equation}
\end{center}
\vspace{1cm}
It is not hard to see (and will prove to be useful) that the intersection matrix of the cycles defined in table \eqref{branepos} is $- \mathds{1}_4 $. I.e.\ writing down \eqref{branepos} amounts to changing coordinates to an orthogonal basis with basis elements $\tE_I$. The complex structure $\Omega_2 = \omega_1 + i\omega_2$ can now be parametrized through the complex structure modulus $S$ of the fiber, the complex structure modulus $U$ of the base and the 16 D7-brane positions $C^a$:
\begin{equation}\label{fullomegaK3}
\Omega_2 = \frac{1}{2}\left( \alpha + (C^2-SU)e_1 + S\beta + Ue_2 + 2 C_I \tE_I \right) ,
\end{equation}
where $C^2 = \sum_I C_I C_I$. The two-forms $\alpha$ and $\beta$ are defined as some linear combination of the $e_i$, the $e^i$ and the $E_I$, such that $\alpha \cdot e_1 = \beta \cdot e_2 = 2$, which leads to the intersection matrix \eqref{intersectionsk3}.

\section{Orientifold Limit of F-Theory}\label{senslimit}
In this appendix we briefly review some aspects about the orientifold limit in F-theory as described in \cite{Sen:1996vd} and refer the reader to \cite{Weigand:2010wm,Denef:2008wq} for a more general introduction to F-theory and its applications to string model building.

We will use the standard description of an elliptic curve, given as a complete intersection in $\mathds{CP}_{2,3,1}$, in terms of three homogeneous coordinates $(x,y,z) \in \mathds{C}^3$ and the equivalence relation
\begin{equation}
(x,y,z) \propto (\lambda^2 x,\lambda^3 y,\lambda z), \qquad \lambda \in \mathds{C}^*.
\end{equation}
The polynomial whose intersection with $\mathds{CP}_{2,3,1}$ defines the elliptic curve reads, in Weierstrass form,
\begin{equation}\label{algtorus}
W(x,y,z) = y^2 - x^3 + fxz^4 + gz^6 = 0 .
\end{equation}
The quantities $f$ and $g$ are, in a given coordinate patch with coordinates $u_i$, $i=1,\dots,3$ of the complex three-dimensional base, complex valued polynomials.
As a mathematical fact, singular points of the elliptic curve are roots of the discriminant $\Delta$:
\begin{equation}
\Delta = 27g^2 + 4f^3 = 0.
\end{equation}
The discriminant enters the SL$(2,\mathds{Z})$-invariant Jacobi $j$-function
\begin{equation} \label{jacobidelta}
j(\tau) = \frac{4(24f)^3}{\Delta}.
\end{equation}
This function makes contact between the torus described by $T^2=\mathds{C}/(\mathds{Z}+\tau\mathds{Z})$ and its algebraic description (\ref{algtorus}), as it relates the complex structure parameter $\tau$ to the polynomials $f$ and $g$:
\begin{equation}\label{jacobij}
j(\tau) = e^{-2 \pi i \tau} + 744 + \mathcal{O}(e^{2 \pi i \tau}).
\end{equation}
The quantity $\tau$ will be the axio-dilaton of the dual Type IIB theory. It undergoes monodromy upon circling around zeroes of the discriminant $\Delta$. The type of monodromy which one encounters at a given such locus determines the type of 7-brane which sits at this point. Those 7-branes are non-perturbative generalizations of D7-branes and O7-planes. We will now review how the latter arise in the orientifold limit of F-theory.

The regime where we trust our Type IIB effective field theory (10D supergravity) is where the axio-dilaton is constant and such that $g_s$ is perturbatively small everywhere in the base manifold. The following construction due to Sen \cite{Sen:1996vd} describes a realization of this limit in F-theory: Suppose that $f$ and $g$ are of the general form
\begin{equation}
f= -3h^2 + \epsilon \eta  , \qquad g= -2h^3 + \epsilon h \eta - \frac{\epsilon^2}{12}\chi ,
\end{equation}
where $\epsilon$ is a constant and $h,\eta,\chi$ are homogeneous polynomials of appropriate degrees in the base-coordinates. Taking $\epsilon$ as a small parameter one finds:
\begin{equation} \label{deltatausen}
\Delta = -9\epsilon^2 h^2 (\eta^2-h\chi) + \ldots \, , \qquad j(\tau) = \frac{24^4}{2}\frac{h^4}{\epsilon^2(\eta^2-h\chi)} + \ldots,
\end{equation}
where the ellipses denote terms which are higher order in $\epsilon$. Thus, in the limit of small $\epsilon$ we have $g_s \sim -\frac{1}{\ln |\epsilon|} \to 0$ except at the point where $h=0$. The singular points of the fibration are given by $h=0$ and $\eta^2=h\chi$. A closer look at the monodromies around these loci reveals that a locus where $h=0$ corresponds to the location of an O7-plane, whereas loci with $\eta^2=h\chi$ give D7-brane positions. The corresponding Calabi-Yau threefold on which the Type IIB theory is defined is then given by the double cover of the base of the elliptic fibration, branched over $h=0$ (i.e.\ branched over the location of the O7-planes).

Using (\ref{jacobij}) and (\ref{deltatausen}) the modular parameter of the fiber torus can be written as
\begin{align}
\tau &= \frac{i}{2 \pi} \ln\left(\frac{288}{\epsilon^2}\right) + \frac{i}{2 \pi}\ln\left(\frac{h^4}{\eta^2-h\chi}\right) + \mathcal{O}(\epsilon) \nonumber  \\
&=: \tau_0 + \frac{i}{2 \pi}\ln\left(\frac{P_{\text{O7}}}{P_{\text{D7}}}\right) + \mathcal{O}(\epsilon), 
\end{align}
where $\tau_0 := \frac{i}{2 \pi} \ln\left(\frac{288}{\epsilon^2}\right)$ is the constant part of the axio-dilaton and $\Im(\tau_0) = g_s^{-1}$. The quantities $P_{\text{O7}}:= h^4$ and $P_{\text{D7}}:= \eta^2-h\chi$ are polynomials which vanish at the locus of the O7-planes and D7-branes, respectively.

\bibliography{ref-inflation}  
\bibliographystyle{utphys}
\end{document}